%% file: main.tex
\documentclass[12pt]{article}
\usepackage{newtxtext,newtxmath,float,pgffor}
\usepackage{graphicx,hyperref,overpic}
\usepackage[section]{placeins}

\usepackage[letterpaper,margin=1in]{geometry}

\usepackage{longtable}

\makeatletter
\renewcommand{\fnum@figure}{\textbf{Fig. \thefigure}}
\renewcommand{\fnum@table}{\textbf{Table \thetable}}
\makeatother

\usepackage{url}

\newcommand{\revise}[1]{#1}

\def\scititle{
	AI-exposed jobs deteriorated before ChatGPT
}
\title{\bfseries \boldmath \scititle}

\author{
	M.~R.~Frank$^{1,2,3\ast}$,
	A.~Javadian~Sabet$^{2,4,5}$,
    L.~Simon$^{6}$,
	S.~H.~Bana$^{2,7}$,
    R.~Yu$^{8,9}$\and
    \small$^{1}$Department of Design and Innovation, Case Western Reserve University, Cleveland, OH 44106, USA.\\
	\small$^{2}$Digital Economy Lab, Stanford University, Stanford, CA 94305, USA.\\
    \small$^{3}$AI Economy Institute, Microsoft, Redmond, WA 98052, USA.\\
    \small$^{4}$Department of Informatics and Networked Systems, University of Pittsburgh, Pittsburgh, PA 15215, USA.\\
    \small$^{5}$Department of Sociology, University of Chicago, Chicago, IL 60637, USA.\\    
    \small$^{6}$Revelio Labs, New York, NY 10010, USA.\\
    \small$^{7}$Department of Management, London School of Economics, London, WC2A 2AE, UK.\\
    \small$^{8}$Teachers College, Columbia University, New York, NY, 10027, USA.\\
    \small$^{9}$Data Science Institute, Columbia University, New York, NY, 10027, USA\\
	\small$^\ast$Corresponding author. Email: mrfrank@case.edu
}

\begin{document} 

\maketitle

\begin{abstract} 
Public debate links worsening job prospects for AI-exposed occupations to the release of ChatGPT in late 2022.
Using monthly U.S. unemployment insurance records, we measure occupation- and location-specific unemployment risk and find that risk rose in AI-exposed occupations beginning in early 2022, months before ChatGPT was released. 
Analyzing millions of LinkedIn profiles, we show that graduate cohorts from 2021 onward entered AI-exposed jobs at lower rates than earlier cohorts, with gaps opening before late 2022. 
Finally, from millions of university syllabi, we find that graduates taking more LLM-exposed curricula had shorter job searches after ChatGPT and higher first-job pay when entering highly exposed occupations. 
\revise{Together, these results---pointing to forces pre-dating generative AI and to the ongoing value of LLM-relevant education---suggest that generative AI may not be the root cause of these labor market hardships and that the launch of ChatGPT is a faulty natural experiment for understanding AI's impact on jobs.}
\end{abstract}

\noindent
Many observers worry that college-educated workers are disproportionately exposed to task automation driven by generative AI~\cite{felten2023occupational,eloundou2024gpts} and thus face job displacement as these tools diffuse~\footnote{``Anthropic CEO warns AI could eliminate half of all entry-level white-collar jobs’’. \href{https://fortune.com/2025/05/28/anthropic-ceo-warning-ai-job-loss/}{Fortune Magazine} (May 28, 2025)}.
Timing is a key implication of this hypothesis: if deteriorating labor market outcomes in recent years were driven primarily by large language models (LLMs), then the sharpest deterioration should occur after the mass-market emergence of LLM applications. 
ChatGPT---launched in November 2022 and rapidly adopted by consumers~\cite{reuters2023chatgpt}---provides a salient inflection point for the broader diffusion of LLM services (e.g., including Claude, Copilot, and Gemini). 
And, consistent with employment concerns, early-career workers in LLM-exposed occupations experienced an employment decline~\cite{brynjolfsson2026canaries,lichtinger2025generative} and/or decreasing salaries~\cite{azar2025ai} while recent college graduates experience elevated unemployment~\cite{nyfed_college_labor_market_2024} during this period.
If LLMs are the cause of these poor outcomes, then perhaps teaching LLM-exposed skills fails to prepare students for the labor market.

\revise{Two features of the past four years complicate this timing logic.
First, LLM capability has improved continuously rather than in a single step.
Frontier models were already improving before ChatGPT's release, and the pace of improvement has increased since. 
The length of task that these models complete with 50\% reliability has doubled about every seven months since 2019, and about every three months since 2024~\cite{metr2026timehorizon}. Benchmark progress accelerated with the introduction of reasoning models~\cite{denain2026accelerated}, and general-purpose systems have since moved from single-turn assistance to longer-horizon agentic work~\cite{bengio2026safety}.
ChatGPT's launch was therefore a discontinuity in public awareness and consumer adoption~\cite{reuters2023chatgpt}---the fastest consumer-technology diffusion on record---rather than a discontinuity in the underlying rate of capability growth.
Second, hiring responds to employers' \emph{expectations} about future capability, not only to the tools available on a given date.
A randomized information experiment finds that employers systematically underestimate the automatability of white-collar work and revise those beliefs upward when shown evidence of AI capability, yet do not correspondingly change hiring plans, instead anticipating higher productivity, new complementary roles, and greater training investment~\cite{brull2025beliefs}.
Together these facts imply that no single release date cleanly separates ``before AI'' from ``after AI.''}

\revise{However, more recent public discourse~\footnote{
\revise{
``Did A.I. Take Your Job? Or Was Your Employer 'A.I.-Washing'?''. \href{https://www.nytimes.com/2026/02/01/business/layoffs-ai-washing.html}{New York Times} (February 1, 2026)
}}\footnote{
\revise{
'Spotting AI Washing: How Companies Overhype Artificial Intelligence''. \href{https://www.forbes.com/sites/bernardmarr/2024/04/25/spotting-ai-washing-how-companies-overhype-artificial-intelligence/}{Forbes} (April 25, 2024)
}} casts doubt on AI as the driver of job market woes and instead points to ``AI washing.'' 
This occurs when corporate leaders cite artificial intelligence as the primary reason for layoffs, restructuring, or hiring freezes when the actual drivers are financial, such as cost-cutting, pandemic-era overhiring, or poor performance.
Other research supports this with studies finding little impact from LLMs on workers' wages or hours worked~\cite{humlum2025large,johnston2026ai,gimbel2025evaluating,iscenko2026looking}.
This null result has persisted as adoption has grown: a second wave of Danish survey and administrative data finds near-universal chatbot use in supported workplaces but no measurable effect on earnings, with pre-existing wage differences between users and non-users predating ChatGPT~\cite{humlum2026still}; a U.S. survey linking occupation-level generative AI use to administrative data finds small \emph{positive} wage effects and no decline in job openings or employment in exposed occupations~\cite{hartley2026labor}; and firm- and industry-level analyses of job postings find no reduction in hiring where AI adoption is highest~\cite{liu2026adoption}.}
It's possible that other macroeconomic and sector-specific forces were already shifting in early 2022 contributing to weakening outcomes in the same AI-exposed occupations.
For example, monetary tightening accelerated through 2022–2023~\cite{FEDFUNDS2025} and may have impacted job stability~\cite{zens2020heterogeneous}.
Also, job postings for software developers declined markedly over 2022–2023~\cite{fred_software_dev_indeed} perhaps as a correction to increased hiring during pandemic years~\cite{mendezcarbajo2023_tech_hiring_bubble}. 
Instead of LLM diffusion, if such forces were central drivers, then deterioration could be detectable before ChatGPT's launch \revise{and worsening labor outcomes may be mis-attributed to AI}.

While research finds mixed evidence, public perception has largely been shaped by negative coverage in the news.
Even in the months after ChatGPT's launch, headlines warned of decreased white collar jobs in the near term~\footnote{``How ChatGPT Will Destabilize White-Collar Work.'' \href{https://www.theatlantic.com/ideas/archive/2023/01/chatgpt-ai-economy-automation-jobs/672767/}{The Atlantic} (January 20, 2023)}.
More recently, similar concerns persist with emphasis on programming jobs~\footnote{``The AI Threat for Coding Jobs Is Becoming Clearer.'' \href{https://www.bloomberg.com/news/newsletters/2025-05-18/the-ai-hiring-pause-is-here}{Bloomberg} (May 18, 2025)} and on the employment prospects of recent college graduates~\footnote{``Goodbye, \$165,000 Tech Jobs. Student Coders Seek Work at Chipotle.'' \href{https://www.nytimes.com/2025/08/10/technology/coding-ai-jobs-students.html}{New York Times} (August 10, 2025)}.
While these articles point to AI as a potential cause, they are informed mostly by subjective interviews and perspectives.
Efforts to settle the academic debate may additionally ground the public discussion and guide decisions by policymakers, educators, employers, and students moving forward.

To investigate AI's role in and the timing of these labor market outcomes, we integrate three large-scale datasets spanning administrative unemployment records, online career histories, and university curricula.
\revise{The unemployment records run monthly from January 2015 through October 2025, the career histories cover graduation cohorts from 2017 through 2024 followed for up to three years after degree completion, with job start dates observed through December 2024, and the syllabi cover courses offered through 2017 (see SI Sections~\ref{si:sec:uiRisk},~\ref{si:sec:jobDelay}, and~\ref{si:sec:aiedu}, respectively).}
First, using monthly unemployment insurance claims linked to occupations and states, we construct a monthly measure of unemployment risk and find that risk rose for LLM exposed occupations beginning in early 2022---months before ChatGPT.
Throughout, we define an occupation's AI or LLM exposure using existing exposure estimates~\cite{eloundou2024gpts} according to the share of its workplace tasks in the U.S. Department of Labor's O*NET database that could be affected by LLMs with simple interfaces and general training (\revise{see SI Section~\ref{si:sec:exposure} for this measure in detail and} SI Section~\ref{si:sec:altExposure} for analyses with alternative exposure scores).
Second, analyzing millions of LinkedIn profiles from Revelio Labs, we show that graduate cohorts from 2021 onward entered LLM exposed jobs at lower rates than earlier cohorts with these gaps opening months or even a year before ChatGPT's launch. 
Finally, using millions of university course syllabi~\cite{javadian2024course}, we measure graduates' educational exposure to LLM-related content and find that higher exposure predicts higher first-job salaries and shorter job searches after ChatGPT. 
Together, these results separate pre-existing labor-market weakening from later LLM diffusion and underscore the ongoing value of LLM-relevant skill development during higher education.
These findings warn against using ChatGPT's 2022 launch as a clean natural experiment and \revise{cast doubt on LLMs as the cause of the worsening outcomes that recent graduates and AI-exposed workers experienced through 2025.}

\subsection*{Unemployment risk for AI-exposed occupations}
\label{sec:uirisk}

We first focus on unemployment for the workers of AI exposed occupations that are most likely to experience change because of AI automation.
Because unemployment data are reported at the major occupation group level (two-digit codes in the Standard Occupation Classification (SOC) system), we measure each group's exposure by averaging the LLM exposure scores of its detailed occupations (six-digit SOC codes).
\revise{That is, a detailed occupation's exposure is the share of its O*NET workplace tasks that are exposed to LLM automation~\cite{eloundou2024gpts}, and a major group's exposure $llm_{soc}$ is the equal-weighted mean of the exposure scores of the detailed occupations nested within it.
By this measure construction and extraction (SOC 47) is the least exposed major group ($llm_{soc}=0.069$) and computer and mathematical occupations (SOC 15) the most exposed ($llm_{soc}=0.741$).
SI Section~\ref{si:sec:exposure} gives this aggregation in full and reports the exposure score of every detailed occupation (SI Table~\ref{si:tab:detailedSocExposure}) and of every major occupation group (SI Table~\ref{si:tab:majorSocExposure}).}
All other analyses use exposure at the six-digit level directly.
Next, we incorporate monthly administrative records from the ETA 203 report from the Employment and Training Administration (ETA) within the U.S. Department of Labor.
This report aggregates monthly data from each U.S. state's unemployment insurance program detailing the most recent major occupation group of each continuing claimant. 
After combining with annual occupation employment data from the BLS Occupational Employment and Wage Statistics (OEWS) program, we calculate an occupation's \emph{unemployment risk}~\cite{frank2025ai} \revise{which approximates} the probability of claiming unemployment in each state each month given a worker's occupation \revise{(see SI Section~\ref{si:sec:uiRisk} for a description of unemployment risk, SI Section~\ref{si:sec:etaCps} for a comparison between the ETA 203 report and the US Census Current Population Survey (CPS), and SI Section~\ref{si:sec:uieq} for the calculation of unemployment risk).}
\revise{In cases where data is not available for a state-occupation pair, that pair is not included in any aggregation of unemployment risk scores.}
\revise{Note that unemployment risk is meant to describe one of the worst career outcomes following a job loss (i.e., unemployment by occupation) but other negative outcomes may occur and be missed by unemployment risk (e.g., workers switch to lower paying jobs without claiming unemployment).}

\begin{figure}[p]
	\centering
	\includegraphics[width=\textwidth]{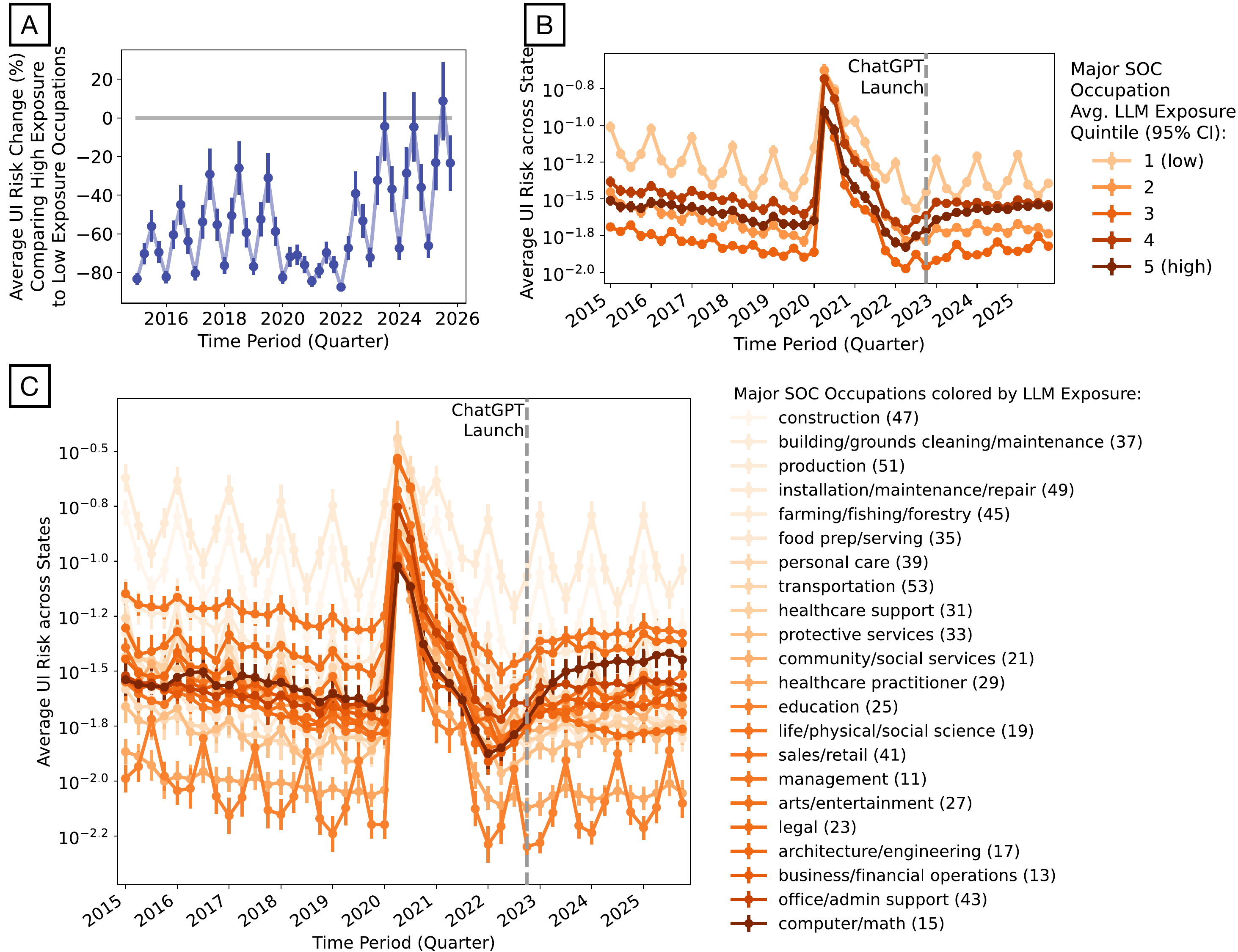}
	\caption{
    Unemployment \revise{insurance (UI)} risk for AI (LLM) exposed workers increased starting before the launch of ChatGPT.
    {\bf (A)} Percent difference in unemployment risk between high- and low-exposure occupations \revise{(see SI Section~\ref{si:sec:uieq} for calculation)}.
    {\bf (B)} Unemployment risk for occupations grouped into exposure quintiles.
    {\bf (C)} Unemployment risk by major occupation.
    Lines are colored by major occupation's AI exposure.
    We compute unemployment risk monthly across states and report quarterly series as the mean of monthly values across all states within each quarter (see SI Fig.~\ref{si:fig:feltenUiRisk}-\ref{si:fig:anthropicAugUiRisk} for similar plots with alternative AI exposure scores \revise{and SI Section~\ref{si:sec:uiMonth} for an analogous monthly time series}).
    \revise{Panels {\bf B} and {\bf C} use a logarithmic vertical axis; SI Table~\ref{si:tab:majorSocUiRisk} reports the same quantity for every major occupation group as a percentage.}
    \revise{Quarterly aggregation serves two purposes here: it keeps 22 overlapping series legible, and averaging the three monthly values within each quarter smooths month-to-month noise, so the series plotted here are already smoothed relative to the underlying monthly data. The unsmoothed monthly series are reported in SI Section~\ref{si:sec:uiMonth}, and the formal timing analysis in SI Section~\ref{si:sec:uiBreaks} is performed on those monthly series rather than on these quarterly averages.}
    }
	\label{fig:uiRisk} 
\end{figure}

\revise{For each quarter, we regress log unemployment risk on quarter fixed effects interacted with major-group LLM exposure $llm_{soc}$ and plot the implied percent change in unemployment risk when moving from an unexposed ($llm_{soc}=0$) to a fully exposed ($llm_{soc}=1$) major group (see Fig.~\ref{fig:uiRisk}A); the observed range runs from construction ($0.069$) to computer/math ($0.741$).}
Typically, unemployment risk is 20-80\% lower in LLM-exposed occupations (e.g., computer/mathematical) compared to less exposed occupations (e.g., construction). 
This gap widens during 2020–2021, consistent with pandemic-era differences in the ability to work remotely. 
Beginning in early 2022, however, the gap narrows sharply with some quarters of \revise{2023-2025} exhibiting no difference between high- and low-exposure groups.

Is this deterioration because exposed occupations have more unemployment risk after the launch of ChatGPT?
We group occupations into quintiles of LLM exposure and plot average unemployment risk over time (Fig.~\ref{fig:uiRisk}B). 
Across the period, lower-exposure quintiles exhibit higher levels and stronger seasonality \revise{(i.e., because of the seasonality in construction and building/grounds maintenance, see Fig~\ref{fig:uiRisk}C)}, while all quintiles show a pandemic spike in 2020 followed by a trough in early 2022. 
Importantly, unemployment risk in the most exposed quintiles begins rising after this early-2022 trough---well before ChatGPT's November 2022 launch---and, rather than accelerating, then stabilizes in the quarters following the launch.

Disaggregating by major occupation group (see Fig.\ref{fig:uiRisk}C \revise{and SI Section~\ref{si:sec:uiMajorSoc} for each of the 22 major groups plotted separately and without normalization}), many high-exposure occupations, including computer and math (SOC 15), office/admin support (43), business/financial operations (13), and architecture/engineering (17), exhibit large increases in unemployment risk during \revise{2022–2025}. 
However, these increases again begin prior to ChatGPT's launch and flatten or decrease afterward. 
Overall, the timing is inconsistent with a deterioration that begins only after ChatGPT and suggests that pre-existing macroeconomic and sectoral forces contributed materially to the early-2022 rise in unemployment risk.
\revise{Rather than impose a treatment date, we also let the data locate structural breaks in each exposure quintile's unemployment risk series and ask which candidate dates they recover (see SI Section~\ref{si:sec:uiBreaks}).
We further ask how much of the movement in unemployment risk is accounted for by macroeconomic conditions rather than by AI (see SI Section~\ref{si:sec:uiMacro}).
The narrowing of the exposure gap tracks the federal funds rate: the interaction between LLM exposure and the interest rate is positive and significant, and it is large enough that the gap implied at a $5\%$ policy rate is most of the way to the narrowing visible in Fig.~\ref{fig:uiRisk}A.
Once that interaction is in the model, the terms specific to LLM exposure after the launch of ChatGPT add less than $0.001$ to $R^2$ and are not significant.
Without the interest rate interaction, the post-launch exposure term is large and highly significant.
That is, dating treatment to November 2022 without controlling for the monetary tightening that began eight months earlier produces a large post-launch effect of LLM exposure.}

Another potential concern is that an earlier vintage of AI was impacting jobs before the launch of ChatGPT in ways that overlap with exposure to modern LLMs. 
For example, GitHub's Copilot became publicly available in June 2022, months before the launch of ChatGPT, and acted as an auto-complete tool for writing computer code.
Even if this tool saw widespread adoption among computer programmers (i.e., within computer/math occupation (SOC 15)), its use likely did not diffuse into the work of the other high-exposure occupations, such as office admin/support, architecture/engineering, and legal occupations, that exhibited a similar early rise in unemployment risk.
\revise{To the best of our knowledge, no other AI technology pre-dates the launch of ChatGPT, would be correlated with occupation exposure to LLMs, and had the potential to impact jobs across so many occupations; we therefore conclude that this early rise in unemployment risk for AI exposed occupations is likely not because of an earlier AI vintage and, instead, is more likely explained by external factors.
Beyond the pre-trends issue, we also find that unemployment risk for AI-exposed occupations does not continue to worsen after the launch of ChatGPT (i.e., through 2023, 2024, and 2025)---which is hard to square with AI being the cause of rising unemployment risk if there were anticipatory effects in some corners of the economy, since we would then still expect worsening unemployment risk after the public launch of ChatGPT (and related tools) as AI diffused through the economy.}

\revise{Our main analysis uses AI exposure scores that are widely used---even by AI companies like Anthropic~\cite{massenkoff2026labor}---to understand how AI is impacting jobs.}
But these patterns also hold under several alternative AI exposure scores~\cite{felten2018method,felten2023occupational} including a measure using actual Claude AI usage data from Anthropic~\cite{appelmccrorytamkin2025geoapi} (see SI Section~\ref{si:sec:altExposure}).
\revise{SI Section~\ref{si:sec:uiMonth} provides monthly versions of Figure~\ref{fig:uiRisk}B\&C and provides a break-point analysis for the monthly exposure quintile time series (SI Section~\ref{si:sec:uiBreaks}).} 
While these patterns hold nationally, a small number of states show post-launch increases in computer and math occupation unemployment risk (e.g., AK, CA, and WA; Figs.~\ref{fig:allUiRiskTimeSeries_AK}, \ref{fig:allUiRiskTimeSeries_CA}, and \ref{fig:allUiRiskTimeSeries_WA}, respectively\revise{; see SI Section~\ref{si:sec:uiStateMajorSoc} for all 50 states}).
In these cases, timing alone cannot rule out a contribution from LLM diffusion. 
\revise{Additionally, California, Connecticut, Delaware, and Nebraska account for most of the state-occupation pairs with no reported claimants, but omitting these states from the analysis does not change the overall result (see Fig.~\ref{fig:uiRiskOmit_CA}-\ref{fig:uiRiskOmit_CACTDENE}).}
However, on aggregate, post-launch unemployment risk remains comparable to pre-pandemic levels, suggesting limited movement in the longer-run equilibrium.

\subsection*{Labor market outcomes for recent college graduates}
\label{sec:firstJob}
Unemployment risk is built from unemployment insurance claims, so it captures only workers who have lost a job and can claim benefits.
Recent college graduates may not be eligible to claim unemployment benefits or may not have a most-recent occupation to include in their unemployment claim and would therefore be absent from our analysis of unemployment risk.
Meanwhile, public discourse and other recent studies~\cite{brynjolfsson2026canaries} suggest that AI most impacts early-career workers competing for LLM exposed jobs.
Thus, while unemployment risk reflects career outcomes in the workforce generally, it may not reflect the outcomes of workers who experience the bulk of AI's impact.

\begin{figure}[p]
	\centering
	\includegraphics[width=\textwidth]{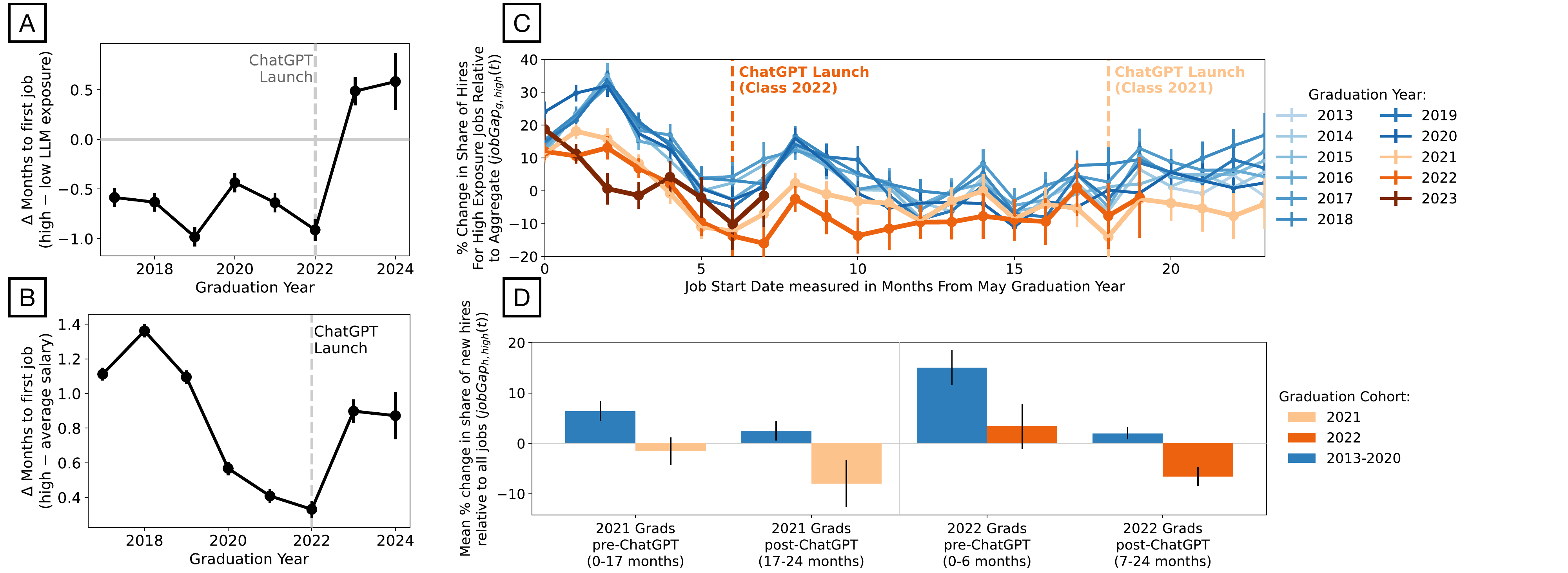}
	\caption{
    College graduates from 2022 onward spend more time securing jobs with high LLM exposure, but poor labor market performance starts months before ChatGPT.
    We consider graduates first job within three years after completing their terminal degree.
    {\bf (A)} Graduate's whose first job had high exposure typically spend less time job seeking compared to peers from the same graduation cohort, but this trend reverses after 2022 \revise{in the full sample; on a common observation window the advantage narrows but does not reverse (SI Fig.~\ref{si:fig:jobDelayWindow})}.
    {\bf (B)} As a comparison, graduates whose first job was high salary also experienced increasing job delays after 2022, but delays that were still comparable to those in previous years.
    In (A) \& (B), points show regression-adjusted differences; bars are 95\% confidence intervals.
    {\bf (C)} For each graduation cohort by month from May of their graduation year, the share of jobs with high LLM exposure started compared to the overall share of jobs.
    Dashed lines indicate the launch of ChatGPT for graduation cohorts 2021 and 2022.
    \revise{{\bf (D)} Similar to (C), comparing the average share of jobs with high exposure (top decile) compared to the overall share of jobs for graduate cohorts around the launch of ChatGPT compared to earlier cohorts.
      In each time period, the class of 2021 or 2022 was significantly less successful at securing high-exposure jobs compared to earlier graduation cohorts.
    In (C) \& (D), vertical lines indicate 95\% confidence intervals and positive values indicate that graduates within a cohort are securing LLM-exposed jobs at a higher rate than less exposed jobs. }
    }
	\label{fig:jobDelay} 
\end{figure}

We address this shortcoming with 10,584,980 individuals' LinkedIn profiles provided by Revelio Labs.
Each profile contains a worker's education history, including degrees obtained, field of study, graduation year, and degree-granting university, as well as their career history, including job titles (i.e., mapped to six-digit SOC code), employers, job start dates, and job locations.
Additionally, Revelio estimates the salary for each job using a proprietary machine learning model.
This salary model uses position-specific information, such as job title, seniority level, company, and location, as well as user-specific information such as the number of years an individual has worked at the company, and is trained on salaries found in publicly-available visa application data, self-reported data, and job postings.
We are not able to directly verify the estimated salaries against workers' actual salaries.
Therefore, we additionally consider college graduates' months to first observed job within three years post-graduation; individuals without an observed job within the window are excluded.
This measure requires only that individual's update their profiles when they start a new job and again corresponds to friction in the job market if graduates spend more time job seeking. 
\revise{
Graduates who have not started a job within three years of graduation are excluded from our analysis, so delays are right-censored at three years and our analysis can miss the students who search longest or who leave the labor market entirely.
Our estimates of job delays are therefore a lower bound.
Because job start dates end in December 2024, the most recent cohorts are also observed for less than three years, so we repeat the job-delay analyses counting only first jobs that start within 12 months (or 6 months) of May of the graduation year and keeping only the cohorts whose data extend through that entire window (SI Section~\ref{si:sec:jobDelayRobust}).
}

We examine job seeking for graduation cohorts while controlling for the monthly job opening rate in the state and month of their first job, fixed effects for the job's state and sector, and fixed effects for the individuals' education characteristics including terminal degree type, field of study, and degree-granting university (see SI Section~\ref{si:sec:jobDelay}).
\revise{We identify a job's LLM exposure using Revelio Labs' mapping of raw job titles to six-digit SOC occupation codes which we reference against \cite{eloundou2024gpts}.}
In all years before the launch of ChatGPT, on average, graduates whose first job had high exposure spent less time job seeking than their peers (see Fig.~\ref{fig:jobDelay}A).
However, graduates from 2023 and 2024 whose first job was exposed instead spent more time job seeking.
\revise{This reversal does not survive a common observation window. When every cohort is held to first jobs starting within 12 months of May of the graduation year, graduates of 2023 whose first job was exposed still found jobs faster than their peers, by a margin similar to the 2017--2019 cohorts; the advantage is largest for the 2020--2022 cohorts and narrows afterward, and within a 6-month window the 2024 cohort shows no significant difference (SI Section~\ref{si:sec:jobDelayRobust}).
The apparent reversal therefore reflects the shorter observation window available for the most recent cohorts rather than a post-ChatGPT penalty for exposed jobs.
We retain the full-sample series in Fig.~\ref{fig:jobDelay}A because it describes what recent graduates entering exposed fields actually experienced, and thus the experience that shapes public perception of early-career prospects in technology: recent graduates in computer science and related majors have faced elevated unemployment in these years relative to graduates of many other fields~\cite{nyfed_college_labor_market_2024}.}

\revise{LLM exposure and pay are positively but only moderately related across occupations: exposure correlates $0.42$ with the log of an occupation's mean annual wage, and the most exposed occupations are middle- rather than top-earning (SI Section~\ref{si:sec:exposureWage}). Eloundou et al. report the same overall pattern, finding both low-wage occupations with high exposure and high-wage occupations with low exposure alongside a positive trend~\cite{eloundou2024gpts}.}
If the increase in job search times were specific to LLM-exposed occupations, an alternative salary-based comparison might produce a different result.
Thus, we additionally examine the time to first job of graduates whose first job was high salary compared to their peers earning average salaries from the same graduation cohort (see Fig.~\ref{fig:jobDelay}B).
Consistent with our hypothesis, we observe a different trend for pre-pandemic graduates who spend more time job seeking. 
But the salary-based analysis moves in the same direction post-2022, suggesting a broader early-career tightening that is not specific to LLM-exposed workers.
\revise{These descriptive analyses encourage us to look more carefully at the temporality of graduates' job market performance.}

Did recent cohorts' shortfall in securing LLM-exposed jobs begin with the launch of ChatGPT?
If so, then each cohort's share of newly started LLM-exposed jobs should fall only after the launch. 
For graduation cohort $g$, we therefore measure its over- or under-representation among newly started first jobs in LLM exposure decile $llm$ as
\begin{equation}
    \text{jobGap}_{g,llm}(t) = 100\cdot\frac{p(g \mid t, llm) - p(g \mid t)}{p(g \mid t)},
    \label{eq:jobGap}
\end{equation}
where $p(g \mid t, llm)$ denotes the share of jobs with LLM exposure decile $llm$ started by graduates of class $g$ in month $t$ since May of their graduation year, and $p(g \mid t)$ denotes the corresponding share across all jobs (see SI Section~\ref{si:sec:jobGap}).

Across exposure deciles, graduation cohorts 2013 through 2020 each exhibit nearly identical $\text{jobGap}_{g,llm}$ trends in their first two years after graduation. 
For jobs started within each of the first four months after graduation, graduates secure a larger share of LLM-exposed jobs compared to the share of all jobs started.
This difference in job finding rates gradually disappears around six months after May of the cohort's graduation year.

However, only graduate cohorts 2021, 2022, and 2023 would have experienced job markets with generative AI adoption within two years after degree completion.
Compared to earlier cohorts, these graduates exhibit lower $\text{jobGap}_{g,llm}$ for the most LLM-exposed jobs (i.e., decile $llm=10$. See Fig.~\ref{fig:jobDelay}C)) in their first few months in the labor market and then again at roughly one year after graduation. 
And this low performance for 2021 and 2022 graduates starts months before the launch of ChatGPT (note 2023 graduates entered the workforce only after ChatGPT's launch). 
\revise{Specifically, we calculate the average $\text{jobGap}_{g,high}(t)$ in the periods before and after the launch of ChatGPT and compare each with the 2013 through 2020 cohorts averaged over the same months since their own graduation; a two-sample $t$-test finds significantly lower scores in all four comparisons (see Fig.~\ref{fig:jobDelay}D).}
As a placebo test, we additionally consider graduate cohorts' $\text{jobGap}_{g,llm}$ trends for the least LLM exposed jobs (i.e., decile $llm=1$) and observe no period of significant difference (see \revise{SI Fig.\ref{fig:gradJobGap_exposure_1}} and SI Section~\ref{si:sec:jobGap} for statistical results).
We provide similar analyses for all LLM-exposure deciles and, alternatively, for job salary deciles in SI Sections~\ref{si:sec:jobGapAi} and \ref{si:sec:jobGapSalary}, respectively.
\revise{These analyses show that the shortfall is not specific to LLM-exposed work. Recent cohorts are under-represented in the upper deciles of both exposure and salary, and the gradient across salary deciles is if anything the cleaner of the two: after the launch the 2022 cohort is significantly under-represented in salary deciles four through ten. What neither margin shows is a sharp movement at the launch itself; the change in each cohort's gap from before to after the launch is small and mostly indistinguishable from zero in both cases (see the summary figures in SI Sections~\ref{si:sec:jobGapAi} and \ref{si:sec:jobGapSalary}). The deterioration these graduates experienced is broad and it is already underway when ChatGPT arrives.}
We provide an alternative analysis of the share of jobs by LLM-exposure demonstrating consistent results in SI Section~\ref{si:sec:gradJobShare}.

\subsection*{Educational exposure to LLM-related skills}
\label{sec:eduExposure}

\revise{Whatever AI's role, recent college graduates faced a difficult job market.}
And students, educators, employers, and policymakers debate how higher education should change---both in terms of pedagogy~\cite{wang2025effect} and workforce preparedness.
For example, should students continue to learn skills that LLMs might automate, such as writing, coding, or information lookup?
If LLMs primarily reduced the market value of LLM-adjacent skills, we might expect educational exposure to those skills to produce negative career outcomes after ChatGPT.

\begin{table}[p]
	\centering
	\caption{
        Education exposure to LLM exposed tasks is associated with shorter job searches after ChatGPT and partially offsets salary penalties in highly exposed occupations. 
        $gpt=1$ if job starts on or after December 2022.
        \revise{$edu$ is education exposure, the exposure-weighted similarity between a graduate's coursework and the O*NET task space, averaged over matched syllabi; it is not a count of tasks (see SI Sections~\ref{si:sec:aiedu} and~\ref{si:sec:eduScale}).}
        $llm$ is first-job occupation six-digit SOC code LLM exposure~\cite{eloundou2024gpts}.
        Continuous predictors and dependent variables are centered and standardized (i.e., z-scores), \revise{so each coefficient is the change in standard deviations of the dependent variable associated with a one standard deviation increase in the predictor, and each interaction term is the product of its standardized components}.
        All models include fixed effects for the state and two-digit NAICS code of the job as well as for the terminal degree type (e.g., bachelors or masters), graduation year, college field of study (i.e., two-digit CIP Code), and degree-granting university.
    }
	\label{tab:eduExposure} 

    \renewcommand{\arraystretch}{.7}
	\begin{tabular}{l|ll|ll}
        \hline
        Dependent Variable: & \multicolumn{2}{c|}{Log Salary (Real 2015 \$)} & \multicolumn{2}{c}{Months to First Job} \\ \hline
        Variable & Model 1 & Model 2 & Model 3 & Model 4 \\
        \hline Job Opening Rate & \revise{0.038}$^{***}$ & \revise{0.038}$^{***}$ & \revise{0.621}$^{***}$ & \revise{0.621}$^{***}$ \\
        Major SOC Log Annual Wage (2020) & \revise{0.295}$^{***}$ & \revise{0.278}$^{***}$ & \revise{-0.015}$^{***}$ & \revise{-0.014}$^{***}$ \\
        \revise{Work-From-Home Score~\cite{hansen2023remote}} & \revise{0.074}$^{***}$ & \revise{0.044}$^{***}$ & \revise{-0.009}$^{***}$ & \revise{-0.006}$^{***}$ \\
        Post ChatGPT Indicator ($gpt$) & 0.035$^{***}$ & 0.031$^{***}$ & 2.335$^{***}$ & 2.338$^{***}$ \\
        \revise{Education Exposure} ($edu$) & -0.000 & \revise{-0.001} & \revise{0.005}$^{***}$ & \revise{0.005}$^{***}$ \\
        LLM Exposure ($llm$) &  & \revise{0.089}$^{***}$ &  & \revise{-0.014}$^{***}$ \\
        $gpt\times llm$ &  & -0.062$^{***}$ &  & 0.032$^{***}$ \\
        $edu \times gpt$ & 0.007$^{***}$ & 0.004$^{*  }$ & \revise{-0.024}$^{***}$ & -0.024$^{***}$ \\
        $edu \times llm$ &  & -0.019$^{***}$ &  & 0.005$^{***}$ \\
        $edu \times llm \times gpt$ &  & 0.011$^{***}$ &  & -0.009$^{***}$ \\
        \hline
        State FE & Yes & Yes & Yes & Yes \\
        Sector FE & Yes & Yes & Yes & Yes \\
        Terminal Degree Type FE & Yes & Yes & Yes & Yes \\
        Graduation Year FE & Yes & Yes & Yes & Yes \\
        College Major FE & Yes & Yes & Yes & Yes \\
        University FE & Yes & Yes & Yes & Yes \\
        \hline
        $R^2$ & 0.438 & 0.442 & 0.375 & 0.375 \\
        adj. $R^2$ & 0.437 & 0.442 & 0.374 & 0.374 \\
        \revise{Observations} & \revise{1,258,068} & \revise{1,258,068} & \revise{1,258,068} & \revise{1,258,068} \\
        \hline \multicolumn{5}{c}{$p_{val}<0.1^*$, $p_{val}<0.01^{**}$, $p_{val}<0.001^{***}$} \\ \hline
    \end{tabular}

\end{table}

To investigate, we again turn to LinkedIn profiles and examine graduates' first jobs within three years after degree completion.
However, using profiles alone, our best proxy for education exposure to LLMs is workers' self-reported field of study (i.e., college major) according to two-digit codes from the Classification of Instructional Programs (CIP) used by the National Center for Education Statistics \revise{which, alone, omits any variation by university or graduation year}.
Thus, we augment worker profiles with 3 million U.S. higher education course syllabi~\cite{javadian2024course} to assess each worker's education LLM exposure $edu$, \revise{defined from the semantic similarity between a course's stated learning objectives and each O*NET task, weighted by that task's LLM exposure~\cite{eloundou2024gpts} and averaged over} the syllabi associated with their graduation year, degree-granting university, and education field of study \revise{(the most recent syllabi are from 2017, so graduates after 2017 are assigned their program's 2014--2017 syllabi; }see SI Section~\ref{si:sec:aiedu}).
\revise{In other words, this measure captures how much of the LLM-exposed O*NET task space is covered by the courses associated with a graduate's inferred education experience.
It is a similarity-weighted measure rather than a tally of tasks, so its values are interpreted in standard deviations or as a share of the exposed task space rather than as a number of tasks (SI Section~\ref{si:sec:eduScale})}.

\revise{Next, we introduce an indicator variable $gpt$ to identify when a first job start date occurred after the launch of ChatGPT.}
After controlling for local labor-demand conditions (e.g., by state and sector~\cite{filippucci2025aggregate}), \revise{the share of job postings from 2017-2019 for the same job title as the first job that offered a work-from-home option~\cite{hansen2023remote,lambert2026broken},} and detailed education and job fixed effects, educational LLM exposure ($edu$) is not meaningfully associated with first-job salary or time to first job for pre-ChatGPT job starts (Models 1 \& 3). 
In contrast, for post-ChatGPT job starts, higher $edu$ predicts higher salaries and shorter job searches (positive $edu\times gpt$ for salary; negative $edu\times gpt$ for job seeking time). 
Jobs with higher occupation (i.e., six-digit SOC code) LLM exposure ($llm$) are associated with weaker outcomes after ChatGPT (See Models 2 \& 4 containing negative $gpt\times llm$ for salary; positive $gpt\times llm$ for months). 
However, the three-way interactions ($edu\times llm\times gpt$) indicates that educational exposure may partially offset these post-ChatGPT penalties in more-exposed jobs after the widespread adoption of LLMs, though the implied effect sizes are modest.
\revise{Because $gpt$ is set by the start date of the first job, it is partly determined by the time-to-first-job outcome itself: within a graduation cohort, a graduate whose first job starts after the launch has necessarily searched longer. SI Section~\ref{si:sec:jobDelayRobust} therefore re-estimates Models 3 and 4 with post-ChatGPT status assigned by graduation cohort. The negative $edu\times gpt$ and positive $gpt\times llm$ associations hold, including when the sample is restricted to the 2019--2023 cohorts and to first jobs starting within 12 months of May of the graduation year, a window that the data cover in full for each of these cohorts, whereas the three-way interaction does not (SI Table~\ref{si:tbl:eduCohortGpt}).}
\revise{SI Fig.~\ref{si:fig:eduInteraction} plots the outcomes implied by Models 2 and 4 as a function of $edu$, separately for jobs before and after the launch and for jobs of high and low LLM exposure, so that the three-way interaction can be read as a change in slope rather than as a coefficient.
Because the sample is large enough that significance alone discriminates poorly between findings, SI Section~\ref{si:sec:eduEffectSize} reports these associations as a percentage of salary and as days of job search, gives the variance explained by each block of terms, and re-estimates both models on random subsamples of $50{,}000$ graduates (SI Table~\ref{si:tab:eduSubsample}).}
The representativeness of each graduation cohort (e.g., by field of study) in the LinkedIn data is a limitation of this analysis; thus, we conduct additional analyses weighting worker profiles to mimic the graduates by field of study in the U.S. National Science Foundation's National Survey of College Graduates and find consistent results (see SI Section~\ref{si:sec:graduateReweight}). 
\revise{We also repeat the graduation-cohort job gap analysis by decile of educational exposure ($edu$) rather than by decile of first-job LLM exposure (see SI Section~\ref{si:sec:jobGapEdu}).}
\revise{See SI Table~\ref{si:tab:eduExposure} for a version of the regression analysis reporting the same models with the dependent variables and predictors in their original units.}

If LLMs were to blame for graduates' poor job market performance, then, after ChatGPT's launch, we would expect to see that education exposure indicates redundant skills that do not add value when job seeking.
\revise{Graduates who land AI-exposed jobs after the launch of ChatGPT earn less of a salary premium and spend more time job seeking than before, but these penalties are smaller for graduates with greater educational AI exposure.
Taken on its own, the exposure of the first job ($llm$) shows the penalty one would expect if LLMs were to blame. 
The salary premium associated with higher occupational LLM exposure is sharply attenuated after December 2022: a one-SD increase in exposure is associated with $0.089$ SD higher log salary before the launch and $0.027$ SD after, a reduction of roughly $70\%$. 
Time to first job tells a similar story: before the launch, graduates entering more exposed occupations found jobs slightly faster than their peers, by $0.014$ SD per SD of exposure, and after the launch they found jobs slightly slower, by $0.018$ SD, although both magnitudes are small (Table~\ref{tab:eduExposure}, Models 2 and 4).}
Educational exposure, by contrast, is associated with less severe losses. 
Graduates whose coursework covered more AI-exposed skills had shorter job searches after ChatGPT's launch and higher salaries when entering highly LLM-exposed occupations.
These associations are difficult to reconcile with the view that LLM-relevant education became less valuable after ChatGPT.
\revise{One caveat to our analysis is our inability to control for students' self-selection into college majors, universities, or occupations upon graduation.
Nevertheless, while not causal, our results associate LLM-relevant preparation with better, not worse, early-career outcomes in the post-ChatGPT period.}

\section*{Discussion}

While news media and some academic studies point to AI for the job market woes of early-career workers, our results indicate that worsening labor-market outcomes in \revise{2022–2025} for LLM-exposed workers and graduates were already underway prior to the mass-market emergence of LLM applications.
Unemployment risk in highly exposed occupations rose beginning in early 2022---well before ChatGPT---and in most occupations and states we observe no discrete break coincident with its introduction. 
Early-career workers were affected disproportionately: graduates from the 2021–2023 cohorts entered highly exposed jobs at lower rates and experienced longer observed delays to their first job than earlier cohorts, with gaps opening, again, before ChatGPT's launch in late 2022. 
At the same time, LLM-relevant education was associated with better outcomes within this environment, particularly in highly exposed occupations. 
Graduates whose programs exhibited greater pre-2020 curricular emphasis on LLM-exposed tasks earned higher first-job pay and reached their first jobs more quickly in the post-ChatGPT period, particularly when entering highly exposed occupations. 
Together, these patterns suggest that LLM diffusion occurred in a labor market already shaped by macroeconomic and sectoral forces, while LLM-complementary skills retained---and possibly increased---their labor-market value.
\revise{These findings thus cast doubt on the use of ChatGPT's launch as a natural experiment to study AI's impact on jobs and suggest that the worsening outcomes for recent graduates and LLM-exposed workers in 2022--2025 may be misattributed to AI.}

\revise{If not LLM diffusion, then what?
Our design identifies timing rather than causes, but the early-2022 onset we document coincides with several forces that plausibly acted on precisely the white-collar occupations that exposure measures rank as most AI-exposed.
Monetary tightening began in March 2022 and was the steepest in four decades~\cite{FEDFUNDS2025}, and contractionary policy is known to fall unevenly across the labor market~\cite{zens2020heterogeneous}.
Simultaneously, the information sector unwound an extraordinary pandemic-era hiring expansion: software developer postings peaked in early 2022 and then fell sharply~\cite{fred_software_dev_indeed}, consistent with correction of a hiring bubble rather than with automation~\cite{mendezcarbajo2023_tech_hiring_bubble}.
A third candidate is fiscal: beginning in tax year 2022, Section 174 of the U.S. Internal Revenue Code required firms to amortize domestic research and experimentation expenditures---including software developer salaries---over five years rather than deduct them immediately, raising the effective cost of technical headcount exactly when our series turn~\footnote{``Section 174 and the state of software engineering hiring.'' \href{https://www.theregister.com/2024/01/12/us_tax_research/}{The Register} (January 12, 2024). We flag this mechanism as a hypothesis; to our knowledge its employment effects have not been estimated in the peer-reviewed literature.}.
Finally, the reorganization of work itself shifted: early-career hiring is concentrated in roles whose viability changed with remote and hybrid arrangements~\cite{lambert2026broken,hansen2023remote}, and firms have adjusted labor demand as much by redesigning job content as by reallocating hiring across occupations~\cite{wang2026reorganization}.
We cannot adjudicate among these channels here, and we do not claim any one of them is decisive.
Our claim is narrower: each of these forces was already operating before November 2022, each affected the same occupations that score highest on AI exposure, and a research design that dates treatment to ChatGPT's release cannot separate any of these forces from LLM diffusion.}

\revise{Independent evidence has since converged on this timing.
Analyzing Lightcast vacancy data with a different exposure measure, Federal Reserve Bank of New York researchers likewise find that the relative decline in postings for AI-exposed occupations predates ChatGPT's release, and that postings for junior and senior roles within highly exposed occupations declined at similar rates---contrary to the entry-level-first prediction~\cite{audoly2026postings}.
Our results are nonetheless in tension with the most prominent contrary finding.
Using high-frequency payroll records through mid-2026, Brynjolfsson et al. report that employment of 22--25 year-olds in AI-exposed occupations now sits roughly 19\% below its counterfactual, operating through reduced hiring rather than separations, and they show this gap survives excluding technology firms and controlling for interest-rate exposure and remotability~\cite{brynjolfsson2026canaries}.
Two observations reconcile the accounts more than they oppose them.
First, that gap is dated to a widening trend rather than to a level shift at November 2022, which is consistent with our finding of pre-ChatGPT onset.
Second, the effect attenuates once education is controlled---which is what one would expect if the operative margin is the market for new college graduates broadly, as our cohort analysis suggests, rather than LLM substitution within occupations.
Distinguishing these interpretations requires linked worker--firm data on actual LLM adoption, not exposure proxies tied to a release date.
The closest existing design, which compares junior employment at firms that post dedicated generative-AI roles with that at firms that do not, finds a relative decline at adopters beginning in 2023~\cite{lichtinger2025generative}; yet its aggregate series show junior employment flattening against senior employment from mid-2022, before ChatGPT's release.}

These findings have implications for research and policy.
First, they caution against treating ChatGPT's launch as a clean natural experiment for AI's labor-market impact: designs that attribute post-2022 labor-market weakness primarily to LLMs risk confounding AI diffusion with concurrent macroeconomic shifts (\revise{possible examples include monetary policy, responses to work-from-home~\cite{lambert2026broken}, sectoral demand, and/or post-pandemic adjustment}).
\revise{This caution is not merely methodological. When firms attribute layoffs to AI, and researchers date treatment to a product launch, the two reinforce a narrative that the underlying measurements do not support: adoption studies covering the period since ChatGPT continue to find little or no effect of LLM use on earnings, hours, or vacancies~\cite{humlum2026still,hartley2026labor,liu2026adoption,johnston2026ai} (but see~\cite{lichtinger2025generative} for junior employment), even as AI is named in a growing share of announced job cuts.
Mis-attribution is costly in two ways. 
It directs policy attention toward technology-displacement remedies while leaving monetary, fiscal, and sectoral drivers unaddressed. 
And it may discourage students from undertaking the training that our third analysis associates with better outcomes.}
Second, they suggest that higher education and workforce programs should not abandon skills often labeled ``AI-exposed,'' such as writing, coding, and information synthesis.
Instead, curricula that teach these skills in LLM-complementary ways (e.g., by emphasizing verification, evaluation, and effective human–LLM collaboration) may improve graduates' resilience in an uncertain labor market.
\revise{The experimental literature supports this reading of our correlational result.
Across randomized and field experiments, generative AI raises output substantially---roughly 20--60\% in controlled trials and 15--30\% in the field~\cite{delriochanona2025review}---in writing~\cite{noy2023experimental}, software development~\cite{peng2023impact}, and customer support~\cite{brynjolfsson2025generative}, with the largest gains accruing to less-experienced workers on well-specified tasks.
But those gains are not uniform. Dell'Acqua et al. show that AI assistance sits on a ``jagged technological frontier'': on tasks inside the frontier consultants improved sharply, while on a task just outside it AI-assisted consultants were markedly \emph{less} likely to reach the correct answer than unassisted peers~\cite{dell2025navigating}.
The competence that distinguishes the two cases is the ability to judge which side of that frontier a given task falls on---to evaluate, verify, and redirect model output rather than accept it.
That is a curricular skill, and it can be taught in exactly the courses our exposure measure flags as LLM-exposed.
Our finding that graduates with more LLM-exposed coursework saw shorter job searches and higher entry pay in highly exposed occupations is consistent with this literature, and it supports teaching these skills more deliberately rather than retreating from them.}
Our analysis of education exposure is purely descriptive but covers millions of educational experiences and graduate outcomes in the labor market and, to the best of our knowledge, it is the only study that connects LLM exposure in education and labor market outcomes in the wake of LLMs at such a large scale.

Our analyses rely on administrative unemployment insurance records, online career histories, task-based exposure measures, and syllabus-derived proxies for curricular content.
Each source involves measurement and selection error. 
In particular, our educational exposure measure is derived from syllabi that predate 2020 and therefore captures baseline differences across programs rather than curricular responses to ChatGPT. 
It is also not comprehensive: the syllabi cover a subset of the courses and institutions a graduate may have encountered, so a graduate's inferred coursework is a sample of what their program taught.
Our time-to-first-job analyses exclude individuals whose first observed job begins more than three years after degree completion, which may understate delays among graduates facing the greatest barriers to entry. 
We do not identify the causal effect of LLMs on labor-market outcomes. 
Nonetheless, by triangulating across independent data sources, we document timing patterns that are difficult to reconcile with a sudden post-ChatGPT collapse in exposed occupations, shifting attention toward broader labor-market dynamics and toward expanding access to AI-complementary education.
\revise{Our analysis does not rule out the potential for AI to create job loss in the future and so it is imperative that we continue to improve data, analysis, and our understanding of AI in the workplace.
This caveat has grown more pressing since these data end: model capability on long-horizon tasks continues to double on a timescale of months~\cite{metr2026timehorizon,denain2026accelerated}, and studies comparing exposure to outcomes over successive windows already report associations that were not detectable earlier~\cite{dominski2025advancing}.
A finding that the 2022--2025 deterioration predates LLM diffusion is therefore a descriptive statement about the past, not a forecast.}
Future work combining direct measures of LLM adoption with linked worker–firm data will be critical for distinguishing displacement from productivity gains and for identifying which workers, regions, and institutions benefit---or fall behind---in the next phase of diffusion.

\bibliography{main} 
\bibliographystyle{unsrt}

\section*{Acknowledgments}
We are grateful for input from Daniel Rock, Esteban Moro, Nathan Goldshlag, and Yong-Yeol Ahn. 

\paragraph*{Funding:}
M.~R.~F. was funded by a grant from the Alfred P. Sloan Foundation, through a gift from Microsoft's AI Economy Institute, and this work is supported by the University
of Pittsburgh Center for Research Computing.

\paragraph*{Author contributions:}
All authors designed the research and drafted the manuscript.
M.R.F. and A.J.S. gathered data and conducted the analysis. 

\paragraph*{Competing interests:}
There are no competing interests to declare.

\paragraph*{Data and materials availability:}
Upon publication, all code will be publicly-available for download through a GitHub repository. 
Underlying worker profile data was provided by Revelio Labs and is not publicly available.
However, data can be requested or purchased from Revelio Labs.




\clearpage


\renewcommand{\thefigure}{S\arabic{figure}}
\renewcommand{\thetable}{S\arabic{table}}
\renewcommand{\theequation}{S\arabic{equation}}
\renewcommand{\thepage}{S\arabic{page}}
\renewcommand{\theHfigure}{S\arabic{figure}}
\renewcommand{\theHtable}{S\arabic{table}}
\renewcommand{\theHequation}{S\arabic{equation}}
\renewcommand{\thesection}{S\arabic{section}}
\renewcommand{\thesubsection}{S\arabic{section}.\arabic{subsection}}
\renewcommand{\thesubsubsection}{S\arabic{section}.\arabic{subsection}.\arabic{subsubsection}}
\renewcommand{\theHsection}{S\arabic{section}}
\renewcommand{\theHsubsection}{S\arabic{section}.\arabic{subsection}}
\renewcommand{\theHsubsubsection}{S\arabic{section}.\arabic{subsection}.\arabic{subsubsection}}
\setcounter{section}{0}
\setcounter{figure}{0}
\setcounter{table}{0}
\setcounter{equation}{0}
\setcounter{page}{1} 


\begin{center}
\section*{Supplementary Materials for\\ \scititle}

M.R. Frank$^{\ast}$,
A. Javadian Sabet,
L. Simon,
S.H. Bana,
R. Yu\\ 
\small$^\ast$Corresponding author. Email: mrfrank@case.edu\\
\end{center}

\newpage



\clearpage
\section{\revise{Exploring Eloundou et al. AI Exposure}}
\label{si:sec:exposure}

\revise{Throughout the manuscript, an occupation's exposure to AI is measured using the LLM exposure scores of Eloundou et al.~\cite{eloundou2024gpts}.
Those scores are defined at the level of the individual workplace task.
For each task $\tau$ in the U.S. Department of Labor's O*NET database, Eloundou et al. assign an exposure value $e_\tau\in\{0,0.5,1\}$ indicating whether access to an LLM (or an LLM-powered system) could reduce the time required to complete that task by at least half.
An occupation's exposure is then the average of $e_\tau$ over the set $T_{soc6}$ of O*NET tasks belonging to that detailed (six-digit) SOC occupation,
\begin{equation}
    llm_{soc6} = \frac{1}{|T_{soc6}|}\sum_{\tau\in T_{soc6}} e_\tau,
    \label{eq:detailedExposure}
\end{equation}
so that $llm_{soc6}\in[0,1]$ can be read as the share of an occupation's workplace tasks that are exposed to LLM automation.
Table~\ref{si:tab:detailedSocExposure} reports $llm_{soc6}$ for all 923 detailed occupations, together with $|T_{soc6}|$, the number of O*NET tasks contributing to each average.
Exposure ranges from $0.000$ for occupations such as pile driver operators, dishwashers, and meat packers to $1.000$ for mathematicians, with computer, office/administrative support, and writing-intensive occupations concentrated at the top of the distribution.}

\revise{Most analyses in this study use $llm_{soc6}$ directly.
The exception is the unemployment risk analysis, because the ETA 203 report records only the \emph{major} occupation group (two-digit SOC code) of each continuing claimant and therefore cannot be linked to detailed occupations.
For that analysis we aggregate exposure to the major occupation group level by taking the equal-weighted mean of the detailed exposure scores of the occupations nested within each group,
\begin{equation}
    llm_{soc} = \frac{1}{|S_{soc}|}\sum_{soc6\in S_{soc}} llm_{soc6},
    \label{eq:majorExposure}
\end{equation}
where $S_{soc}$ is the set of detailed SOC codes belonging to major occupation group $soc$.
The average is equal-weighted across detailed occupations rather than weighted by employment, so that $llm_{soc}$ describes the exposure of the \emph{kinds of work} performed within a major group rather than the exposure of the particular mix of workers currently employed in it; this keeps the exposure measure fixed over time and independent of the employment counts that enter unemployment risk in Eq.~\ref{eq:uiRisk}.
Table~\ref{si:tab:majorSocExposure} reports the resulting $llm_{soc}$ for all 22 civilian major occupation groups, together with $|S_{soc}|$.
These are the values that determine the ordering and coloring of occupations in Fig.~\ref{fig:uiRisk}C, the assignment of occupations to exposure quintiles in Fig.~\ref{fig:uiRisk}B, and the variable $llm_{soc}$ entering Eq.~\ref{eq:uiRiskRegression}.
Major group exposure ranges from $0.069$ for construction and extraction to $0.741$ for computer and mathematical occupations.
Analyses repeating our results under alternative AI exposure measures appear in SI Section~\ref{si:sec:altExposure}.}

\input{beta_exposure_table.tex}

\begin{table}[H]
\centering
\caption{\revise{\textbf{Average LLM exposure by major occupation group.}
    Average LLM exposure~\cite{eloundou2024gpts} by two-digit SOC major group, computed from Eq.~\ref{eq:majorExposure} as the equal-weighted mean across the detailed (six-digit) SOC codes within each group, sorted from least to most exposed.
    The final column reports the number of detailed SOC codes contributing to each average.}
}
\label{si:tab:majorSocExposure}
\small
\begin{tabular}{llrr}
\hline
Major SOC Code & Title & \shortstack[r]{Avg. LLM\\Exposure} & \shortstack[r]{Num. Detailed\\SOC Codes} \\[2pt]
\hline
47 & Construction and Extraction & 0.069 & 61 \\
37 & Building and Grounds Cleaning and Maintenance & 0.121 & 8 \\
51 & Production & 0.128 & 107 \\
49 & Installation, Maintenance, and Repair & 0.133 & 50 \\
45 & Farming, Fishing, and Forestry & 0.134 & 12 \\
35 & Food Preparation and Serving Related & 0.146 & 16 \\
39 & Personal Care and Service & 0.206 & 31 \\
53 & Transportation and Material Moving & 0.217 & 52 \\
31 & Healthcare Support & 0.239 & 19 \\
33 & Protective Service & 0.284 & 26 \\
21 & Community and Social Service & 0.318 & 14 \\
29 & Healthcare Practitioners and Technical & 0.341 & 89 \\
25 & Educational Instruction and Library & 0.433 & 62 \\
19 & Life, Physical, and Social Science & 0.461 & 60 \\
41 & Sales and Related & 0.462 & 22 \\
11 & Management & 0.463 & 56 \\
27 & Arts, Design, Entertainment, Sports, and Media & 0.471 & 40 \\
23 & Legal & 0.485 & 7 \\
17 & Architecture and Engineering & 0.491 & 56 \\
13 & Business and Financial Operations & 0.524 & 48 \\
43 & Office and Administrative Support & 0.602 & 51 \\
15 & Computer and Mathematical & 0.741 & 36 \\
\hline
\end{tabular}
\end{table}

\clearpage
\section{Unemployment Risk}
\label{si:sec:uiRisk}
The ETA 203 (officially titled ``Distribution of Characteristics of the Insured Unemployed'') is a monthly report submitted by state Unemployment Insurance (UI) agencies to the Employment \& Training Administration (ETA) of the United States Department of Labor.
The ETA 203 is a demographic \& characteristics snapshot of continued UI claimants (specifically, the ``insured unemployed'') across each state and for the nation as a whole.
It is reported on a monthly basis, with data capturing the week including the 12th of the month which corresponds to the similarly-timed survey week for the U.S. Bureau of Labor Statistics (BLS) and U.S. Census's Current Population Survey (CPS).
The data are submitted by states to the national office of ETA via the Office of Workforce Security.
\revise{Our unemployment risk panel covers January 2015 through October 2025, the most recent month available at the time of this revision, at monthly frequency for each major occupation group in each of the 51 reporting jurisdictions (the 50 states and the District of Columbia).}
\revise{Because the ETA 203 is an administrative tabulation of the insured unemployed rather than a sample survey, its coverage is properly described by how completely states map claimants onto the Standard Occupation Classification System (SOC), not by a sampling error.
Across the major occupation group $\times$ state $\times$ month cells of the January 2020 through October 2025 window tabulated in Table~\ref{tab:uiMissingData}, 41 of the 51 reporting jurisdictions record at least one continuing claimant in every major occupation group in every month.
All 2,901 zero-claimant cells fall in the ten jurisdictions listed in that table, and two of them---Delaware and California---account for 2,231 of those cells (77\%).
Zero-claimant cells are fewer than 5\% of the cells in that window; excluding Delaware and California, they number 670 in total.}

\begin{figure}[t]
    \centering
    \includegraphics[width=\textwidth]{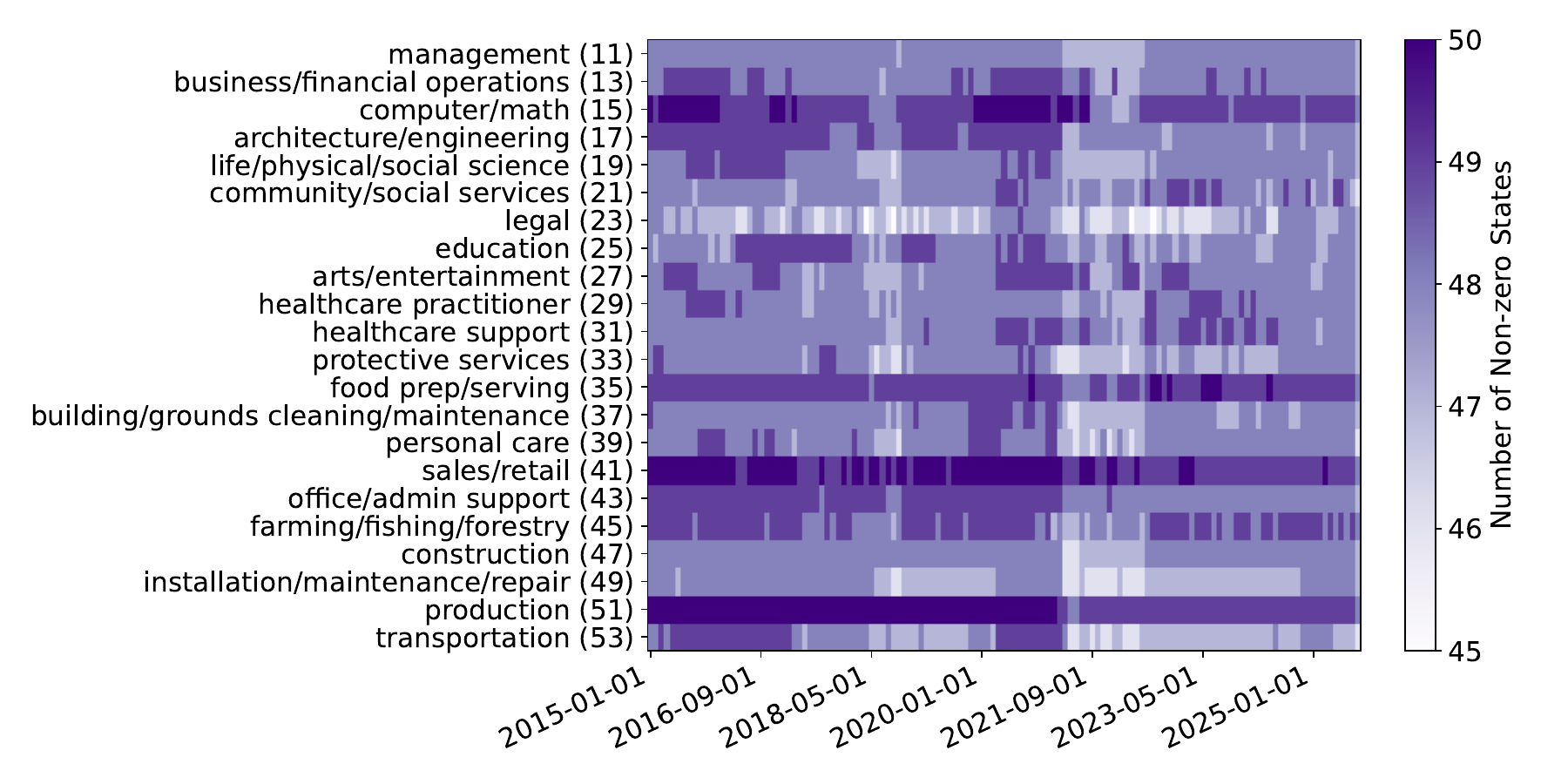}
    \caption{
        \textbf{Number of states with non-zero unemployment claims by major occupation group and month.}
        A state reporting zero claimants may be accurate or may reflect a failure to connect a state's internal occupation classification system with the Standard Occupation Classification (SOC) used by the U.S. Department of Labor.
    }
    \label{fig:uiRiskSocCoverage}
\end{figure}

\revise{Two distinguishable patterns generate these zeros.
A jurisdiction reporting zero claimants for a given major occupation group in nearly every month of the panel indicates a failure to align that state's internal occupation classification system with the SOC rather than a genuine absence of claimants; Delaware exhibits this pattern for most occupation groups, and California exhibits it for business/financial operations, management, and education occupations while still reporting counts for computer/math occupations (see Table~\ref{tab:uiMissingData}).
Isolated zero months in a small state for a small occupation group---in the worst case, four months in which Legal occupations (Major SOC 23) report zero claimants in five states (see Fig.~\ref{fig:uiRiskSocCoverage})---are instead consistent with genuinely small claimant counts.
In either case the affected state-occupation pairs are excluded from aggregation rather than imputed.}
You can download the complete data set here: \url{https://oui.doleta.gov/unemploy/csv/ar203.csv}

\subsection{ETA 203 Report compared to Census Current Populatiom Survey (CPS) Unemployment}
\label{si:sec:etaCps}

The ETA 203 report (``Characteristics of the Insured Unemployed'') is an administrative unemployment insurance (UI) tabulation compiled from state UI systems that describes the demographic, industry, and occupation characteristics of the \textit{insured unemployed}. 
In the ETA 203 framework, the insured unemployed are defined as \textit{regular state UI continued weeks claimed} (intrastate plus interstate agent), excluding extended and other compensation programs \cite{DOL_ETA203_Handbook401_2001}. 
Importantly for comparisons to survey-based unemployment statistics, ETA guidance aligns the ETA 203 ``survey week'' to the Current Population Survey (CPS) reference week (typically the week containing the 12th of the month), operationalized through continued weeks claimed filed during the week containing the 19th, and explicitly notes that ETA 203 characteristics can be compared to characteristic figures for the CPS \textit{total unemployed} \cite{DOL_ETA203_Handbook401_2001,DOL_OUI_DataDownloads_2025,DOL_UIPL22_11_2011}. 
By contrast, CPS unemployment is a population-representative household survey construct rather than a program-based count, and cannot be inferred from UI rolls because many unemployed individuals are ineligible for UI, do not apply, or exhaust benefits \cite{BLS_HowMeasuresUnemployment_2014}.

Empirical benchmarks underscore both the value and the limitations of ETA 203--CPS comparisons. 
Cortes and Forsythe compare CPS-predicted UI recipients to actual UI recipients reported to the U.S.\ Department of Labor (DOL) via the ETA 203 system during April 2020--February 2021, emphasizing that the DOL administrative compilation aggregates state reports of varying quality and exhibits substantial missingness for some variables (e.g., race and occupation) \cite{CortesForsythe2021_CARES_Inequality}. 
Their comparisons show close agreement for some characteristics (e.g., female and Hispanic shares) alongside larger discrepancies for others (e.g., Black and older-age shares), which they interpret as potentially reflecting CPS classification issues in certain contexts and/or reporting differences across sources \cite{CortesForsythe2021_CARES_Inequality}. 
Taken together, these resources suggest that ETA 203--CPS comparisons are most defensible when framed explicitly as contrasts between \textit{insured unemployment} (receipt/claims-based) and \textit{total unemployment} (labor-force-status-based), with careful attention to timing alignment, program scope (regular UI only), and item-specific data quality in administrative tabulations \cite{DOL_ETA203_Handbook401_2001,CortesForsythe2021_CARES_Inequality,BLS_HowMeasuresUnemployment_2014}.

\revise{This comparison also supplies the standard against which the occupational coverage reported in Section~\ref{si:sec:uiRisk} should be judged.
First, no survey-based series provides a benchmark at the same resolution: the CPS monthly sample is far too small to support estimates by occupation within state within month, so at the granularity our analysis requires the ETA 203 is the more complete of the two sources, and the relevant contrast remains one between insured and total unemployment rather than between two measurements of the same quantity \cite{BLS_HowMeasuresUnemployment_2014}.
Second, the item-specific pattern we observe is the pattern the published assessment of this series reports: Cortes and Forsythe identify occupation and race as precisely the ETA 203 characteristics exhibiting substantial missingness, while other characteristics track their CPS analogues closely \cite{CortesForsythe2021_CARES_Inequality}.
The occupational coverage of our panel is therefore typical of this administrative series rather than anomalous, and---unlike a survey nonresponse problem, which is diffuse---it is concentrated in a small number of identifiable jurisdictions.}

\subsection{Calculating Unemployment Risk}
\label{si:sec:uieq}
Combined with employment statistics, we calculate an occupation's \emph{unemployment risk} according to 
\begin{equation}
    p(\text{unemp} | soc,s,t) = \frac{4\cdot n(\text{unemp} | soc, s,t)}{4\cdot n(\text{unemp} | soc, s,t)+n(\text{emp} | soc,s,t)}
    \label{eq:uiRisk}
\end{equation}
where $\text{emp}$ indicates a worker is employed, $\text{unemp}$ indicates a worker is unemployed, $soc$ denotes an occupation's two-digit Standard Occupation Classification (SOC) code, $s$ denotes a state, $t$ denotes a time period (i.e., year and month), and $n$ captures the number of employed or unemployed workers by occupation, state, and time period.
$n(\text{unemp} | soc, s,t)$ comes from the detailed unemployment data while $n(\text{emp} | soc, s,t)$ comes from annual BLS OEWS.
We multiply $n(\text{unemp} | soc, s,t)$ by $4$ to roughly scale from one week of data to one month under the assumption that the week including the 12th of the month is representative of the whole month.
Employment by occupation within a state $n(\text{emp} | soc, s,t)$ varies annually and so we use it here missing monthly variation.
Given a state and time, $p(\text{unemp} | soc,s,t)$ is the probability that a worker with occupation $soc$ claims continuing unemployment benefits.
Unemployment risk varies by occupation (e.g., construction workers versus transportation workers) while controlling for states' total labor market size and occupations' local employment share (e.g., more retail workers may receive unemployment benefits in a state where retail workers are a larger share of employment).

Many studies focus on an occupation's change in employment following a technological disruption, so one might assume that an occupation's rising employment would directly correspond to falling unemployment risk.
Similarly, a more typical analysis of a state's total unemployment rate may sufficiently explain variation in unemployment risk as being largely driven by systemic economic factors rather than factors varying by occupation.
Using Bayes' Theorem, we can relate an occupation's unemployment risk to its local employment share and the state's total unemployment rate according to
\begin{equation}
    p(\text{unemp}|soc,s,t) = \frac{p(soc|\text{unemp},s,t)\cdot p(\text{unemp}|s,t)}{p(soc|\text{emp} \cup \text{unemp},s,t)}
    \label{eq:bayes}
\end{equation}
where $p(\text{unemp}|s,t)$ is the total unemployment rate from BLS LAUS and $p(soc|\text{emp} \cup \text{unemp},s,t)$ is the share of the local labor force associated with $soc$.
However, as Figure 1 in \cite{frank2025ai} shows, an occupation's employment share over time within a state has no consistent empirical relationship with the occupation's unemployment risk.
And, an occupation-state pair's unemployment risk co-varies with the state's total unemployment rate over time but not always.

\begin{figure}[th]
    \centering
    \includegraphics[width=.7\textwidth]{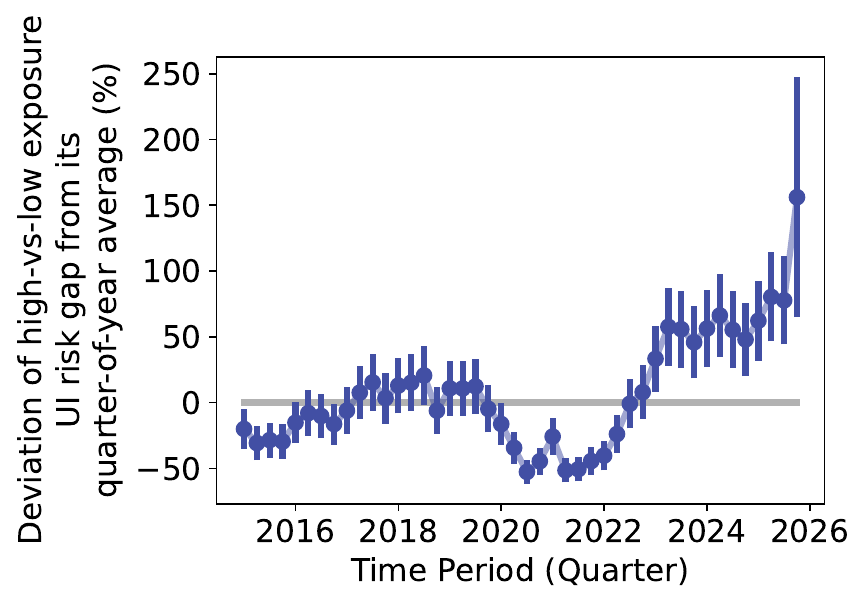}
    \caption{
        \revise{\textbf{The high- versus low-exposure contrast of Figure~\ref{fig:uiRisk}A after adjusting for quarter-of-year effects.}
        For each quarter, the estimated contrast is expressed as a deviation from the average contrast for that same quarter across all years in the panel.
        The vertical axis is therefore a deviation and not a level: zero indicates that the exposure gap in that quarter equals the typical gap for that quarter, not that unemployment risk is equal across high- and low-exposure occupations.
        The level of the gap should be read from Fig.~\ref{fig:uiRisk}A or, in levels, from Fig.~\ref{fig:uiRisk}B.
        Note also that long-term annual trends, such as the COVID-19 pandemic in 2020 and 2021, are not controlled for here; see SI Section~\ref{si:sec:uiBreaks}.}
    }
    \label{si:fig:highLowLlmAdjusted}
\end{figure}

In Figure~\ref{fig:uiRisk}A, we estimate differences in unemployment risk for low- and high-exposure occupations using the interaction terms $\beta_t$ from the following ordinary least squares regression model:
\begin{equation}
    \log\left(p(\text{unemp}|soc,s,t)\right) \sim \sum_{t\in T} \alpha_t\cdot I_t+\beta_t\cdot I_t\cdot llm_{soc}
    \label{eq:uiRiskRegression}
\end{equation}
where $T$ indexes over quarters each year, $llm(soc)\in[0,1]$ is the average AI-exposure~\cite{eloundou2024gpts} of the major occupation group $soc\in \text{SOC}$, and $I_t$ is an indicator variable for time $t$.
As a result, $100\cdot(e^{\beta_t}-1)$ describes the percentage change in unemployment risk in a given quarter as $llm_{soc}$ varies from $llm_{soc}=0$ to $llm_{soc}=1$.
\revise{Because unemployment claims are seasonal, we also report this estimate after adjusting for quarter-of-year effects (see Fig.~\ref{si:fig:highLowLlmAdjusted}).
The adjustment expresses the estimated high- versus low-exposure contrast in a given quarter as a deviation from the average contrast for that same quarter across all years in the panel.
This removes the recurring seasonal component, but it also removes the average level of the contrast, so the vertical scale of Fig.~\ref{si:fig:highLowLlmAdjusted} is a deviation rather than a level.
A value of zero in that figure therefore indicates that the exposure gap in that quarter equals the typical gap for that quarter, not that unemployment risk is equal across high- and low-exposure occupations.
Questions about the level of the gap---including whether unemployment risk is higher or lower in LLM-exposed occupations---should be read from Fig.~\ref{fig:uiRisk}A, which reports the unadjusted contrast, or from Fig.~\ref{fig:uiRisk}B, which reports unemployment risk in levels by exposure quintile.
What Fig.~\ref{si:fig:highLowLlmAdjusted} adds is that the narrowing of the exposure gap beginning in 2022 is not an artifact of seasonality: it persists once quarter-of-year effects are removed.
Note also that this adjustment does not control for longer-term annual trends such as the COVID-19 pandemic in 2020 and 2021; the structural break analysis in SI Section~\ref{si:sec:uiBreaks} addresses the pandemic explicitly.}

\revise{While unemployment risk could in principle also be calculated by sector rather than occupation, we focus on occupation because labor economics research consistently shows that technology reshapes work at the task level, and workplace tasks are organized around occupations rather than industries. The same occupation---for example, secretary---appears across many disparate industries while performing a largely similar set of workplace tasks regardless of industry context; this is reflected in databases such as O*NET (U.S. Bureau of Labor Statistics), which represent each occupation once rather than separately by industry. As a result, a technology that automates the tasks performed by secretaries may register as only a minor factor within any single industry's aggregate exposure, while still carrying substantial implications for secretaries as an occupational group across the economy. For this reason, the majority of AI exposure studies measure exposure at the occupation and/or task level rather than at the sector level, and we adopt the same approach here.}

\begin{table}[!t]

	\centering
	\caption{\textbf{Missing data from the ETA 203 report.}
	   For each major occupation group and state, the number of months with zero continuing unemployment claimants \revise{between January 2020 and October 2025 (i.e., 70 is the maximum).
	   Each cell counts the individual months in which that state reported no continuing claimants for that major occupation group.}
       States with no missing data are omitted. 
    }
	\label{tab:uiMissingData} 

    \small
    
    \begin{tabular*}{\linewidth}{@{\extracolsep{\fill}}lrrrrrrrrrr}
    Major Occupation Group& \multicolumn{10}{c}{State} \\ \hline
     & DE & CT & CA & NE & KY & NH & VT & AK & KS & SD \\ 
    life/physical/social science (19) & 70 & 2 & 55 & 23 & 0 & 0 & 0 & 0 & 0 & 0 \\
    management (11) & 70 & 0 & 54 & 30 & 0 & 0 & 0 & 0 & 0 & 0 \\
    transportation (53) & 70 & 1 & 53 & 6 & 2 & 0 & 0 & 0 & 0 & 50 \\
    healthcare support (31) & 70 & 2 & 28 & 15 & 2 & 0 & 0 & 0 & 0 & 0 \\
    education (25) & 70 & 12 & 52 & 7 & 5 & 0 & 0 & 0 & 0 & 0 \\
    food prep/serving (35) & 0 & 0 & 36 & 29 & 3 & 0 & 0 & 0 & 0 & 0 \\
    personal care (39) & 70 & 4 & 55 & 22 & 1 & 0 & 0 & 0 & 0 & 0 \\
    sales/retail (41) & 0 & 0 & 41 & 0 & 3 & 0 & 0 & 0 & 0 & 0 \\
    production (51) & 0 & 0 & 53 & 0 & 4 & 0 & 0 & 0 & 0 & 0 \\
    computer/math (15) & 0 & 3 & 48 & 10 & 1 & 0 & 0 & 0 & 3 & 0 \\
    construction (47) & 70 & 0 & 54 & 30 & 3 & 0 & 0 & 0 & 0 & 0 \\
    community/social services (21) & 70 & 13 & 22 & 25 & 2 & 0 & 0 & 0 & 0 & 0 \\
    legal (23) & 70 & 37 & 46 & 29 & 3 & 0 & 2 & 17 & 0 & 0 \\
    architecture/engineering (17) & 70 & 4 & 54 & 0 & 3 & 0 & 0 & 0 & 0 & 0 \\
    installation/maintenance/repair (49) & 70 & 0 & 54 & 13 & 3 & 58 & 0 & 0 & 0 & 0 \\
    healthcare practitioner (29) & 70 & 0 & 36 & 30 & 3 & 0 & 0 & 0 & 0 & 0 \\
    business/financial operations (13) & 70 & 0 & 43 & 10 & 3 & 0 & 0 & 0 & 0 & 0 \\
    protective services (33) & 70 & 17 & 55 & 28 & 5 & 0 & 0 & 0 & 0 & 0 \\
    building/grounds cleaning/maintenance (37) & 70 & 7 & 54 & 18 & 4 & 0 & 0 & 0 & 0 & 0 \\
    farming/fishing/forestry (45) & 1 & 7 & 55 & 30 & 13 & 0 & 0 & 0 & 0 & 0 \\
    arts/entertainment (27) & 70 & 2 & 39 & 10 & 1 & 0 & 0 & 0 & 0 & 0 \\
    office/admin support (43) & 69 & 0 & 54 & 0 & 0 & 0 & 0 & 0 & 0 & 0 \\
    \end{tabular*}
\end{table}

\clearpage
\section{Unemployment Risk by Major Occupation Group}
\label{si:sec:uiMajorSoc}
\revise{Fig.~\ref{fig:uiRisk}C in the main text shows quarterly unemployment risk for selected major occupation groups. Here we show the same series for all 22 civilian major occupation groups individually (see Fig.~\ref{fig:allUiRiskTimeSeries} and Fig.~\ref{fig:allUiRiskTimeSeries2}), so that the trend for any single group can be inspected without the normalization used in the main text.
The panels are drawn three to a row across two pages rather than five to a row on one page, so that each series can be read at the size at which it is printed.}

\revise{Table~\ref{si:tab:majorSocUiRisk} reports the same series as numbers.
For each major occupation group it gives the group's average LLM exposure and its average unemployment risk, in percent, over four windows: the pre-pandemic years, the pandemic years, the months of $2022$ before the launch of ChatGPT, and the period after the launch.
The final column is the change from the last pre-launch window to the post-launch period.
Unemployment risk rises after the launch in $20$ of the $22$ groups, and the increase is largest in the most LLM-exposed groups: $+87\%$ for computer and mathematical occupations, $+51\%$ for arts and entertainment, and $+47\%$ for architecture and engineering, against $+5\%$ for construction and $-3\%$ for farming, fishing and forestry.
Because these are averages over windows rather than a timing analysis, they say nothing about when within a window the increase began; Fig.~\ref{fig:uiRisk}B and SI Section~\ref{si:sec:uiBreaks} address that question, and both place the turn in early $2022$, before the launch.}

\newcounter{plotcounter}
\begin{figure}[H]
    \centering

    \foreach [count=\i] \n in {11,13,15,17,19,21,23,25,27,29,31} {%
        \stepcounter{plotcounter}%
        \includegraphics[width=0.32\textwidth]{figures/siFigures/uiRiskTimeSeries/majorSocUiRiskTimeSeries_centered_\n.pdf}%
        \pgfmathtruncatemacro{\mod}{mod(\i,3)}
        \ifnum\mod=0\par\medskip\fi
    }

    \caption{
        \textbf{Quarterly unemployment risk by major occupation group.}
        Each point is the average unemployment risk across all states and months in each quarter.
        \revise{Major occupation groups $11$ through $31$; the remaining groups are shown in Fig.~\ref{fig:allUiRiskTimeSeries2}, and the same series are reported as numbers in Table~\ref{si:tab:majorSocUiRisk}.}
    }
    \label{fig:allUiRiskTimeSeries}
\end{figure}

\begin{figure}[H]
    \centering

    \foreach [count=\i] \n in {33,35,37,39,41,43,45,47,49,51,53} {%
        \stepcounter{plotcounter}%
        \includegraphics[width=0.32\textwidth]{figures/siFigures/uiRiskTimeSeries/majorSocUiRiskTimeSeries_centered_\n.pdf}%
        \pgfmathtruncatemacro{\mod}{mod(\i,3)}
        \ifnum\mod=0\par\medskip\fi
    }

    \caption{
        \revise{\textbf{Quarterly unemployment risk by major occupation group, continued.}
        As in Fig.~\ref{fig:allUiRiskTimeSeries}, for major occupation groups $33$ through $53$.}
    }
    \label{fig:allUiRiskTimeSeries2}
\end{figure}

\begin{table}[H]
    \centering
    \caption{\revise{\textbf{Unemployment risk by major occupation group, in percent.} The series plotted in Fig.~\ref{fig:allUiRiskTimeSeries} and Fig.~\ref{fig:allUiRiskTimeSeries2}, reported as numbers. Each entry is the average of $p(\text{unemp} \mid soc,s,t)$ over the state-months of the window, expressed as a percentage, and $llm$ is the group's average LLM exposure~\cite{eloundou2024gpts} as defined in SI Section~\ref{si:sec:exposure}. The pre-launch window ends with $2022$Q$4$ and the post-launch window begins with $2023$Q$1$, following the convention used throughout the paper; the final column is the change between them. Groups are ordered by LLM exposure. The military group ($55$) is omitted, as elsewhere.}}
    \label{si:tab:majorSocUiRisk}
    \footnotesize
    \setlength{\tabcolsep}{4pt}
    \renewcommand{\arraystretch}{0.9}
    \begin{tabular}{l|r|rrrr|r}
    \hline
     &  & \multicolumn{4}{c|}{Unemployment risk (\%)} & \\
    Major occupation group (SOC) & $llm$ & 2015--19 & 2020--21 & 2022 & 2023--25 & Change \\
    \hline
    computer/math (15) & 0.74 & 3.843 & 6.425 & 2.447 & 4.571 & +87\% \\
    office/admin support (43) & 0.60 & 3.319 & 9.196 & 2.617 & 3.205 & +22\% \\
    business/financial operations (13) & 0.52 & 3.129 & 5.732 & 1.921 & 2.666 & +39\% \\
    architecture/engineering (17) & 0.49 & 3.364 & 7.626 & 1.976 & 2.907 & +47\% \\
    legal (23) & 0.49 & 2.675 & 5.611 & 2.173 & 2.857 & +31\% \\
    arts/entertainment (27) & 0.47 & 4.965 & 14.652 & 3.519 & 5.329 & +51\% \\
    management (11) & 0.46 & 7.010 & 14.709 & 4.249 & 5.747 & +35\% \\
    sales/retail (41) & 0.46 & 3.007 & 10.896 & 2.268 & 2.791 & +23\% \\
    life/physical/social science (19) & 0.46 & 3.618 & 7.306 & 2.436 & 3.271 & +34\% \\
    education (25) & 0.43 & 1.460 & 5.340 & 1.030 & 1.285 & +25\% \\
    healthcare practitioner (29) & 0.34 & 1.261 & 5.015 & 1.001 & 1.077 & +8\% \\
    community/social services (21) & 0.32 & 3.001 & 6.958 & 2.222 & 2.524 & +14\% \\
    protective services (33) & 0.28 & 2.246 & 5.993 & 1.702 & 2.119 & +24\% \\
    healthcare support (31) & 0.24 & 3.493 & 8.898 & 2.082 & 2.258 & +8\% \\
    transportation (53) & 0.22 & 4.631 & 9.535 & 2.797 & 3.582 & +28\% \\
    personal care (39) & 0.21 & 2.549 & 17.656 & 2.801 & 2.503 & -11\% \\
    food prep/serving (35) & 0.15 & 2.705 & 15.495 & 2.247 & 2.457 & +9\% \\
    farming/fishing/forestry (45) & 0.13 & 20.052 & 26.013 & 15.916 & 15.481 & -3\% \\
    installation/maintenance/repair (49) & 0.13 & 3.990 & 9.050 & 2.431 & 2.947 & +21\% \\
    production (51) & 0.13 & 6.852 & 14.496 & 4.202 & 5.340 & +27\% \\
    building/grounds cleaning/maintenance (37) & 0.12 & 4.828 & 12.793 & 3.651 & 4.030 & +10\% \\
    construction (47) & 0.07 & 11.751 & 16.908 & 8.361 & 8.798 & +5\% \\
    \hline
    \end{tabular}
\end{table}


\clearpage
\section{Unemployment risk by State and Major Occupation Group}
\label{si:sec:uiStateMajorSoc}

\begin{figure}[H]
	\centering
	\includegraphics[width=\textwidth]{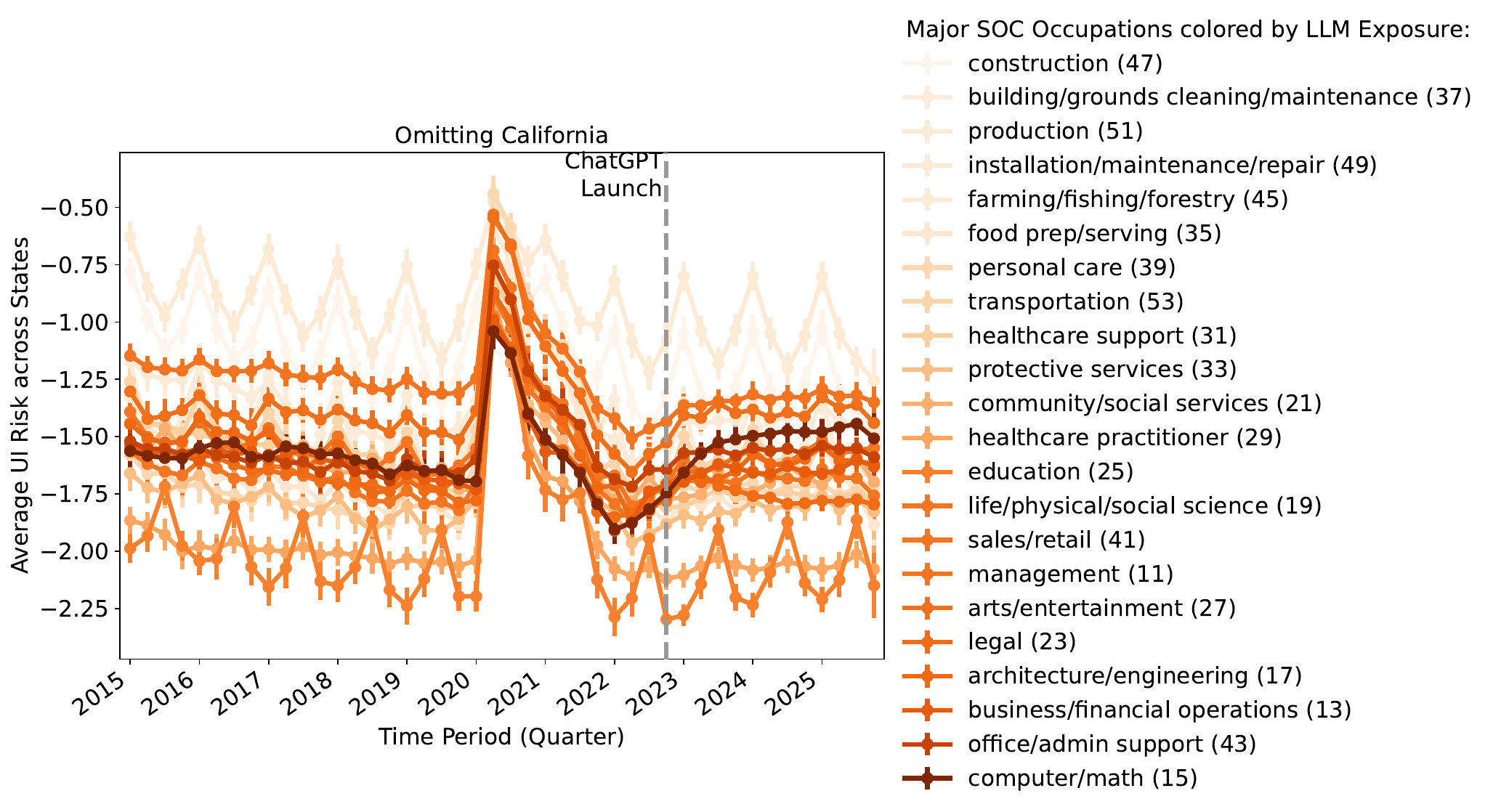}

	\includegraphics[width=\textwidth]{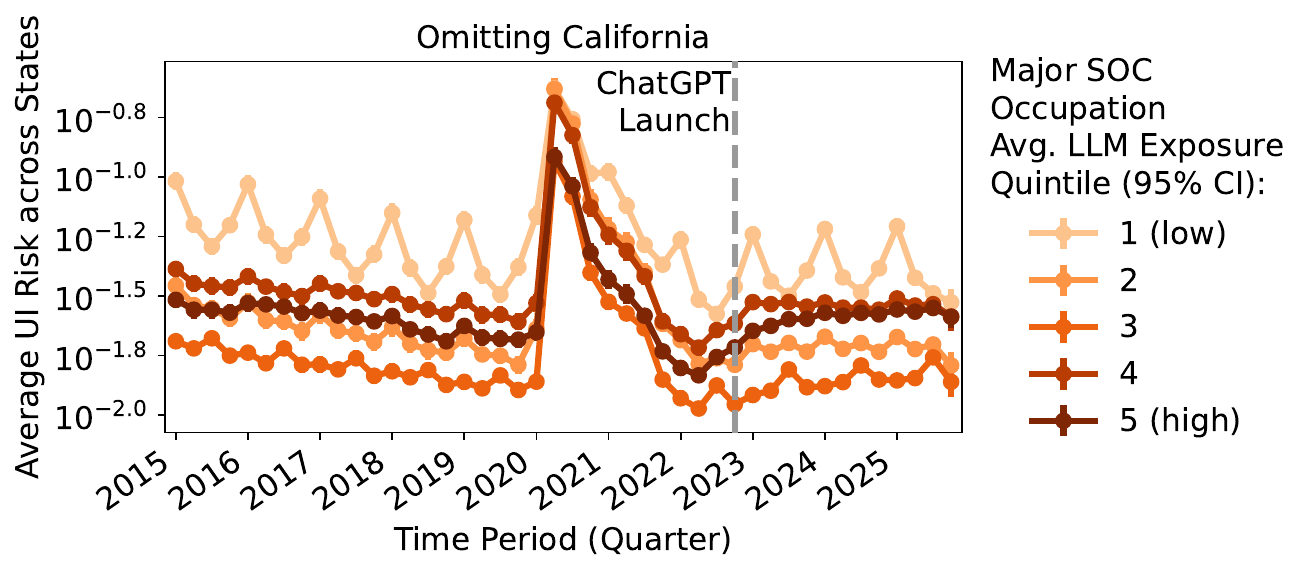}
	\caption{
		\textbf{Same as Figure~\ref{fig:uiRisk}C (upper) and Figure~\ref{fig:uiRisk}B (lower), but omitting data from California.}
		Each point is the average unemployment risk across all remaining states and months in each quarter.
	}
	\label{fig:uiRiskOmit_CA}
\end{figure}

\begin{figure}[H]
	\centering
	\includegraphics[width=\textwidth]{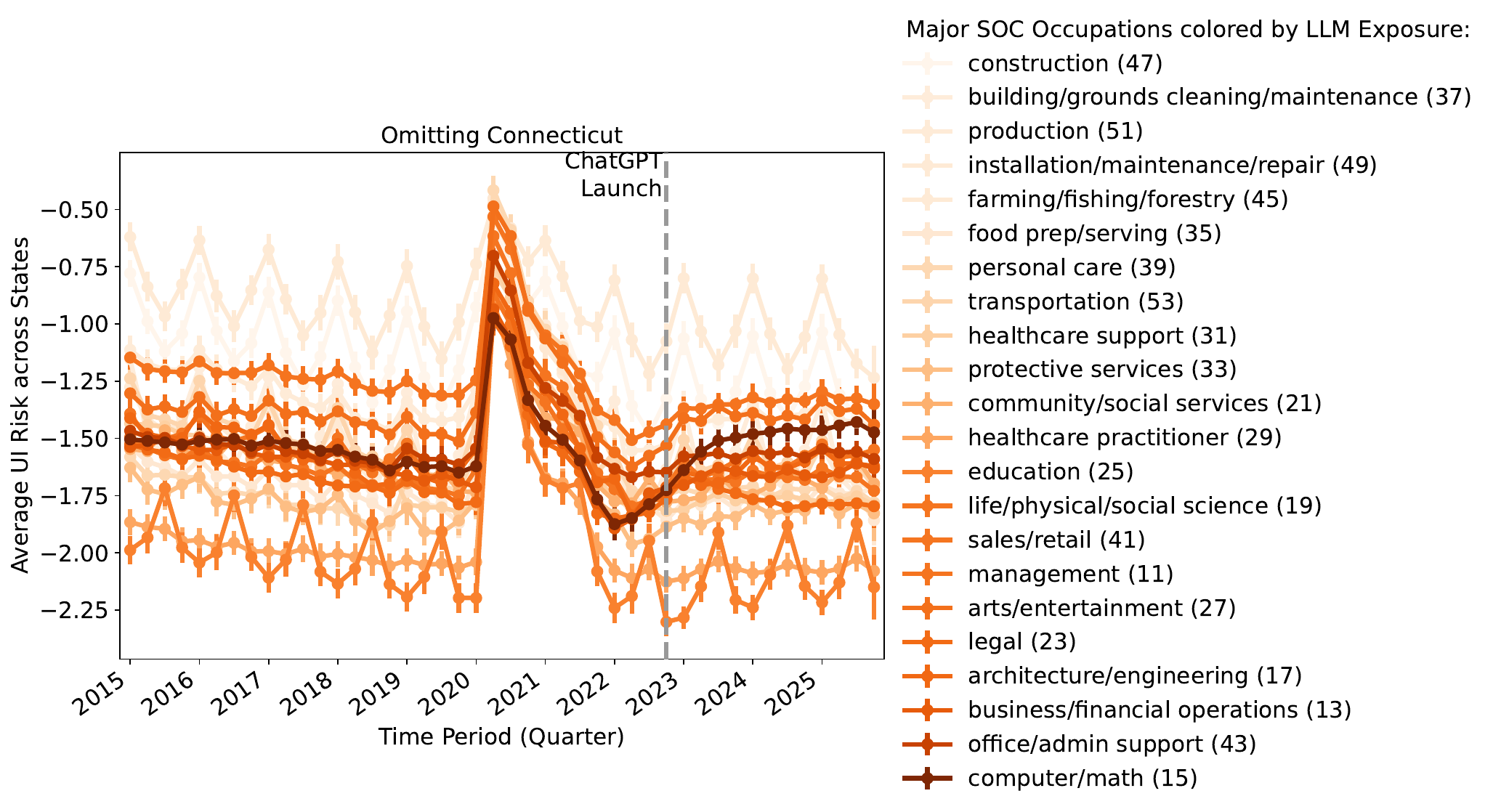}

	\includegraphics[width=\textwidth]{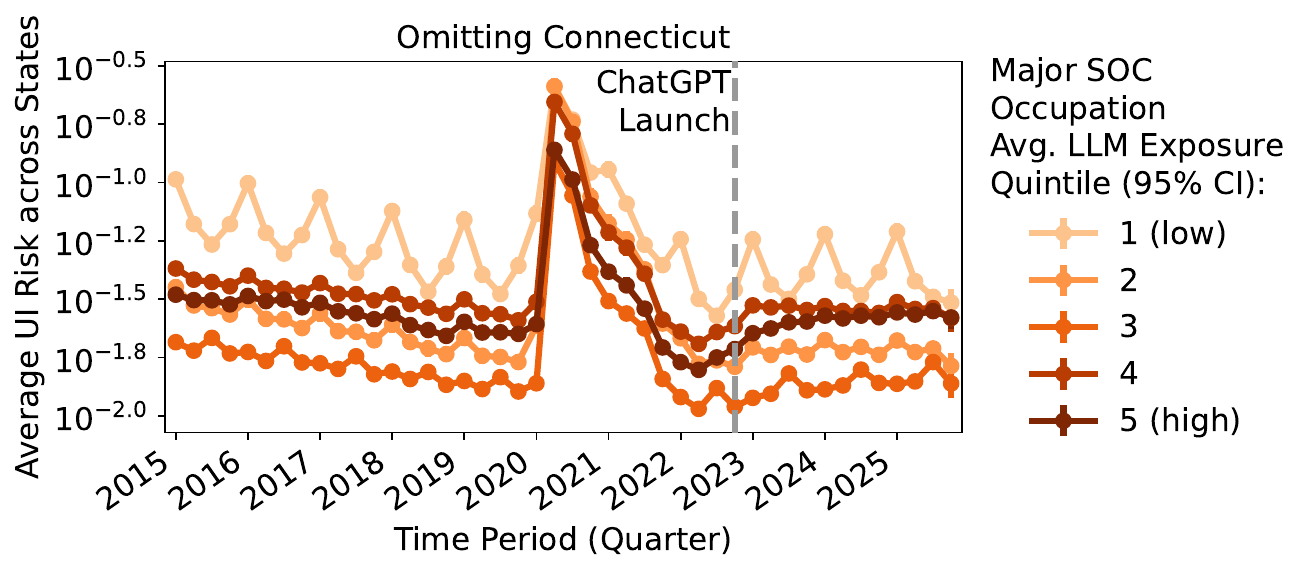}
	\caption{
		\textbf{Same as Figure~\ref{fig:uiRisk}C (upper) and Figure~\ref{fig:uiRisk}B (lower), but omitting data from Connecticut.}
		Each point is the average unemployment risk across all remaining states and months in each quarter.
	}
	\label{fig:uiRiskOmit_CT}
\end{figure}

\begin{figure}[H]
	\centering
	\includegraphics[width=\textwidth]{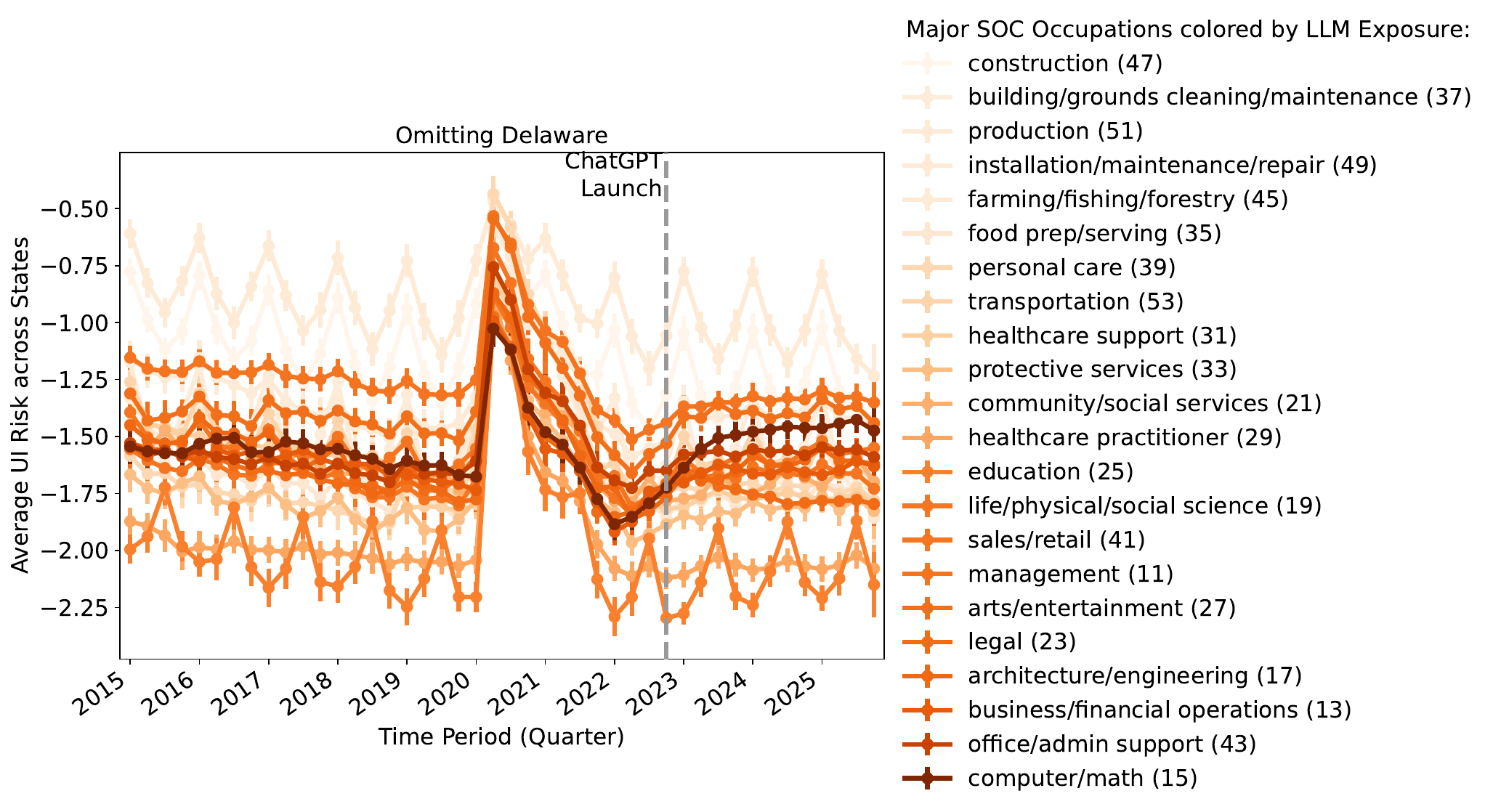}

	\includegraphics[width=\textwidth]{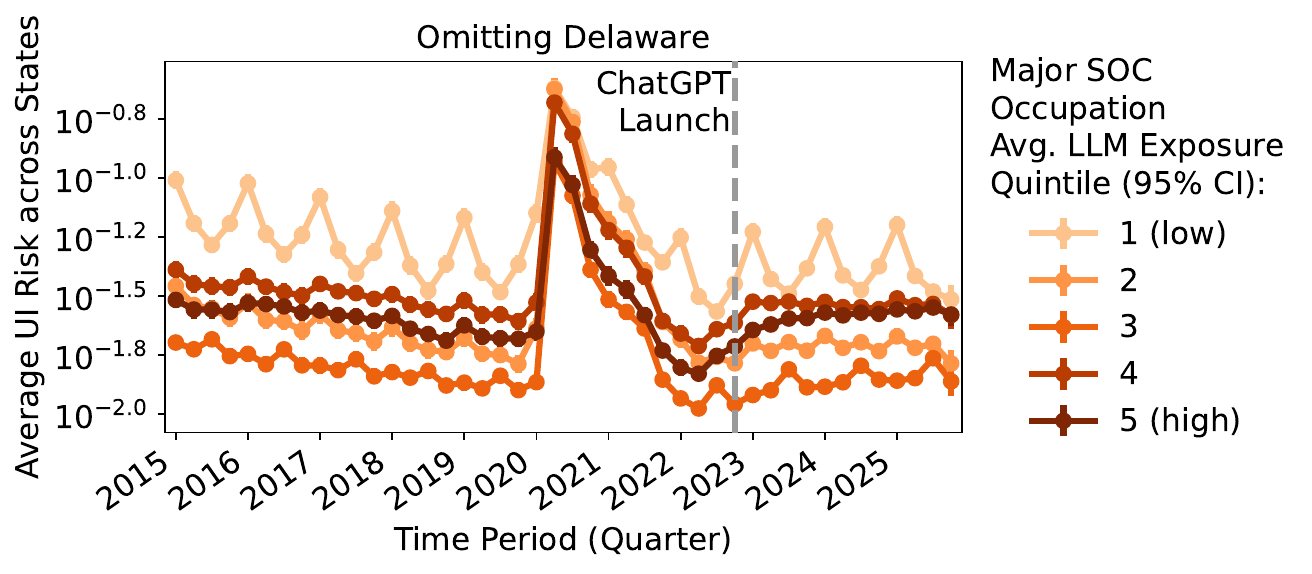}
	\caption{
		\textbf{Same as Figure~\ref{fig:uiRisk}C (upper) and Figure~\ref{fig:uiRisk}B (lower), but omitting data from Delaware.}
		Each point is the average unemployment risk across all remaining states and months in each quarter.
	}
	\label{fig:uiRiskOmit_DE}
\end{figure}

\begin{figure}[H]
	\centering
	\includegraphics[width=\textwidth]{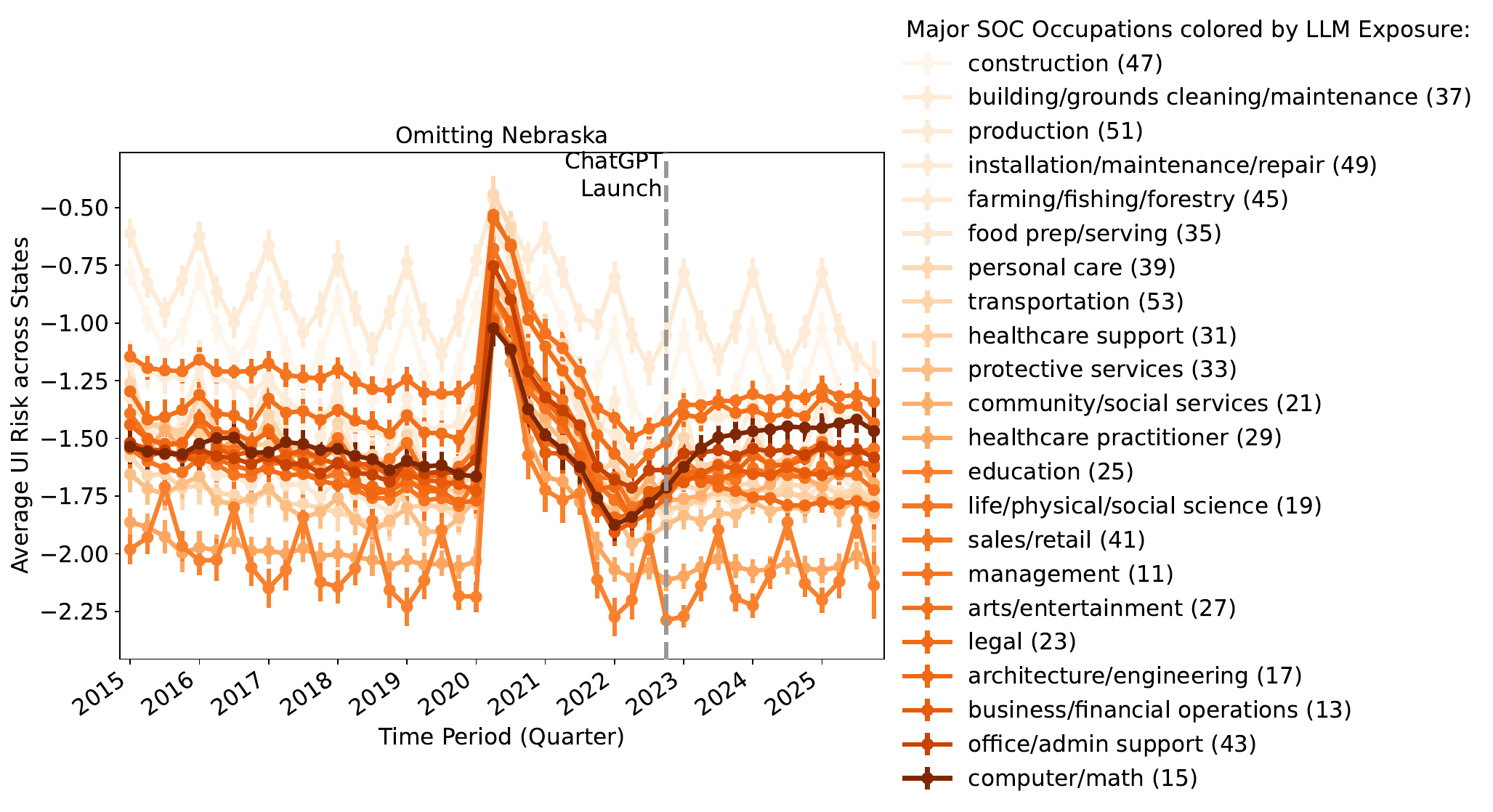}

	\includegraphics[width=\textwidth]{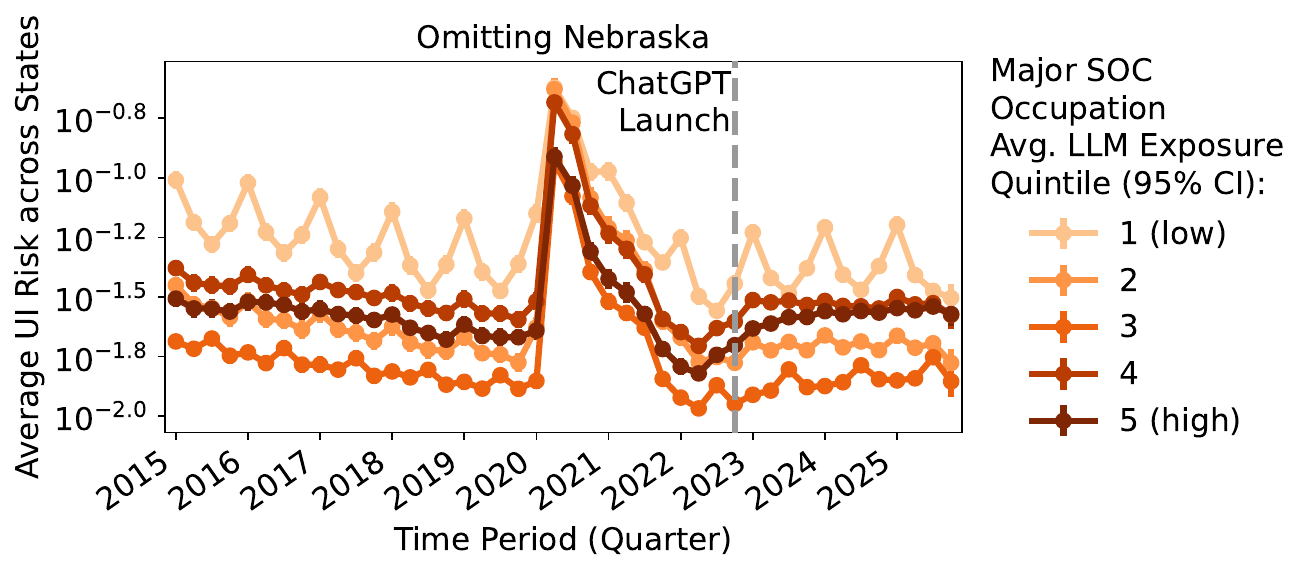}
	\caption{
		\textbf{Same as Figure~\ref{fig:uiRisk}C (upper) and Figure~\ref{fig:uiRisk}B (lower), but omitting data from Nebraska.}
		Each point is the average unemployment risk across all remaining states and months in each quarter.
	}
	\label{fig:uiRiskOmit_NE}
\end{figure}

\begin{figure}[H]
	\centering
	\includegraphics[width=\textwidth]{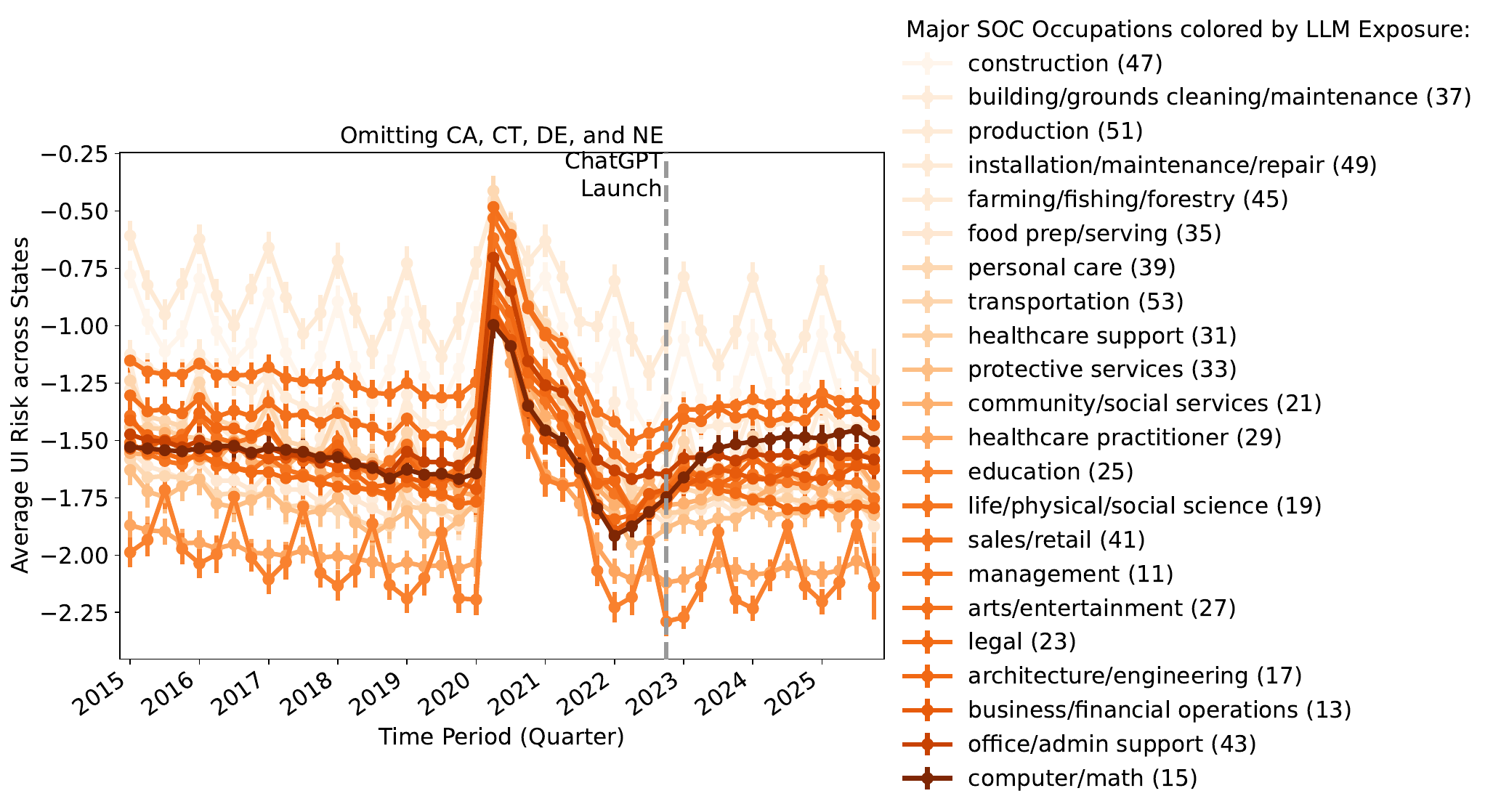}

	\includegraphics[width=\textwidth]{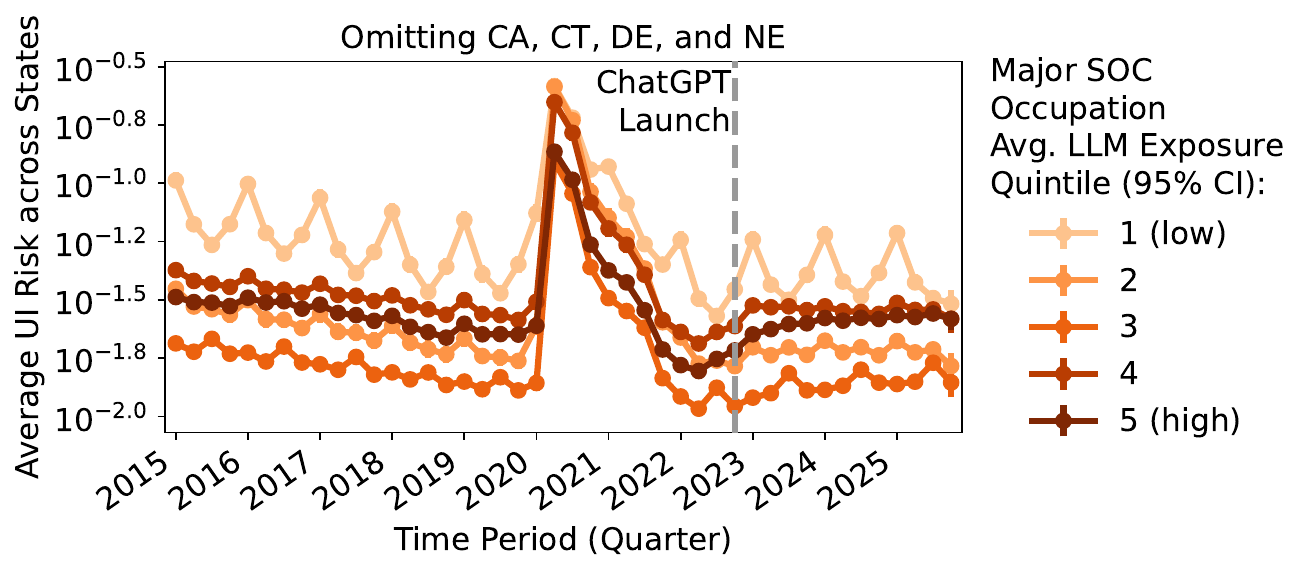}
	\caption{
		\textbf{Same as Figure~\ref{fig:uiRisk}C (upper) and Figure~\ref{fig:uiRisk}B (lower), but omitting data from California, Connecticut, Delaware, and Nebraska.}
		Each point is the average unemployment risk across all remaining states and months in each quarter.
	}
	\label{fig:uiRiskOmit_CACTDENE}
\end{figure}

\revise{Two kinds of state can distort the national aggregates in Fig.~\ref{fig:uiRisk}B\&C.
California and Connecticut show quarterly trends visibly unlike most other states, and California is also one of the two jurisdictions responsible for most of the zero-claimant cells documented in Table~\ref{tab:uiMissingData}; Delaware is the other, and Nebraska is the next largest contributor of zero-claimant cells.
As a robustness check we therefore recompute the national series while omitting each of these states in turn, and then all four together (see Figs.~\ref{fig:uiRiskOmit_CA}--\ref{fig:uiRiskOmit_CACTDENE}).
In every case the increase in unemployment risk for the most LLM-exposed occupations still begins before the launch of ChatGPT, so the timing reported in the main text is not produced by any one state.}

\revise{Following the same layout in each figure, the upper panel corresponds to Fig.~\ref{fig:uiRisk}C (unemployment risk by major occupation group) and the lower panel to Fig.~\ref{fig:uiRisk}B (unemployment risk by LLM exposure quintile).}


\foreach [count=\i] \n in {AK, AL, AR, AZ, CA, CO, CT, DE, FL, GA, HI, IA, ID, IL, IN, KS, KY, LA, MA, MD, ME, MI, MN, MO, MS, MT, NC, ND, NE, NH, NJ, NM, NV, NY, OH, OK, OR, PA, RI, SC, SD, TN, TX, UT, VA, VT, WA, WI, WV, WY} {%
    \begin{figure}[H]
        \centering
        \includegraphics[width=.8\textwidth]{figures/siFigures/stateUiRiskTimeseries/relUiRiskTimeSeries_\n.pdf}    \caption{
            \textbf{Quarterly unemployment risk by major occupation group in state of \n.}
            Each point is the average unemployment risk across months in each quarter.
            Lines are shifted to be relative to the first quarter of 2022
        }
        \label{fig:allUiRiskTimeSeries_\n}
    \end{figure}
}

\clearpage
\section{Trends in Unemployment Risk using Alternative Exposure Scores}
\label{si:sec:altExposure}

\begin{figure}[H]
    \centering
    \begin{overpic}[width=.39\textwidth]{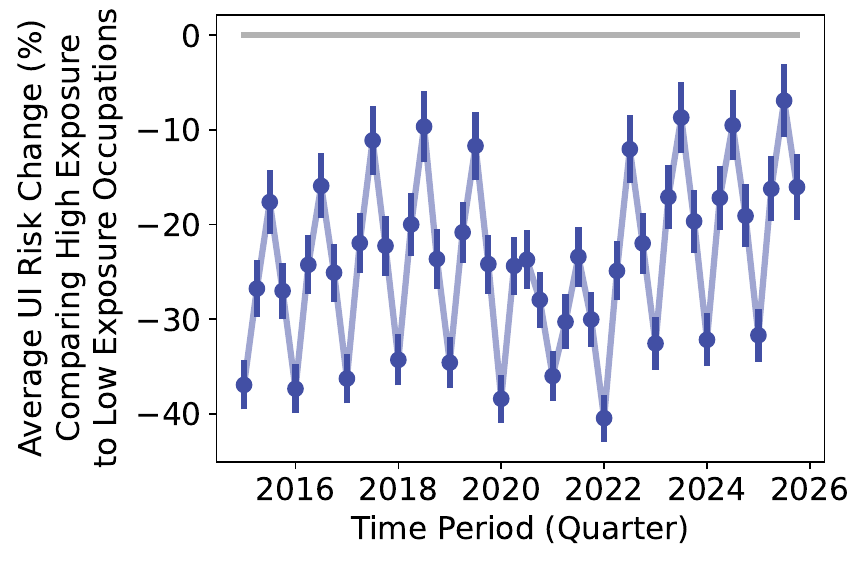}
        \put(26,53){\fbox{\small A}}
    \end{overpic}
    \begin{overpic}[width=.59\textwidth]{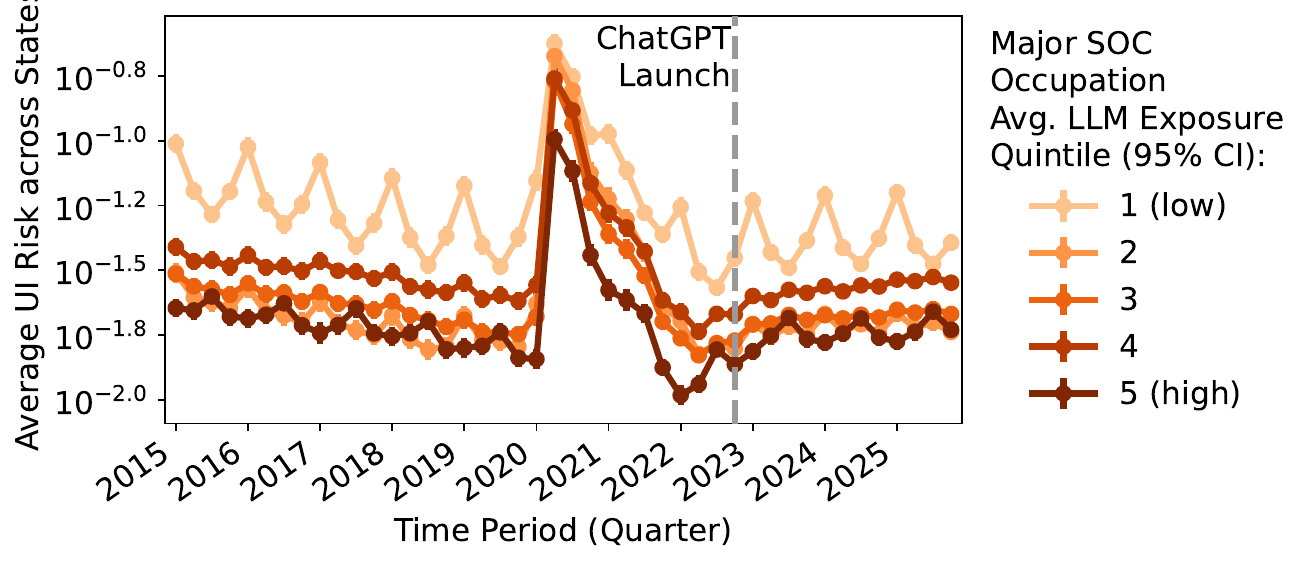}
        \put(14,37){\fbox{\small B}}
    \end{overpic}
    \begin{overpic}[width=.8\textwidth]{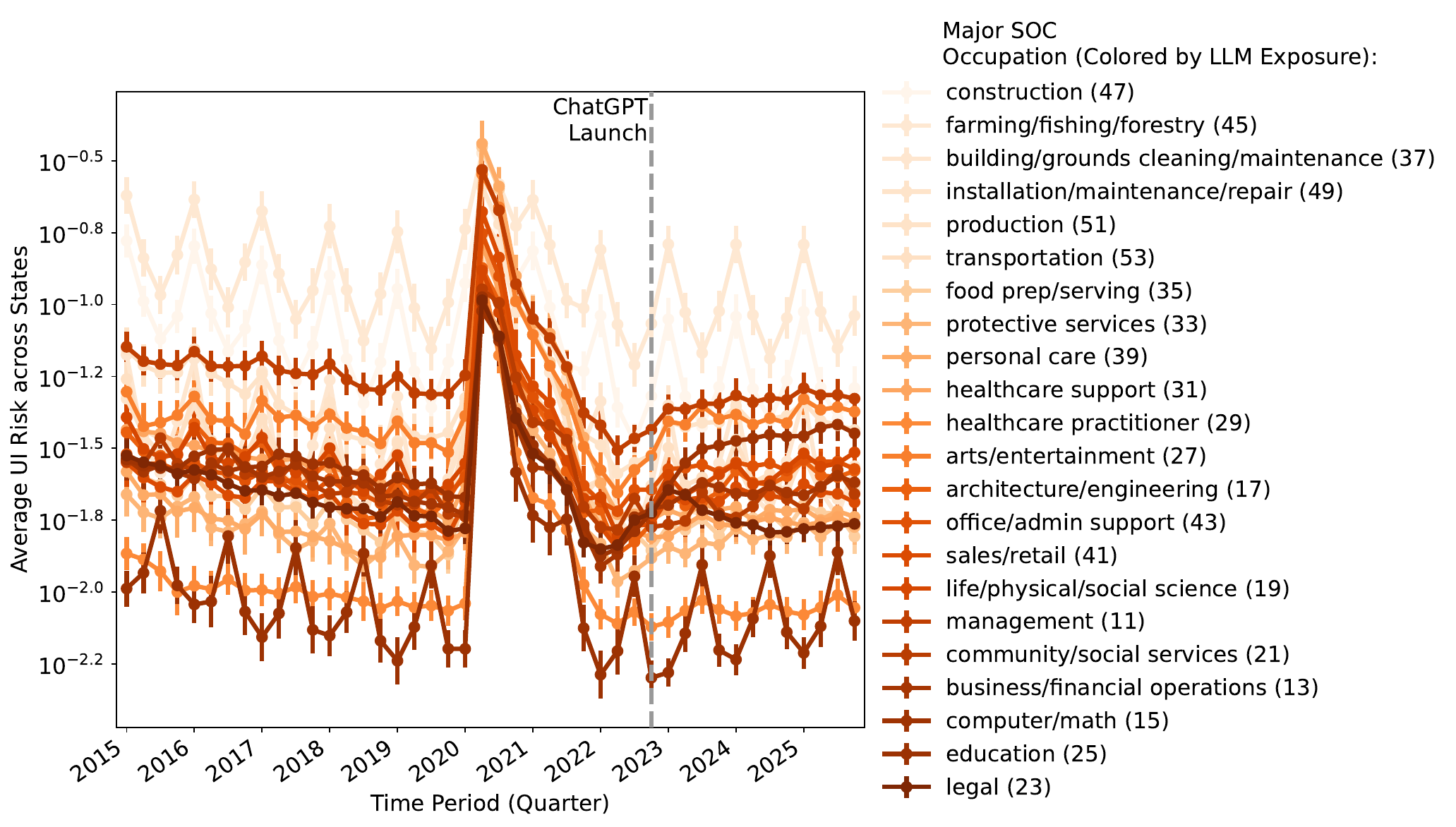}
        \put(10,45){\fbox{\small C}}
    \end{overpic}
    \caption{
        \textbf{Unemployment risk for AI (LLM) exposed workers increased starting before the launch of ChatGPT using an alternative AI exposure score from Felten et al~\cite{felten2018method,felten2023occupational}.}
        {\bf (A)} Unemployment risk for occupations with high exposure compared to occupations with low exposure.
        {\bf (B)} Unemployment risk for occupations grouped into exposure quintiles.
        {\bf (C)} Unemployment risk by major occupation.
        Lines are colored by major occupation's AI-exposure.
        We compute unemployment risk monthly across states and report quarterly series as the mean of monthly values across all states within each quarter.
    }
    \label{si:fig:feltenUiRisk}
\end{figure}

Here we consider some alternative AI exposure scores to the one used in the main text~\cite{eloundou2024gpts}.
The first alternative score we use is from Felten et al~\cite{felten2018method,felten2023occupational}.
That the main results about the early rise of unemployment risk in the main figure remain unchanged (see Fig.~\ref{si:fig:feltenUiRisk}).

Next, we incorporate actual AI usage data from Claude according to Anthropic's Economic Index Report~\cite{appelmccrorytamkin2025geoapi}. 
This report includes O*NET tasks inferred from the conversations of actual Claude AI users. 
However, additional work is needed to turn these statistics into something like a general AI exposure score that can be applied to occupations across the U.S. workforce because Claude users may not represent all workers and Claude users may not represent all LLM users (e.g., AI users may tend toward Claude for certain tasks but other AI systems for other tasks).
Additionally, the report characterizes conversations containing each task as either automating or augmenting.

Let $\mathcal{T}$ denote the set of O*NET work activities (tasks) and let $\text{AI} \in \{0,1\}$ indicate whether an occupation's task is engaged via an AI system (specifically, Claude). 
We seek to estimate $p(\text{automation} \mid \text{occ})$, the probability that a given occupation $occ$ is subject to automative AI use, and $p(\text{augmenting} \mid \text{occ})$, the probability that a given occupation $occ$ is subject to augmenting AI use, drawing on the following empirical distributions:

\begin{itemize}
    \item $p(\text{task} \mid \text{occ})$: the distribution over O*NET tasks within an occupation, sourced from the Anthropic Economic Index, which inherits task--occupation mappings from O*NET;
    \item $p(\text{task} \mid \text{AI})$: the probability that a given O*NET task is detected in Claude user conversations;
    \item $p(\text{automation} \mid \text{task}, \text{AI})$ and $p(\text{augmentation} \mid \text{task}, \text{AI})$: conditional on AI use and task detection, the probability that the interaction is classified as automative or augmentative, respectively;
    \item $p(\text{occ})$: the marginal occupation distribution, estimated from Bureau of Labor Statistics Occupational Employment and Wage Statistics (BLS OEWS);
    \item $p(\text{AI} \mid \text{occ})$: the probability that workers in a given occupation use AI, approximated via \cite{eloundou2024gpts}'s AI exposure measure in the absence of ground truth. Note that this is likely an overestimation since exposure to AI only indicates that AI could do something and not that AI will actually be used to do it.
\end{itemize}


We begin by marginalizing over tasks and AI use:
\begin{equation}
    p(\text{automation} \mid \text{occ})
    = \sum_{\text{AI}} \sum_{\text{task}}
      p(\text{automation} \mid \text{task}, \text{AI}) \cdot
      p(\text{task} \mid \text{AI}, \text{occ}) \cdot
      p(\text{AI} \mid \text{occ}).
    \label{eq:full_marginalization}
\end{equation}
Since $p(\text{automation} \mid \text{task}, \text{AI} = 0) = 0$ by definition---automation risk is only realized through AI-mediated task engagement---the sum over $\text{AI}$ collapses to the $\text{AI} = 1$ term, and we write:
\begin{equation}
    p(\text{automation} \mid \text{occ})
    = \sum_{\text{task}}
      p(\text{automation} \mid \text{task}, \text{AI}) \cdot
      p(\text{task} \mid \text{AI}, \text{occ}) \cdot
      p(\text{AI} \mid \text{occ}).
\end{equation}


The term $p(\text{task} \mid \text{AI}, \text{occ})$ is not directly observed. Applying Bayes' rule yields:
\begin{equation}
    p(\text{task} \mid \text{AI}, \text{occ})
    = \frac{p(\text{task} \mid \text{occ}) \cdot p(\text{AI} \mid \text{task}, \text{occ})}{p(\text{AI} \mid \text{occ})}.
\end{equation}
We invoke the simplifying assumption that AI adoption and occupation are conditionally independent given task---that is, AI use is mediated entirely through the nature of the task rather than the broader occupational context.
That is, $p(\text{AI} \mid \text{task}, \text{occ}) \approx p(\text{AI} \mid \text{task}).$

\noindent Under this assumption:
\begin{equation}
    p(\text{task} \mid \text{AI}, \text{occ})
    \approx \frac{p(\text{task} \mid \text{occ}) \cdot p(\text{AI} \mid \text{task})}{p(\text{AI} \mid \text{occ})}.
    \label{eq:task_given_ai_occ}
\end{equation}


Substituting \eqref{eq:task_given_ai_occ} into \eqref{eq:full_marginalization}, the $p(\text{AI} \mid \text{occ})$ terms cancel:
\begin{equation}
    p(\text{automation} \mid \text{occ})
    \approx \sum_{\text{task}}
      p(\text{automation} \mid \text{task}, \text{AI}) \cdot
      p(\text{task} \mid \text{occ}) \cdot
      p(\text{AI} \mid \text{task}).
    \label{eq:simplified}
\end{equation}


The available data supply $p(\text{task} \mid \text{AI})$ rather than $p(\text{AI} \mid \text{task})$. Applying Bayes' rule:
\begin{equation}
    p(\text{AI} \mid \text{task})
    = \frac{p(\text{task} \mid \text{AI}) \cdot p(\text{AI})}{p(\text{task})},
    \label{eq:bayes_flip}
\end{equation}
where the marginals are estimated from the joint distribution over tasks and occupations:
\begin{equation}
    p(\text{task}) = \sum_{\text{occ}} p(\text{task} \mid \text{occ})\, p(\text{occ}),
    \qquad
    p(\text{AI}) = \sum_{\text{occ}} p(\text{AI} \mid \text{occ})\, p(\text{occ}).
\end{equation}

\begin{figure}[H]
    \centering
    \begin{overpic}[width=.39\textwidth]{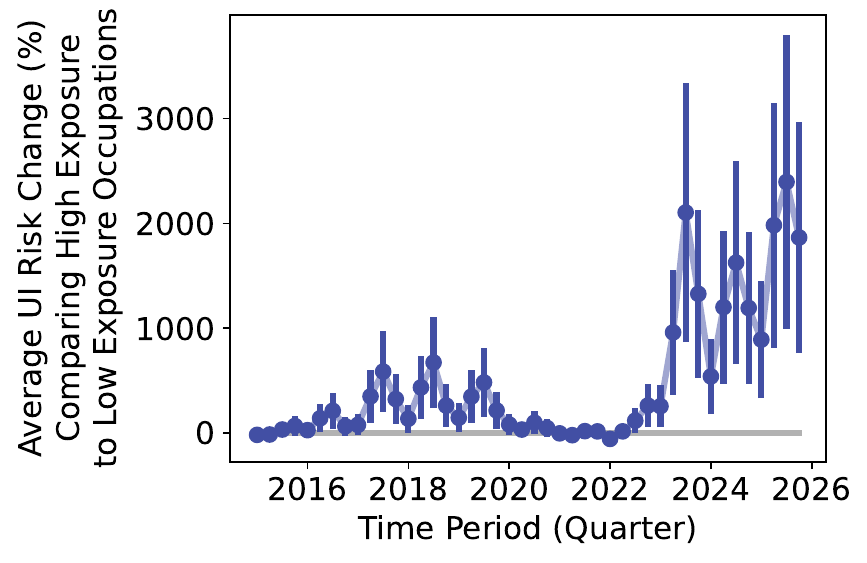}
        \put(26,53){\fbox{\small A}}
    \end{overpic}
    \begin{overpic}[width=.59\textwidth]{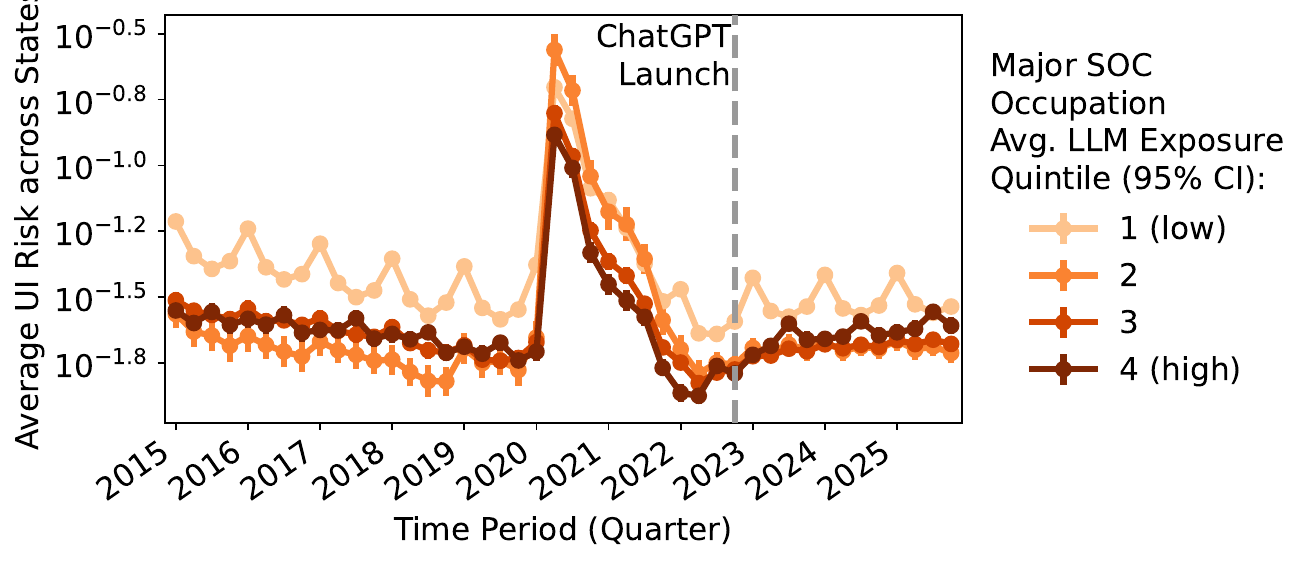}
        \put(14,37){\fbox{\small B}}
    \end{overpic}
    \begin{overpic}[width=.8\textwidth]{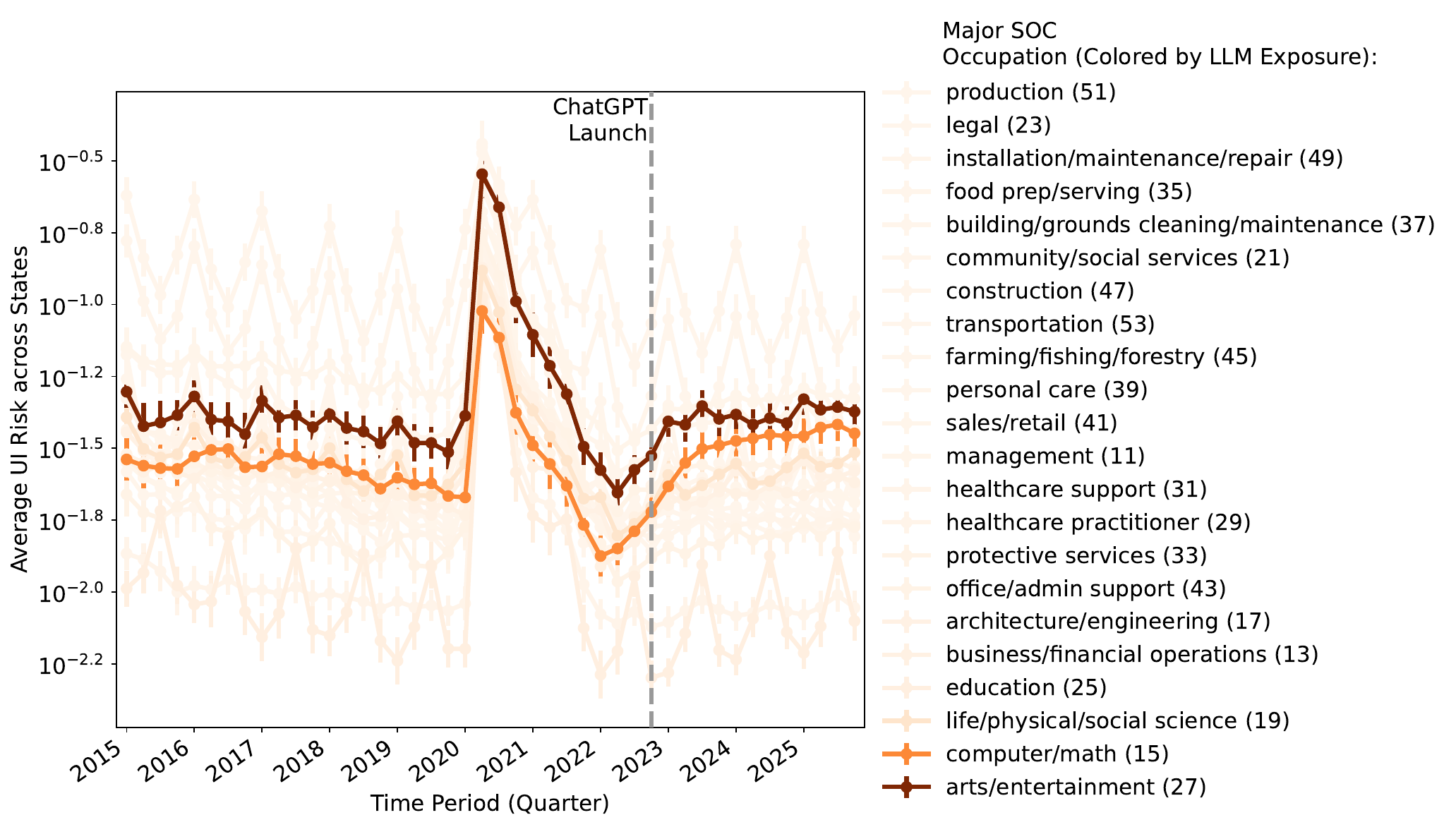}
        \put(10,45){\fbox{\small C}}
    \end{overpic}
    \caption{
        \textbf{Unemployment risk for AI (LLM) exposed workers increased starting before the launch of ChatGPT using an alternative AI exposure score for task automation from Anthropic's Economic Index Report~\cite{appelmccrorytamkin2025geoapi}.}
        {\bf (A)} Unemployment risk for occupations with high exposure compared to occupations with low exposure.
        {\bf (B)} Unemployment risk for occupations grouped into exposure quintiles.
        {\bf (C)} Unemployment risk by major occupation.
        Lines are colored by major occupation's AI-exposure.
        We compute unemployment risk monthly across states and report quarterly series as the mean of monthly values across all states within each quarter.
    }
    \label{si:fig:anthropicAutoUiRisk}
\end{figure}
\begin{figure}[H]
    \centering
    \begin{overpic}[width=.39\textwidth]{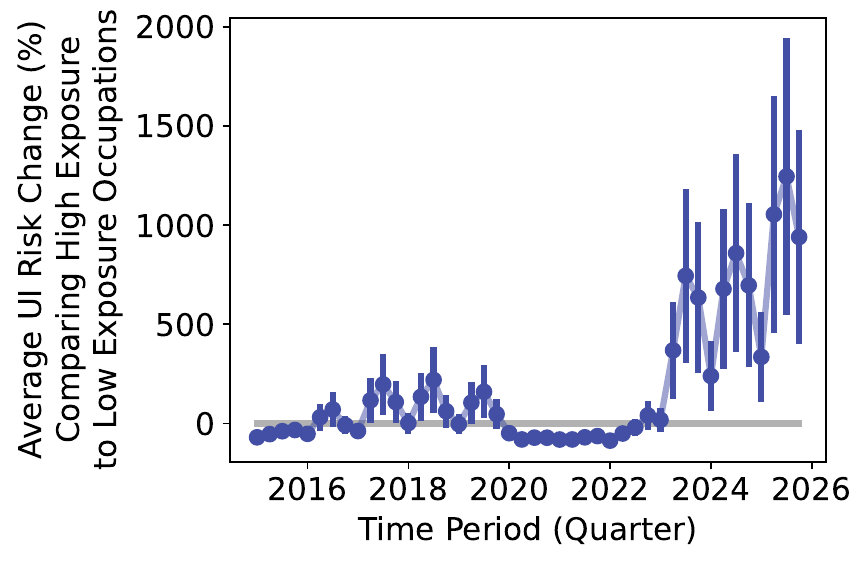}
        \put(26,53){\fbox{\small A}}
    \end{overpic}
    \begin{overpic}[width=.59\textwidth]{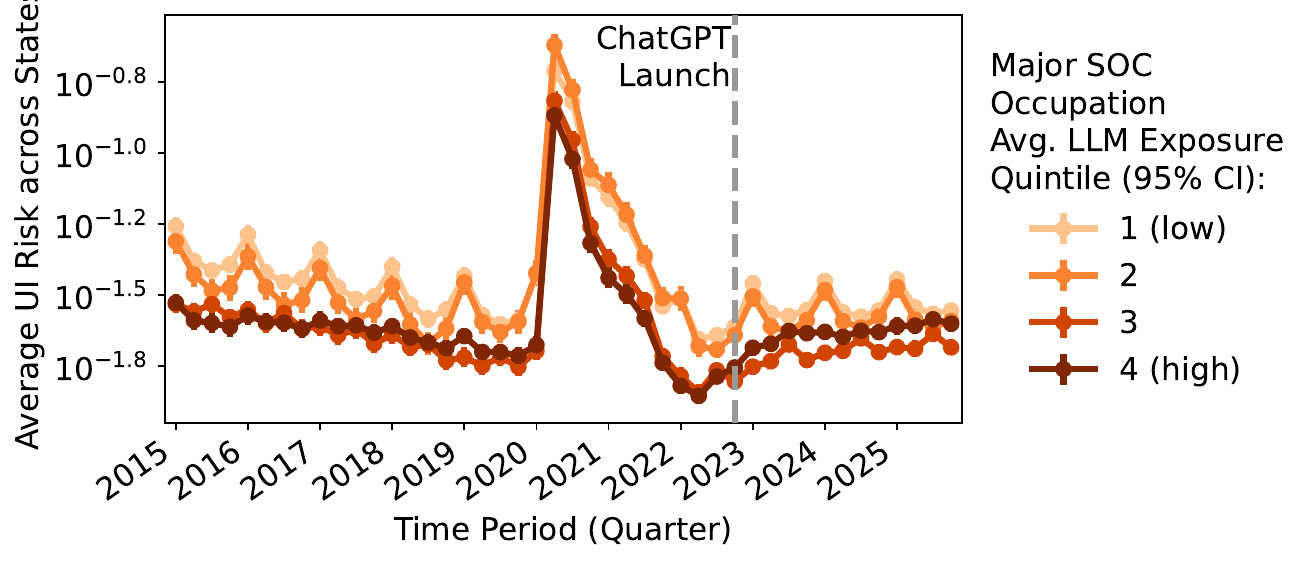}
        \put(14,37){\fbox{\small B}}
    \end{overpic}
    \begin{overpic}[width=.8\textwidth]{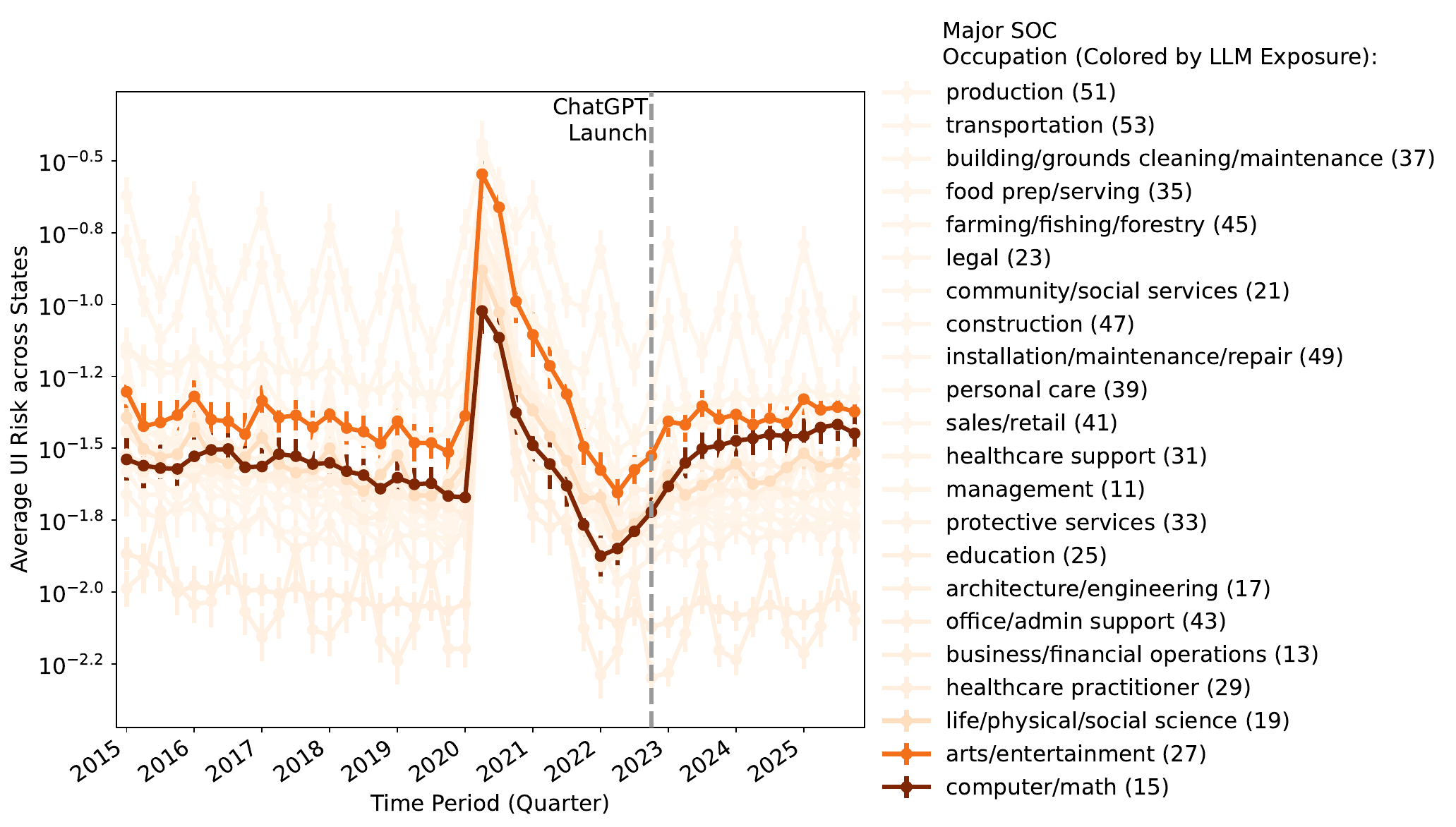}
        \put(10,45){\fbox{\small C}}
    \end{overpic}
    \caption{
        \textbf{Unemployment risk for AI (LLM) exposed workers increased starting before the launch of ChatGPT using an alternative AI exposure score for task augmentation from Anthropic's Economic Index Report~\cite{appelmccrorytamkin2025geoapi}.}
        {\bf (A)} Unemployment risk for occupations with high exposure compared to occupations with low exposure.
        {\bf (B)} Unemployment risk for occupations grouped into exposure quintiles.
        {\bf (C)} Unemployment risk by major occupation.
        Lines are colored by major occupation's AI-exposure.
        We compute unemployment risk monthly across states and report quarterly series as the mean of monthly values across all states within each quarter.
    }
    \label{si:fig:anthropicAugUiRisk}
\end{figure}

Substituting \eqref{eq:bayes_flip} into \eqref{eq:simplified} yields the complete estimator:
\begin{equation}
    p(\text{automation} \mid \text{occ})
    \approx \sum_{\text{task}}
      p(\text{automation} \mid \text{task}, \text{AI}) \cdot
      p(\text{task} \mid \text{occ}) \cdot
      \frac{p(\text{task} \mid \text{AI}) \cdot p(\text{AI})}{p(\text{task})}.
    \label{eq:final}
\end{equation}
Equation~\eqref{eq:final} expresses $p(\text{automation} \mid \text{occ})$ entirely in terms of the empirically available distributions enumerated above, under the conditional independence assumption that AI adoption is fully mediated by task content rather than occupational identity. Applying a similar approach yields $p(\text{augmenting} \mid \text{occ})$.
Using either $p(\text{automation} \mid \text{occ})$ or $p(\text{augmenting} \mid \text{occ})$  produces similar results to those presented in Figure 1 using Eloundou et al. (see Fig.~\ref{si:fig:anthropicAutoUiRisk} and Fig.~\ref{si:fig:anthropicAugUiRisk}).

\clearpage
\section{\revise{Unemployment Risk By Month}}
\label{si:sec:uiMonth}
\begin{figure}[p!]
    \centering
    \begin{overpic}[width=\textwidth]{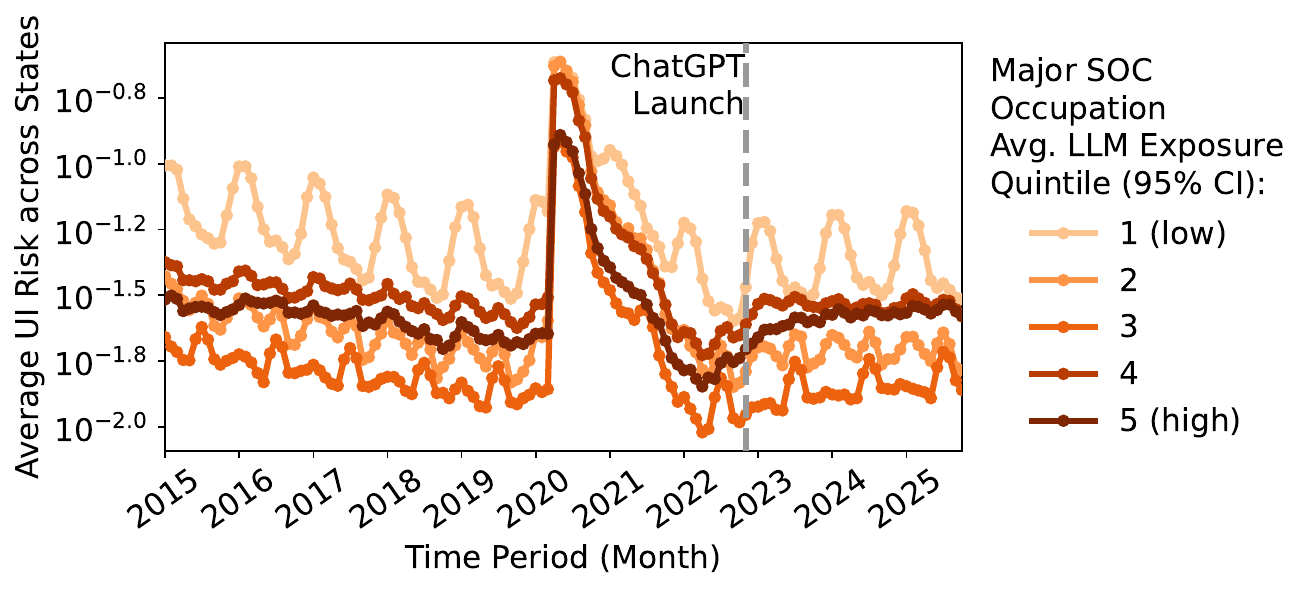}
        \put(15,38){\fbox{\small A}}
    \end{overpic}
    \begin{overpic}[width=\textwidth]{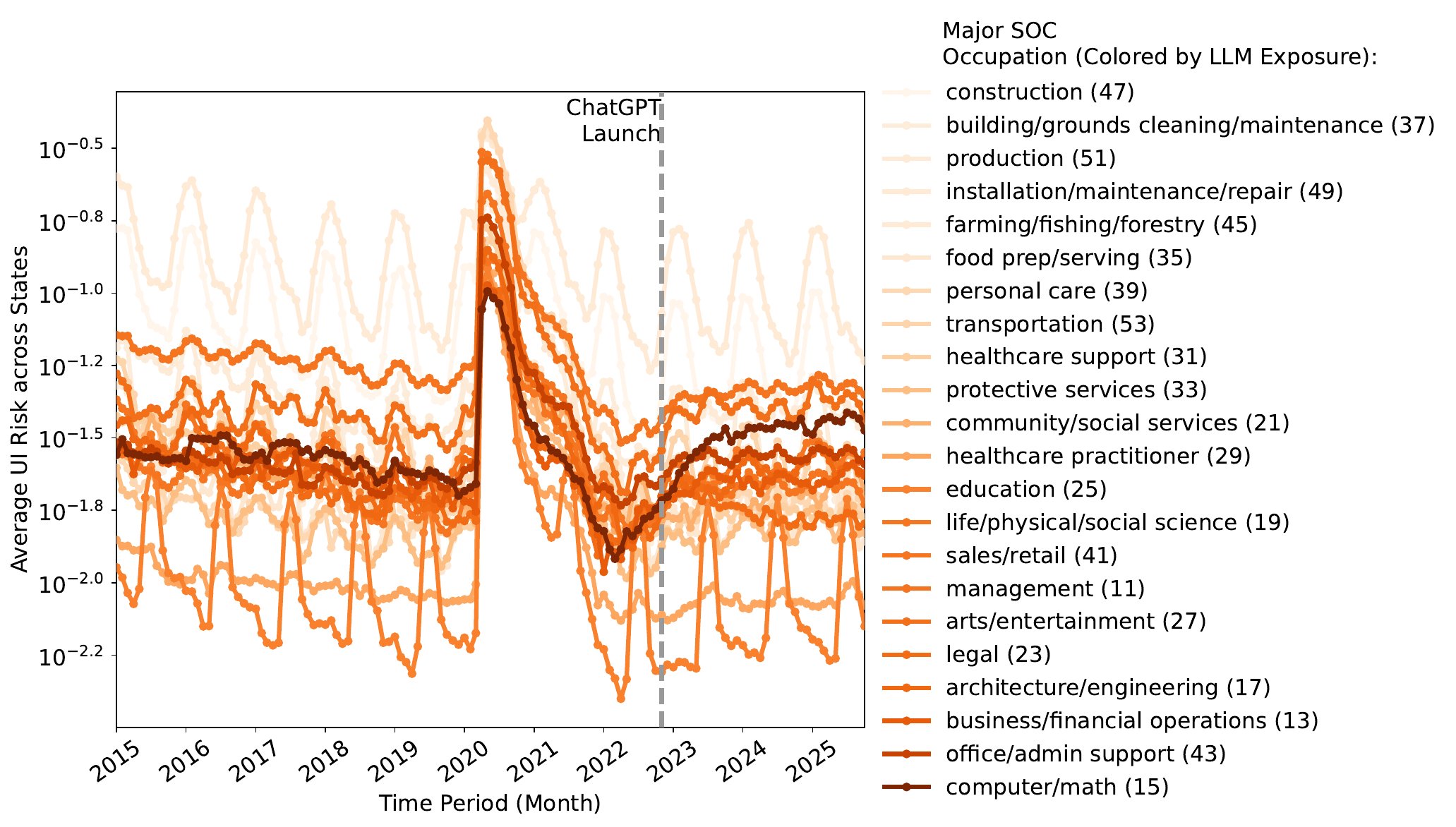}
        \put(10,46){\fbox{\small B}}
    \end{overpic}
    \caption{
        Monthly unemployment risk time series grouping major occupations into AI exposure~\cite{eloundou2024gpts} quintiles {\bf (A)} or as individual time series {\bf (B)}.
    }
    \label{si:fig:monthlyUiRisk}
\end{figure}

\begin{figure}[p!]
    \centering
    \begin{overpic}[width=\textwidth]{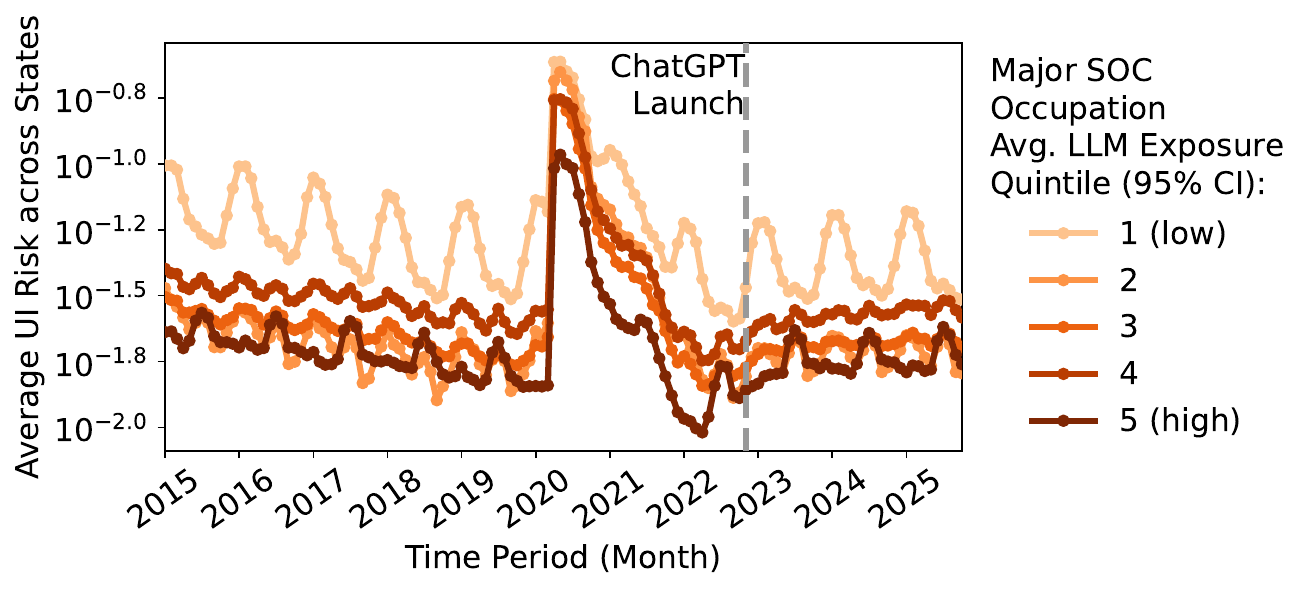}
        \put(15,38){\fbox{\small A}}
    \end{overpic}
    \begin{overpic}[width=\textwidth]{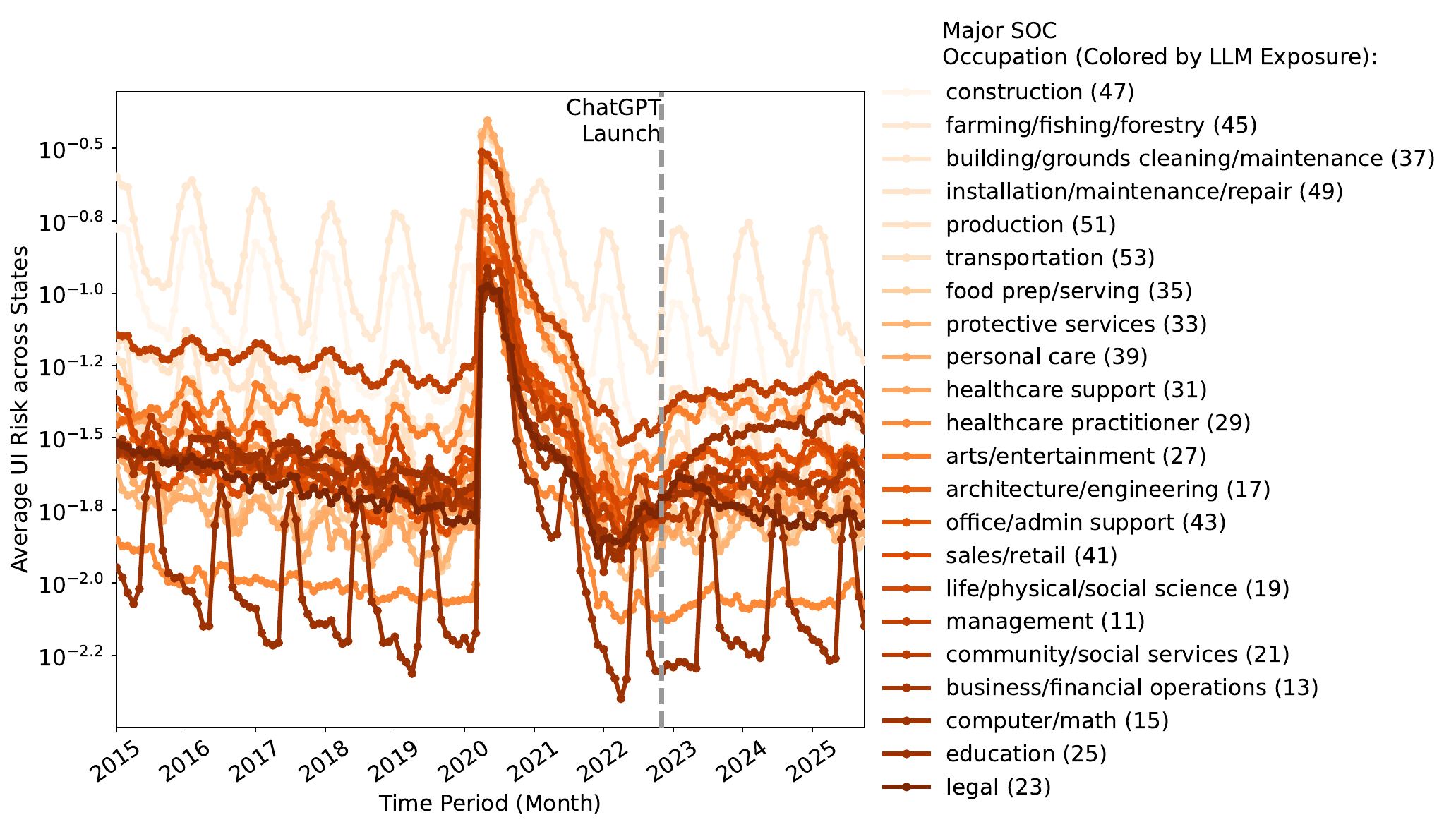}
        \put(10,46){\fbox{\small B}}
    \end{overpic}
    \caption{
        Monthly unemployment risk time series grouping major occupations into AI exposure~\cite{felten2023occupational} quintiles {\bf (A)} or as individual time series {\bf (B)}.
    }
    \label{si:fig:monthlyUiRisk_felten}
\end{figure}

\begin{figure}[p!]
    \centering
    \begin{overpic}[width=\textwidth]{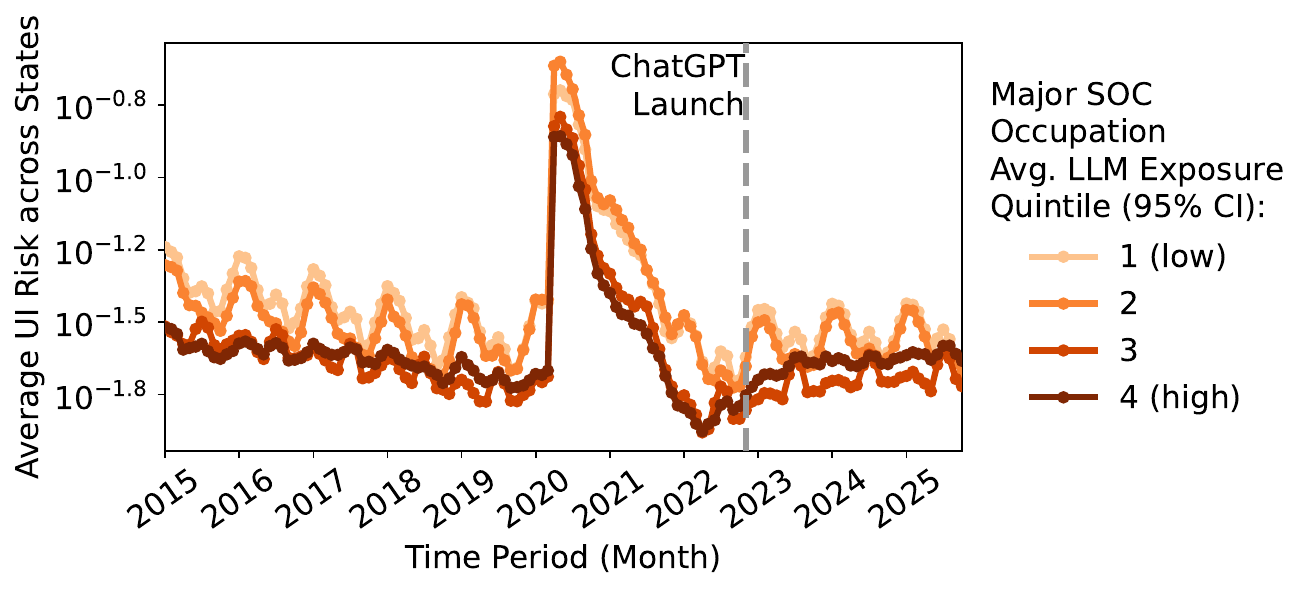}
        \put(15,38){\fbox{\small A}}
    \end{overpic}
    \begin{overpic}[width=\textwidth]{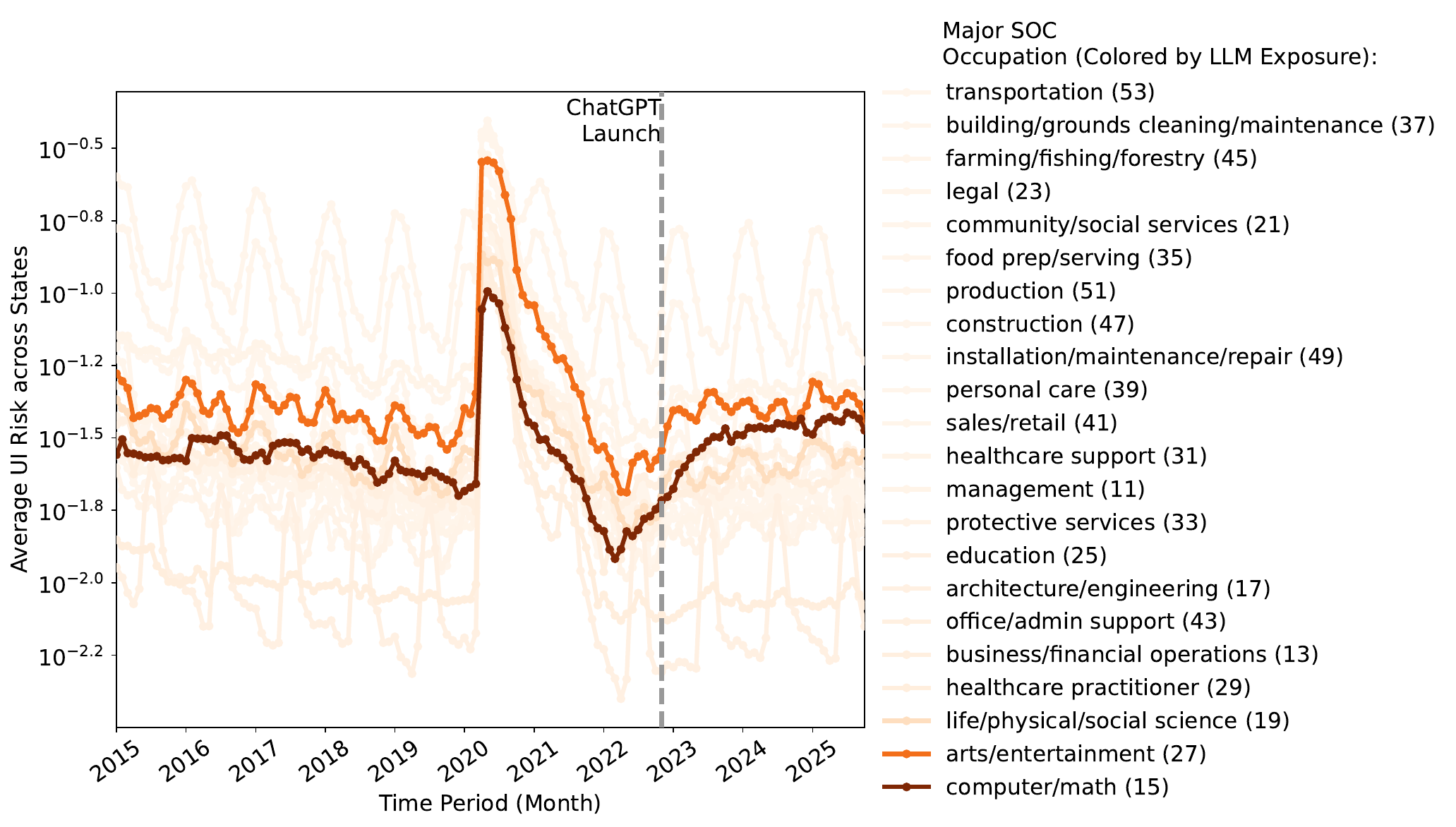}
        \put(10,46){\fbox{\small B}}
    \end{overpic}
    \caption{
        Monthly unemployment risk time series grouping major occupations into AI exposure quintiles {\bf (A)} or as individual time series {\bf (B)} according to their task augmentation from Anthropic's Economic Index Report~\cite{appelmccrorytamkin2025geoapi}.
        See SI Section~\ref{si:sec:altExposure} for exposure calculation.
    }
    \label{si:fig:monthlyUiRisk_anthropicAug}
\end{figure}

\begin{figure}[p!]
    \centering
    \begin{overpic}[width=\textwidth]{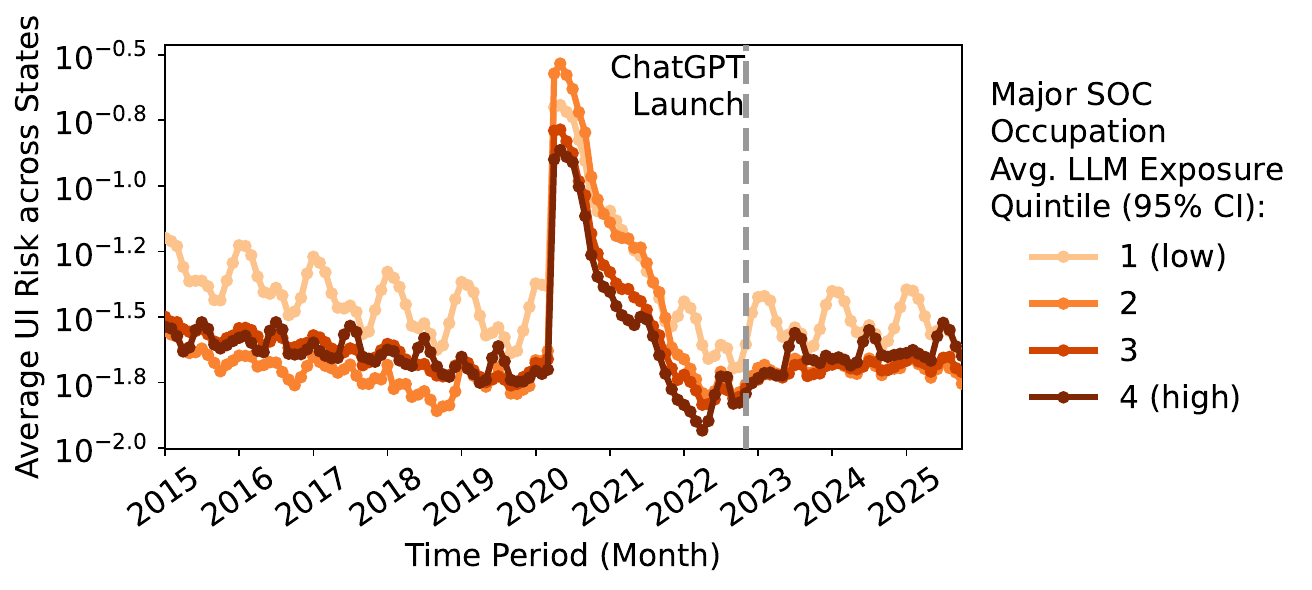}
        \put(15,38){\fbox{\small A}}
    \end{overpic}
    \begin{overpic}[width=\textwidth]{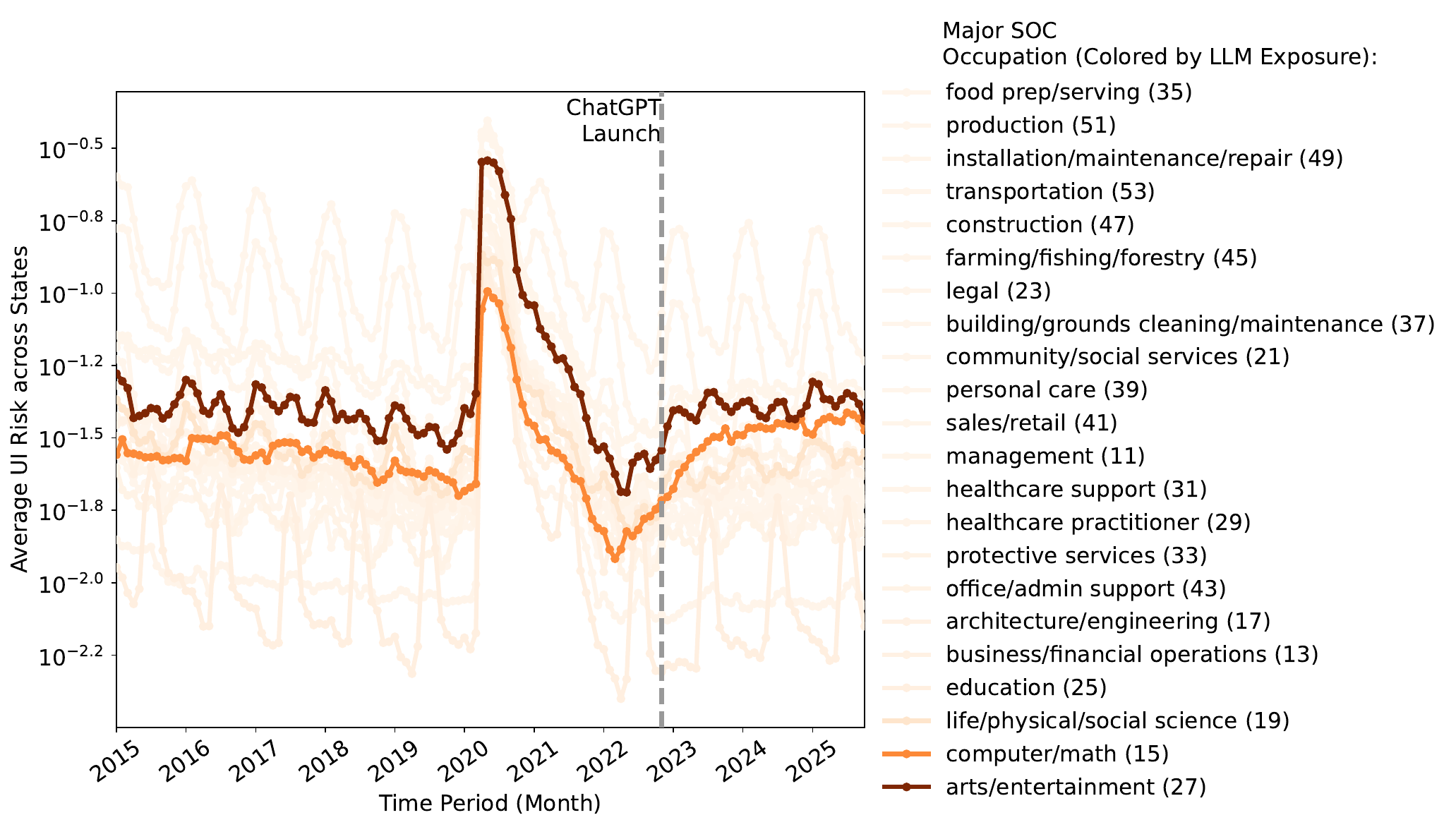}
        \put(10,46){\fbox{\small B}}
    \end{overpic}
    \caption{
        Monthly unemployment risk time series grouping major occupations into AI exposure quintiles {\bf (A)} or as individual time series {\bf (B)} according to their task automation from Anthropic's Economic Index Report~\cite{appelmccrorytamkin2025geoapi}.
        See SI Section~\ref{si:sec:altExposure} for exposure calculation.
    }
    \label{si:fig:monthlyUiRisk_anthropicAuto}
\end{figure}

This section recreates unemployment risk time series from the main text with monthly time points instead of quarterly.
\revise{Figure~\ref{si:fig:monthlyUiRisk} presents the monthly series using the LLM exposure scores of Eloundou et al.~\cite{eloundou2024gpts} used throughout the main text, and Figs.~\ref{si:fig:monthlyUiRisk_felten}--\ref{si:fig:monthlyUiRisk_anthropicAuto} repeat the exercise for each alternative exposure measure considered in SI Section~\ref{si:sec:altExposure}: the AI occupational impact scores of Felten et al.~\cite{felten2018method,felten2023occupational}, and task augmentation and task automation from Anthropic's Economic Index Report~\cite{appelmccrorytamkin2025geoapi}.
Across all four exposure measures, the monthly series reproduce the quarterly patterns reported in the main text, indicating that the quarterly aggregation used in Fig.~\ref{fig:uiRisk} is not responsible for those patterns.}

\clearpage
\subsection{\revise{Finding breaks in the unemployment risk time series}}
\label{si:sec:uiBreaks}
\begin{figure}[t]
    \centering
    \includegraphics[width=\textwidth]{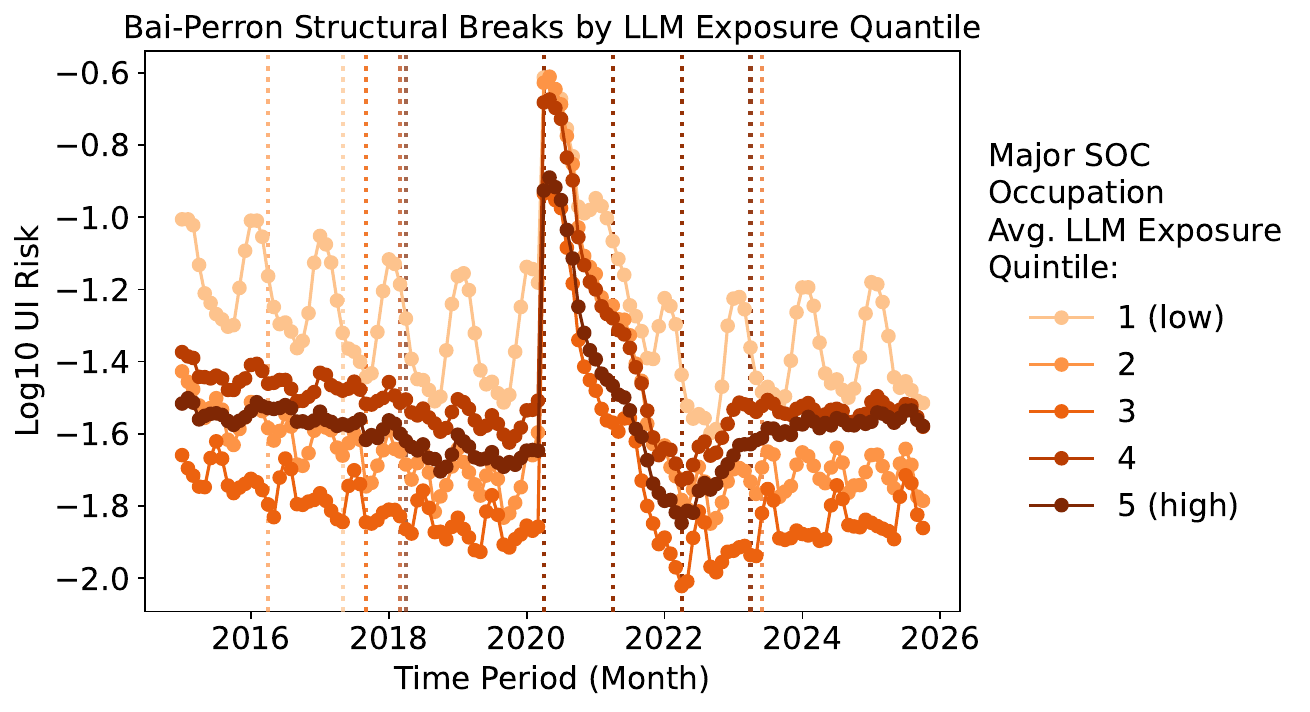}
    \caption{
        Log$_{10}$ monthly unemployment insurance (UI) risk over time by quintile of occupational LLM exposure. 
        Colored points show, for each exposure quintile, the average $\log_{10}$ UI risk across constituent occupations in each month; darker orange denotes higher LLM exposure. 
        Dotted vertical lines mark structural breaks identified independently within each quintile's series by a Bai and Perron (2003)-style multiple structural change procedure.
    }
    \label{fig:uiBreaks}
\end{figure}

\revise{
To identify structural changes in unemployment insurance (UI) risk over time, we applied a Bai and Perron~\cite{bai2003computation}-style multiple structural change procedure independently to the $\log_{10}$-transformed monthly UI risk series for each quintile of occupational LLM exposure (see Figure~\ref{fig:uiBreaks} and Table~\ref{tab:baiperron_breaks}). 
For each quintile, we located, for every candidate number of breaks $k \in \{0, \dots, 10\}$, the exact partition minimizing the within-regime sum of squared residuals via dynamic programming (\texttt{ruptures}, \texttt{Dynp}, $L_2$ cost), enforcing a minimum regime length of 12 months, and selected $k$ by minimizing the Bayesian Information Criterion (BIC).  
All five exposure quintiles independently identify a break in April 2020, consistent with the onset of the COVID-19 pandemic and the associated surge in unemployment insurance claims, a disruption that affected occupations broadly regardless of LLM exposure. 
All five quintiles additionally identify breaks in April 2021 and April 2022, indicating that labor-market disruption from this period persisted across the entire exposure distribution through at least the early stages of pandemic recovery. 
By contrast, only the three highest-exposure quintiles exhibit a further, distinct break in 2023 (April--June), following the public launch of ChatGPT in November 2022---a pattern absent from the two lowest-exposure quintiles. 
However, this late shift is not associated with increasing unemployment risk, but rather flattening or decreasing unemployment risk.
} 

\begin{table}[H]
    \centering  
    \begin{tabular}{clcr}  
        \hline
        Exposure Quintile & Detected Breakpoints (Year-Month) & $N$ Breaks & BIC \\ \hline  
        1 (low)  & 2017-05, 2020-04, 2021-04, 2022-04                 & 4 & $-518.02$ \\  
        2        & 2016-04, 2017-09, 2020-04, 2021-04, 2022-04         & 5 & $-565.64$ \\  
        3        & 2017-09, 2020-04, 2021-04, 2022-04, 2023-06         & 5 & $-572.72$ \\  
        4        & 2018-03, 2020-04, 2021-04, 2022-04, 2023-04         & 5 & $-600.36$ \\  
        5 (high) & 2018-04, 2020-04, 2021-04, 2022-04, 2023-04         & 5 & $-626.34$ \\  
    \end{tabular}  
    \caption{Structural breaks in $\log_{10}$ monthly UI risk by LLM exposure quintile, identified via Bai and Perron (2003)-style multiple structural change detection (minimum 12-month regime length, break count selected by BIC).}  
    \label{tab:baiperron_breaks}  
\end{table}  

\subsection{\revise{Macroeconomic conditions and unemployment risk}}
\label{si:sec:uiMacro}
\revise{
The break analysis above establishes when unemployment risk changed but not what changed it.
This section asks how much of the variation in unemployment risk can be accounted for by ordinary macroeconomic conditions, and how much remains for AI exposure and for the launch of ChatGPT to explain.
We estimate a sequence of nested ordinary least squares models on the monthly panel of major occupation group $\times$ state $\times$ month cells described in SI Section~\ref{si:sec:uiRisk}, covering January 2015 through October 2025.
The dependent variable is $\log_{10} p(\text{unemp}|soc,s,t)$ from Eq.~\ref{eq:uiRisk}.
The macroeconomic condition we use is the effective federal funds rate $fed_t$ in percentage points~\cite{FEDFUNDS2025}, which the monetary tightening cycle beginning in March 2022 makes the most salient candidate for the period in question.
As above, $llm_{soc}\in[0,1]$ is the major group's average LLM exposure~\cite{eloundou2024gpts} and $gpt_t$ is an indicator for months from December 2022 onward.
All models include state and month-of-year fixed effects, the latter absorbing the seasonality of unemployment claims; the final model adds major occupation group fixed effects.
Standard errors are two-way clustered by state and by month.
Results appear in Table~\ref{tab:uiMacroModels}.
}

\revise{
Three findings bear on the timing question.
First, the interest rate accounts for a substantial share of the movement in unemployment risk on its own: adding $fed$ to the fixed effects alone raises $R^2$ from $0.325$ to $0.350$ (Models 1--2).
Second, the interaction $llm\times fed$ is positive and significant ($0.075$, Model 3), which is to say that the exposure gap in unemployment risk narrows as interest rates rise.
The magnitude is large enough to account for the narrowing seen in Fig.~\ref{fig:uiRisk}A without reference to AI.
Evaluated at a zero federal funds rate, Model 3 places unemployment risk $66\%$ lower in a fully exposed occupation than in an unexposed one ($10^{-0.467}=0.34$), consistent with the range reported in the main text; evaluated at a rate of $5\%$, the same model places it only $19\%$ lower ($10^{-0.092}=0.81$).
Third, and most directly relevant to whether the launch of ChatGPT marks a change, the two terms that carry an AI-specific post-launch effect---$llm\times gpt$ and $gpt\times fed\times llm$---raise $R^2$ from $0.399$ to $0.399$, a change of less than $0.001$ (Models 5--6).
Neither term is individually significant at conventional levels, and both carry negative signs, the opposite of what displacement of exposed workers would predict.
}

\revise{
Model 4 is included because it shows what happens when the rate cycle is left out of the interaction structure.
It is the specification a researcher would estimate if they suspected ChatGPT and did not suspect monetary policy: it omits $llm\times fed$ and retains $llm\times gpt$.
In that model $llm\times gpt$ is positive and highly significant ($0.261$), which would ordinarily be read as exposed occupations deteriorating after the launch.
Restoring $llm\times fed$ in Model 6 drives the same coefficient to $-0.110$ and to insignificance, and drives the $gpt$ indicator from $0.366$ to $0.068$, a reduction of $81\%$.
The apparent post-ChatGPT penalty for exposed occupations is therefore not robust to controlling for the interest rate environment; it is what the tightening cycle looks like when the tightening cycle is not in the model.
Model 7 adds major occupation group fixed effects and leaves every interaction essentially unchanged ($llm\times fed$ from $0.119$ to $0.120$, $llm\times gpt$ unchanged at $-0.110$, $gpt\times fed$ unchanged at $0.117$), which indicates that this structure is not an artifact of exposed occupations carrying lower unemployment risk on average.
The $llm$ main effect is absorbed by those fixed effects and is therefore not reported in Model 7.
}

\revise{
We are deliberate about what these models do and do not establish.
They are an accounting of variance rather than an identification of causes, and we do not interpret the federal funds coefficient causally: the Federal Reserve raises rates in response to labor market conditions, so the negative pre-launch rate coefficient reflects policy tightening into a strong labor market rather than tightening reducing unemployment.
The near-collinearity of $gpt$ and $fed$ over this window is itself a limitation and, we would argue, the central point.
The federal funds rate exceeded $3\%$ in only two months before the launch of ChatGPT (October and November 2022), so the two regressors separate only weakly in these data, and the $gpt\times fed$ terms are identified from a thin region of the sample.
This is precisely the identification problem the paper documents: a treatment date placed at the launch of ChatGPT is nearly indistinguishable from the monetary regime that began eight months earlier, and any design that dates treatment to November 2022 will attribute the one to the other.
What we can say is that once the rate cycle is in the model, the terms specific to AI exposure after the launch explain no additional variance.
}

\begin{table}[H]
    \centering
    \caption{\revise{\textbf{Macroeconomic conditions, LLM exposure, and unemployment risk.} Nested ordinary least squares models of $\log_{10}$ monthly unemployment risk across major occupation group $\times$ state $\times$ month cells, January 2015 through October 2025. $fed$ is the effective federal funds rate in percentage points~\cite{FEDFUNDS2025}, $llm\in[0,1]$ is the major occupation group's average LLM exposure~\cite{eloundou2024gpts}, and $gpt$ indicates months from December 2022 onward. Standard errors are two-way clustered by state and by month. Model 4 omits $llm\times fed$ to show the coefficient on $llm\times gpt$ that results when the rate cycle is excluded from the interaction structure; every other model is nested in the one that follows it.}}
    \label{tab:uiMacroModels}
    \small
    \setlength{\tabcolsep}{4pt}
    \begin{tabular}{l|lllllll}
    \multicolumn{8}{c}{Dependent Variable: $\log_{10}$ unemployment risk} \\
    Variable & Model 1 & Model 2 & Model 3 & Model 4 & Model 5 & Model 6 & Model 7 \\
    \hline Federal Interest Rate ($fed$) &  & -0.042$^{***}$ & -0.042$^{***}$ & -0.119$^{***}$ & -0.129$^{***}$ & -0.129$^{***}$ & -0.129$^{***}$ \\
    LLM Exposure ($llm$) &  &  & -0.467$^{***}$ & -0.537$^{***}$ & -0.467$^{***}$ & -0.413$^{***}$ &  \\
    Post ChatGPT Indicator ($gpt$) &  &  &  & 0.366$^{***}$ & 0.068$^{***}$ & 0.068$^{***}$ & 0.068$^{***}$ \\
    $llm \times fed$ &  &  & 0.075$^{***}$ &  & 0.076$^{***}$ & 0.119$^{***}$ & 0.120$^{***}$ \\
    $llm \times gpt$ &  &  &  & 0.261$^{***}$ &  & -0.110 & -0.110 \\
    $gpt \times fed$ &  &  &  &  & 0.117$^{***}$ & 0.117$^{***}$ & 0.117$^{***}$ \\
    $gpt \times fed \times llm$ &  &  &  &  &  & -0.033$^{*  }$ & -0.034$^{*  }$ \\
    State F.E. & Yes & Yes & Yes & Yes & Yes & Yes & Yes \\
    Month F.E. & Yes & Yes & Yes & Yes & Yes & Yes & Yes \\
    Major SOC F.E. & No & No & No & No & No & No & Yes \\
    \hline
    $R^2$ & 0.325 & 0.350 & 0.379 & 0.395 & 0.399 & 0.399 & 0.560 \\
    adj. $R^2$ & 0.325 & 0.349 & 0.378 & 0.395 & 0.398 & 0.399 & 0.560 \\
    \hline \multicolumn{8}{c}{$p_{val}<0.1^*$, $p_{val}<0.01^{**}$, $p_{val}<0.001^{***}$} \\ \hline
    \end{tabular}
\end{table}

\clearpage
\section{Estimating College-Graduate Job Seeking Times}
\label{si:sec:jobDelay}

We first filter to individuals in the Revelio Labs worker profile data who ``graduated'' (i.e., completed their terminal degree) in 2017 through 2024.
\revise{Graduation cohorts 2017 through 2024 therefore define the observation period for the worker profile data, and each graduate is followed for up to three years after degree completion.}
We then filter each individual's work history to their first job after degree completion restricting to within three years of degree completion. 
When an exact graduation is not provided, we assume a May graduation in the year of the individual's terminal degree completion. 
Each job has an salary estimated by Revelio Labs, which we convert to real 2015 dollars, and an AI-exposure score~\cite{eloundou2024gpts}.
Individual $i$'s $\text{jobDelay}_i$ is the number of months between terminal degree completion and start date of each graduate's first post-graduation job.

We study early-career outcomes for college graduates by graduation cohort using an event study focused around the launch of ChatGPT in November 2022.
We estimate an ordinary least-squares regression model of the form
\begin{equation}
\label{si:eq:jobDelay}
\begin{aligned}
    \text{jobDelay}_i &\sim
        \sum_{g\in G}\left(\beta_g\cdot I_g
        + \beta_{g,llm}\cdot I_g\cdot llm_i\right)
        +\beta_{\text{openings}}\cdot \text{jobOpeningRate}_i \\
        &+ \sum_{s\in \text{State}} \beta_s\cdot I_s
        + \sum_{n\in \text{NAICS2}} \beta_n\cdot I_n
        + \sum_{fos\in \text{FieldOfStudy}} \beta_{fos}\cdot I_{fos} \\
        &+ \sum_{d\in \text{Degrees}}\beta_d\cdot I_d
        + \sum_{u\in \text{Universities}}\beta_U\cdot I_u
\end{aligned}
\end{equation}
where $\beta$ represent regression coefficients, $I_g$ is an indicator variable for the year of terminal degree completion, $llm_i$ is the LLM exposure of the first job after graduation (i.e., according to six-digit SOC code), $\text{jobOpeningRate}_i$ is the job opening rate in the state and month of the job according to BLS JOLTS, $I_s$ is an indicator variable for the state of the job, $I_n$ is an indicator variable for the sector of the job (i.e., two-digit NAICS code), $I_{fos}$ is the field of study of the individual's terminal degree (i.e., two-digit codes from the Classification of Instructional Programs (CIP) used by the National Center for Education Statistics), $I_d$ is an indicator variable for the type of terminals degree (i.e., GED, associates, bachelors, masters, MBA, or doctoral), and $I_u$ is an indicator variable for the university or college from which the individual earned their terminal degree (e.g., University of Pittsburgh). 
Each regression contains 2,665,306 individuals (i.e., rows of data).
We conduct an additional but similar regression swapping LLM exposure ($llm_i$) for the estimated salary of the first job ($salary_i$).
\revise{Estimates of both specifications are reported in Table~\ref{si:tbl:jobDelayRegression}.}

\revise{
Because the size of this sample and its coverage bear on how the regression results should be read, we document here how the analytic sample is constructed and what it contains.
The processed Revelio extract contains $40{,}106{,}792$ position records held by $10{,}584{,}980$ workers with a degree record, each record carrying a Revelio salary estimate, a six-digit SOC code, a NAICS sector, a state, a start date, and the degree record of the worker who holds it.
The restrictions described above reduce this to one position per graduate; Table~\ref{si:tbl:revelioSample} reports the number of position records and the number of distinct graduates surviving each restriction in turn.
The binding restrictions are the graduation-cohort window and the requirement that the position begin on or after May of the graduation year, which together remove workers whose terminal degree predates 2017 and positions held before degree completion.
The final two rows are worth separating: $2{,}665{,}306$ graduates hold $3{,}380{,}832$ positions that begin on their earliest post-graduation start date, because a graduate who begins two positions in the same month contributes more than one record, and we retain one position per graduate.
This is the source of the sample size of $2{,}665{,}306$ that each regression reports: it is the number of graduates, not a rounded or sampled figure.
}

\revise{
Table~\ref{si:tbl:revelioContinuous} reports the distribution of every continuous variable entering the regression.
The median graduate begins their first post-graduation position $3.1$ months after May of the graduation year, the mean is $6.0$ months, and the distribution is right-skewed, with a $90$th percentile of $15.3$ months; $13.7\%$ of graduates begin a position in the month of graduation itself.
First-job salaries have a median of \$$51{,}102$ in real 2015 dollars, and first-job LLM exposure~\cite{eloundou2024gpts} has a mean of $0.460$ on the unit interval, close to the midpoint of the scale.
Table~\ref{si:tbl:revelioCohorts} reports the composition of the sample by terminal degree and by graduation cohort.
Bachelor's degree recipients are $69.8\%$ of the sample and master's recipients a further $21.0\%$.
Cohorts are of similar size from 2017 through 2020 and fall off thereafter, from $469{,}100$ graduates in 2017 to $38{,}690$ in 2024, which is the mechanical consequence of the three-year follow-up window: a graduate of 2024 enters the sample only if they begin a position before the end of the data.
The 2024 cohort is therefore selected toward fast job finders, which is visible in its mean job delay of $0.77$ months, and we include graduation-year fixed effects in every specification for this reason.
}

\revise{
Table~\ref{si:tbl:revelioComposition} reports the composition of graduates' first jobs by field of study, sector, and major occupation group.
The sample covers all $50$ states, $25$ two-digit NAICS sectors, $22$ major occupation groups, $335$ distinct six-digit SOC codes, $38$ two-digit CIP fields of study, and $1{,}535$ degree-granting universities and colleges, with a median of $567$ graduates per institution.
Coverage is broad but not uniform.
The ten largest institutions account for $10.1\%$ of graduates, the largest hundred for $46.5\%$, and the largest five hundred for $86.2\%$; the ten largest states account for $58.4\%$.
Business, management, and marketing is the largest field of study at $22.5\%$ of graduates.
Mean first-job LLM exposure varies substantially across these groupings, from $0.281$ in healthcare practitioner occupations to $0.781$ in computer and mathematical occupations, and this variation is what identifies the coefficients on $llm$.
}

\revise{
On representativeness, we compare the field-of-study distribution of Revelio graduates against the National Survey of College Graduates in SI Section~\ref{si:sec:graduateReweight}, and re-estimate our regressions on a sample reweighted to match the survey.
Compared to the NSCG, the Revelio data over-represent business and communications majors and under-represent engineering and education majors, and the reweighted estimates leave our conclusions unchanged (Table~\ref{si:tbl:eduExposureWeighted}).
That comparison is limited to bachelor's degree recipients in the biennial survey years and to the field-of-study margin, so it should be read as evidence that our results are not driven by the composition of majors rather than as a general claim of representativeness.
Three limitations remain and we state them plainly.
Worker profiles are self-reported and are a non-probability sample of the workforce, so graduates who do not maintain a profile are absent and workers in occupations where profiles are less common are under-represented.
Salaries are estimated by Revelio Labs rather than observed, which is why we also report time to first job, a quantity that depends only on the dates a graduate records.
And the profiles carry no demographic information beyond education and employment history, so we can neither weight on nor report descriptive statistics for characteristics such as sex, race, or parental education.
}

\revise{
Table~\ref{si:tbl:jobDelayRegression} reports the estimates of Eq.~\ref{si:eq:jobDelay} alongside those of the parallel specification that replaces LLM exposure with the log of the estimated first-job salary.
The two columns are the models plotted in Fig.~\ref{fig:jobDelay}A and Fig.~\ref{fig:jobDelay}B respectively.
Both are estimated by ordinary least squares on all $2{,}665{,}306$ graduates, and standard errors, shown in parentheses, are the classical ordinary least squares standard errors.
Two conventions are needed to read the interaction coefficients.
First, $llm_i$ enters the interaction on its original scale, so $I_g\times llm$ is the difference in months to first job between a graduate of cohort $g$ whose first job is fully LLM-exposed ($llm=1$) and one whose first job is unexposed ($llm=0$); this is the quantity plotted in Fig.~\ref{fig:jobDelay}A.
Second, $\log(salary_i)$ is centered at its sample mean before it is interacted, so $I_g\times\log(salary)$ is the additional months to first job associated with a first-job salary one log point above the sample mean for cohort $g$.
The graduation-year indicators and the five fixed-effect blocks are each entered as complete sets of indicators with no intercept, so their levels are identified only up to constants that offset across blocks; we therefore report them as included rather than as individual coefficients.
The interaction terms and the job opening rate are not affected by this and are reported in full.
The five fixed-effect blocks contribute the bulk of the estimated parameters, the large majority of them university indicators.
}

\revise{
The estimates reproduce the pattern shown in Fig.~\ref{fig:jobDelay}.
Within every cohort that graduated before the launch of ChatGPT, graduates whose first job was LLM-exposed reached that job sooner than their classmates, by between $0.44$ months (2020) and $0.98$ months (2019).
The sign reverses for the 2023 and 2024 cohorts, for whom a fully exposed first job is associated with $0.49$ and $0.58$ additional months of job seeking.
The salary specification shows no comparable reversal: the coefficient on $I_g\times\log(salary)$ is positive in every cohort and its 2023 and 2024 values ($0.90$ and $0.87$) are close to those of 2017 through 2019 ($1.11$, $1.36$, and $1.09$).
The job opening rate enters with a coefficient of about $2.3$ months per percentage point in both specifications, so the state-month labor market conditions a graduate faces account for more variation in job seeking time than exposure does.
Both models explain a modest share of the variance in individual job seeking times, with $R^2$ of $0.148$ and $0.151$; job delay is dominated by individual circumstances that administrative and profile data do not record, and we make no claim to predict it.
}

\revise{
The number of observations is identical across specifications because it is the count of graduates in the analytic sample rather than a draw from it, as Table~\ref{si:tbl:revelioSample} traces.
A sample of this size does raise the question of whether ordinary least squares remains an appropriate model.
Consistency and unbiasedness do not deteriorate as $N$ grows; what changes is that standard errors shrink until statistical significance stops discriminating between findings, so the interpretive burden shifts from significance to magnitude.
We therefore read these coefficients in months rather than in stars.
Against a mean job delay of $6.04$ months with a standard deviation of $6.66$ months (Table~\ref{si:tbl:revelioContinuous}), a pre-launch exposure advantage of roughly half a month to one month, and a post-launch penalty of roughly half a month, are small relative to the dispersion of individual outcomes and are best read as shifts in a cohort-level average.
The quantity that carries the paper's argument is not the size of any single coefficient but the reversal of sign between cohorts, and the year-to-year movement before 2023 (for example from $-0.98$ in 2019 to $-0.44$ in 2020) is itself a useful gauge of how much these estimates move for reasons unrelated to LLMs.
}

\revise{
To check directly that the estimates are not an artifact of the sample size, we re-estimate both specifications on $50$ independent random subsamples of $50{,}000$ graduates each, drawn without replacement from the analytic sample.
Each draw is roughly $1.9\%$ of the data and refits the identical specification, including the full set of university indicators, which is thinly populated at this sample size: the median institution contributes about eleven graduates to a draw, and the number of estimated parameters falls by about a fifth, to a mean of $1{,}403$, because a draw contains fewer distinct institutions than the full sample.
Draws are seeded so that the results are reproducible.
Table~\ref{si:tbl:jobDelaySubsample} reports, for each graduation cohort, the full-sample estimate, the mean and standard deviation of the estimates across draws, the mean of the standard errors estimated within draws, the share of draws whose estimate carries the same sign as the full-sample estimate, and the share significant at the $5\%$ level.
}

\revise{
Three features of that table answer the question.
First, the subsample estimates are centered on the full-sample estimates: across all sixteen interaction terms, the mean across draws differs from the full-sample estimate by at most $0.11$ months, and the mean of the $2023$ LLM-exposure interaction is $0.40$ months against a full-sample value of $0.49$.
Second, the standard errors behave as sampling theory requires.
A sample $53$ times smaller should inflate standard errors by $\sqrt{2{,}665{,}306/50{,}000}=7.30$, and the mean within-draw standard error is $7.38$ times the full-sample standard error in the LLM-exposure specification and $7.37$ times it in the salary specification; the dispersion of estimates across draws is of the same order.
The standard errors reported in Table~\ref{si:tbl:jobDelayRegression} are therefore small because the sample is large, not because the model understates uncertainty.
Third, the pattern the paper reports survives resampling even though individual coefficients often do not reach significance in a draw: every pre-$2023$ LLM-exposure interaction is negative in $82\%$ of draws, the $2023$ interaction is positive in $90\%$ and the $2024$ interaction in $92\%$, the $2023$ estimate exceeds every pre-$2023$ estimate in $94\%$ of draws, and the average of the $2023$ and $2024$ estimates exceeds the average of the $2017$ through $2022$ estimates in every one of the $50$ draws.
The salary specification shows no such reversal in any draw.
}

\revise{
Two caveats belong with this exercise.
Individual significance falls sharply in the subsamples, as it must: the $2023$ LLM-exposure interaction is significant at the $5\%$ level in only $4\%$ of draws and the $2024$ interaction in none, because an effect of half a month cannot be resolved against a standard error of roughly half a month.
This is the expected consequence of estimating the same effect on $1.9\%$ of the data and is the reason we do not rest the argument on any single coefficient.
And fitting roughly $1{,}400$ parameters to $50{,}000$ observations makes the design nearly singular in places: $R^2$ rises from $0.148$ in the full sample to a mean of $0.174$ across draws, which is overfitting rather than improved explanation, and one of the $900$ subsample coefficient estimates had no computable standard error.
Neither affects the full-sample estimates, which are fit on all $2{,}665{,}306$ observations.
}

\subsection{\revise{LLM exposure and occupational wages}}
\label{si:sec:exposureWage}
\revise{
The comparison in Fig.~\ref{fig:jobDelay}B places first-job salary beside LLM exposure, so it is worth stating how the two are related across occupations rather than leaving a reader to infer it.
We take the O*NET-occupation exposure scores of SI Section~\ref{si:sec:exposure} and aggregate them one step further, to six-digit SOC occupations, by an equal-weighted mean across the O*NET occupations sharing a SOC code; the wage of an occupation is the mean annual wage published for it in the May 2021 BLS Occupational Employment and Wage Statistics, the same wage source used by Eloundou et al.~\cite{eloundou2024gpts}, converted to real 2015 dollars.
Of the $923$ O*NET occupations carrying an exposure score, $767$ match an OES detailed occupation, together covering $130.6$ million employed workers; $321$ of those occupations are entered by graduates in our sample, accounting for $95.2\%$ of graduates.
Table~\ref{si:tbl:exposureWageDist} reports the distribution of both variables and Table~\ref{si:tbl:exposureWage} reports how they move together.
}

\revise{
Exposure and pay are positively but only moderately related.
Across the $767$ occupations, the correlation between exposure and the log of the mean annual wage is $0.417$ (Spearman $0.484$), and weighting occupations by employment raises it to $0.485$.
Within the occupations our graduates enter the correlation is $0.376$, or $0.486$ when occupations are weighted by the number of graduates entering them, and at the level of individual graduates the correlation between first-job exposure and log first-job salary is $0.339$.
For the same $321$ occupations, the OES mean annual wage and the median Revelio first-job salary correlate $0.751$, which is a useful external check on the salary estimates discussed above: the two measures order occupations similarly although one is an administrative average over all incumbents and the other an estimate for new graduates.
}

\revise{
The relationship is also not monotonic at the top of either variable.
Mean exposure rises across wage deciles from $0.182$ in the lowest decile to $0.501$ in the ninth and then falls to $0.446$ in the highest (Table~\ref{si:tbl:exposureWage}, Panel B).
Read the other way, median wage rises with exposure up to the $0.4$--$0.5$ band, where it peaks at \$$75{,}302$, and then falls back to \$$54{,}510$ among the most exposed occupations, close to the \$$50{,}070$ median across all occupations (Panel A).
The most exposed occupations are therefore middle-earning rather than top-earning, which is the pattern Eloundou et al. describe when they note that their data contain both low-wage occupations with high exposure and high-wage occupations with low exposure alongside an overall positive trend.
Our aggregation to the occupation level is an equal-weighted mean across tasks, whereas theirs gives core tasks twice the weight of supplemental ones, so small differences from their figure are expected.
}

\revise{
This moderate correlation is what makes the salary comparison in Fig.~\ref{fig:jobDelay}B informative rather than redundant.
If exposure and pay were close to interchangeable, the salary panel would restate the exposure panel and its different behavior after 2022 could not be interpreted.
Because the two orderings of occupations agree only in part, the two panels are genuinely distinct comparisons, and the fact that the exposure comparison reverses after 2022 while the salary comparison returns to its pre-pandemic level is informative about which margin moved.
}

\begin{table}[H]
    \centering
    \caption{\revise{\textbf{Construction of the analytic sample of recent graduates.} Each row applies the restriction named in addition to all restrictions above it, and reports the number of position records and the number of distinct graduates that survive. Position counts exceed graduate counts until the final row because a graduate may hold several qualifying positions.}}
    \label{si:tbl:revelioSample}
    \small
    \begin{tabular}{l|rr}
    \hline
    Restriction & Position records & Graduates \\
    \hline
    Processed Revelio position records & 40,106,792 & 10,584,980 \\
    \quad with a Revelio salary estimate & 40,106,792 & 10,584,980 \\
    \quad with a state-month JOLTS job opening rate & 40,106,726 & 10,584,967 \\
    \quad with a 2020 OES occupation mean wage & 40,097,579 & 10,583,651 \\
    \quad terminal degree completed 2017--2024 & 23,093,100 & 5,289,250 \\
    \quad position begins within three years of degree completion & 21,646,015 & 5,249,311 \\
    \quad position begins on or after May of the graduation year & 5,882,259 & 2,665,306 \\
    \quad earliest such position & 3,380,832 & 2,665,306 \\
    \hline
    Analytic sample: one position per graduate & 2,665,306 & 2,665,306 \\
    \hline
    \end{tabular}
\end{table}

\begin{table}[H]
    \centering
    \caption{\revise{\textbf{Continuous variables in the analytic sample of $2{,}665{,}306$ recent graduates.} $\text{jobDelay}$ is measured in months from May of the graduation year. Salaries are the Revelio Labs estimate for the first post-graduation position, converted to real 2015 dollars. LLM exposure is the exposure of the first job's six-digit SOC code~\cite{eloundou2024gpts}. The job opening rate is the BLS JOLTS rate in the state and month of the position.}}
    \label{si:tbl:revelioContinuous}
    \small
    \setlength{\tabcolsep}{5pt}
    \begin{tabular}{l|rrrrrr}
    \hline
    Variable & Mean & S.D. & $p_{25}$ & Median & $p_{75}$ & $p_{90}$ \\
    \hline
    Months to first job ($\text{jobDelay}$) & 6.04 & 6.66 & 1.03 & 3.07 & 8.17 & 15.27 \\
    First-job salary (real 2015 \$) & 58,266 & 31,010 & 36,347 & 51,102 & 72,211 & 95,721 \\
    Log first-job salary (real 2015 \$) & 10.85 & 0.48 & 10.50 & 10.84 & 11.19 & 11.47 \\
    First-job LLM exposure ($llm$) & 0.460 & 0.198 & 0.350 & 0.467 & 0.556 & 0.706 \\
    State-month job opening rate (\%) & 5.17 & 1.24 & 4.20 & 4.80 & 6.10 & 7.00 \\
    \hline
    \end{tabular}
\end{table}

\begin{table}[H]
    \centering
    \caption{\revise{\textbf{Composition of the analytic sample by terminal degree and graduation cohort.} Salary is the median first-job salary in real 2015 dollars; LLM exposure is the mean exposure of the first job~\cite{eloundou2024gpts}. Cohort sizes decline after 2020 because each graduate is followed for at most three years, so later cohorts enter the sample only if they begin a position before the end of the data.}}
    \label{si:tbl:revelioCohorts}
    \small
    \setlength{\tabcolsep}{5pt}
    \begin{tabular}{l|rrrrr}
    \hline
     & Graduates & \% of sample & Mean $\text{jobDelay}$ & Median salary & Mean $llm$ \\
    \hline
    \multicolumn{6}{l}{\emph{Panel A: terminal degree}} \\
    Bachelor & 1,860,686 & 69.8 & 5.78 & 47,186 & 0.453 \\
    Master & 560,733 & 21.0 & 6.30 & 61,399 & 0.488 \\
    Doctor & 92,791 & 3.5 & 5.69 & 74,558 & 0.444 \\
    MBA & 77,161 & 2.9 & 8.33 & 65,377 & 0.491 \\
    Associate & 62,326 & 2.3 & 9.21 & 37,536 & 0.396 \\
    High School & 11,609 & 0.4 & 6.48 & 45,894 & 0.376 \\
    \hline
    \multicolumn{6}{l}{\emph{Panel B: graduation cohort}} \\
    2017 & 469,100 & 17.6 & 7.07 & 52,373 & 0.473 \\
    2018 & 463,466 & 17.4 & 6.52 & 52,386 & 0.470 \\
    2019 & 454,668 & 17.1 & 6.53 & 52,579 & 0.467 \\
    2020 & 421,293 & 15.8 & 7.26 & 52,628 & 0.459 \\
    2021 & 373,347 & 14.0 & 5.45 & 50,890 & 0.448 \\
    2022 & 285,052 & 10.7 & 4.24 & 48,537 & 0.443 \\
    2023 & 159,690 & 6.0 & 2.95 & 44,411 & 0.432 \\
    2024 & 38,690 & 1.5 & 0.77 & 41,645 & 0.457 \\
    \hline
    \end{tabular}
\end{table}

\begin{table}[H]
    \centering
    \caption{\revise{\textbf{Composition of the analytic sample by field of study, sector of first job, and major occupation group of first job.} The ten largest categories are shown in each panel. Codes are two-digit CIP codes for field of study, two-digit NAICS codes for sector, and two-digit SOC codes for major occupation group. Mean $llm$ is the mean LLM exposure of the first job~\cite{eloundou2024gpts}.}}
    \label{si:tbl:revelioComposition}
    \footnotesize
    \setlength{\tabcolsep}{5pt}
    \renewcommand{\arraystretch}{0.9}
    \begin{tabular}{l|l|rrr}
    \hline
    Code & Category & Graduates & \% of sample & Mean $llm$ \\
    \hline
    \multicolumn{5}{l}{\emph{Panel A: field of study of terminal degree (CIP)}} \\
    52 & Business, management, and marketing & 599,420 & 22.5 & 0.487 \\
    51 & Health professions & 267,285 & 10.0 & 0.342 \\
    14 & Engineering & 213,092 & 8.0 & 0.502 \\
    30 & Multi/interdisciplinary studies & 207,590 & 7.8 & 0.620 \\
    45 & Social sciences & 155,950 & 5.9 & 0.468 \\
    09 & Communication and journalism & 155,286 & 5.8 & 0.516 \\
    50 & Visual and performing arts & 150,350 & 5.6 & 0.446 \\
    42 & Psychology & 119,974 & 4.5 & 0.385 \\
    26 & Biological and biomedical sciences & 114,408 & 4.3 & 0.375 \\
    13 & Education & 99,890 & 3.7 & 0.379 \\
    \hline
    \multicolumn{5}{l}{\emph{Panel B: sector of first job (NAICS)}} \\
    54 & Professional, scientific, and technical services & 369,490 & 13.9 & 0.532 \\
    62 & Health care and social assistance & 326,870 & 12.3 & 0.346 \\
    61 & Educational services & 315,559 & 11.8 & 0.425 \\
    51 & Information & 234,102 & 8.8 & 0.587 \\
    92 & Public administration & 223,162 & 8.4 & 0.411 \\
    52 & Finance and insurance & 205,287 & 7.7 & 0.555 \\
    33 & Manufacturing & 170,012 & 6.4 & 0.511 \\
    45 & Retail trade & 118,947 & 4.5 & 0.429 \\
    81 & Other services & 96,920 & 3.6 & 0.426 \\
    71 & Arts, entertainment, and recreation & 86,268 & 3.2 & 0.387 \\
    \hline
    \multicolumn{5}{l}{\emph{Panel C: major occupation group of first job (SOC)}} \\
    11 & Management & 300,847 & 11.3 & 0.468 \\
    13 & Business and financial operations & 278,231 & 10.4 & 0.535 \\
    27 & Arts, design, entertainment, sports, and media & 271,772 & 10.2 & 0.550 \\
    15 & Computer and mathematical & 252,639 & 9.5 & 0.781 \\
    25 & Educational instruction and library & 214,719 & 8.1 & 0.370 \\
    43 & Office and administrative support & 206,185 & 7.7 & 0.616 \\
    29 & Healthcare practitioners and technical & 183,047 & 6.9 & 0.281 \\
    41 & Sales and related & 171,377 & 6.4 & 0.407 \\
    17 & Architecture and engineering & 166,172 & 6.2 & 0.493 \\
    19 & Life, physical, and social science & 135,256 & 5.1 & 0.491 \\
    \hline
    \end{tabular}
\end{table}

\begin{table}[H]
    \centering
    \caption{\revise{\textbf{Estimates of the graduate job seeking time models.} Ordinary least squares estimates of Eq.~\ref{si:eq:jobDelay} (LLM exposure) and of the parallel specification that replaces LLM exposure with the log of the estimated first-job salary, which are the models plotted in Fig.~\ref{fig:jobDelay}A and Fig.~\ref{fig:jobDelay}B. The dependent variable is $\text{jobDelay}_i$ in months. Classical standard errors appear in parentheses. LLM exposure enters on its original $[0,1]$ scale and $\log(salary)$ is centered at its sample mean, as described in the text. The graduation-year indicators and the five fixed-effect blocks are entered as complete sets of indicators without an intercept and are reported as included rather than individually; they are dominated by the roughly fifteen hundred university indicators.}}
    \label{si:tbl:jobDelayRegression}
    \scriptsize
    \setlength{\tabcolsep}{5pt}
    \renewcommand{\arraystretch}{0.9}
    \begin{tabular}{l|rr}
    \hline
    \multicolumn{3}{c}{Dependent variable: $\text{jobDelay}_i$ (months)} \\
    \hline
    Variable & LLM exposure & First-job salary \\
    \hline
    Job opening rate ($\text{jobOpeningRate}$) & 2.298$^{***}$ (0.005) & 2.289$^{***}$ (0.005) \\
    $I_{2017}\times llm$ & -0.585$^{***}$ (0.048) & \\
    $I_{2018}\times llm$ & -0.632$^{***}$ (0.048) & \\
    $I_{2019}\times llm$ & -0.982$^{***}$ (0.048) & \\
    $I_{2020}\times llm$ & -0.436$^{***}$ (0.049) & \\
    $I_{2021}\times llm$ & -0.637$^{***}$ (0.051) & \\
    $I_{2022}\times llm$ & -0.911$^{***}$ (0.057) & \\
    $I_{2023}\times llm$ & 0.486$^{***}$ (0.074) & \\
    $I_{2024}\times llm$ & 0.582$^{***}$ (0.146) & \\
    $I_{2017}\times \log(salary)$ & & 1.112$^{***}$ (0.019) \\
    $I_{2018}\times \log(salary)$ & & 1.361$^{***}$ (0.019) \\
    $I_{2019}\times \log(salary)$ & & 1.095$^{***}$ (0.020) \\
    $I_{2020}\times \log(salary)$ & & 0.568$^{***}$ (0.020) \\
    $I_{2021}\times \log(salary)$ & & 0.409$^{***}$ (0.022) \\
    $I_{2022}\times \log(salary)$ & & 0.332$^{***}$ (0.025) \\
    $I_{2023}\times \log(salary)$ & & 0.898$^{***}$ (0.034) \\
    $I_{2024}\times \log(salary)$ & & 0.872$^{***}$ (0.070) \\
    \hline
    Graduation-year indicators ($I_g$) & Yes & Yes \\
    State F.E. & Yes & Yes \\
    Sector (NAICS) F.E. & Yes & Yes \\
    Field of study (CIP) F.E. & Yes & Yes \\
    Degree F.E. & Yes & Yes \\
    University F.E. & Yes & Yes \\
    \hline
    $R^2$ & 0.148 & 0.151 \\
    adj. $R^2$ & 0.147 & 0.150 \\
    Observations & 2,665,306 & 2,665,306 \\
    \hline \multicolumn{3}{c}{$p_{val}<0.1^*$, $p_{val}<0.01^{**}$, $p_{val}<0.001^{***}$} \\ \hline
    \end{tabular}
\end{table}

\begin{table}[H]
    \centering
    \caption{\revise{\textbf{The job seeking time models re-estimated on random subsamples of $50{,}000$ graduates.} Both specifications of Table~\ref{si:tbl:jobDelayRegression} are refit on $50$ independent subsamples of $50{,}000$ graduates drawn without replacement from the analytic sample, using the identical specification and seeded draws. For each graduation cohort the table reports the full-sample estimate with its standard error, the mean and standard deviation of the estimates across draws, the mean of the standard errors estimated within draws, the share of draws whose estimate carries the same sign as the full-sample estimate, and the share of draws in which the estimate is significant at the $5\%$ level with that sign. A sample $53$ times smaller implies standard errors $\sqrt{2{,}665{,}306/50{,}000}=7.30$ times larger; the realized ratios are $7.38$ (Panel A) and $7.37$ (Panel B). One of the $900$ subsample estimates had no computable standard error and is excluded from the standard error column of its row.}}
    \label{si:tbl:jobDelaySubsample}
    \footnotesize
    \setlength{\tabcolsep}{4pt}
    \renewcommand{\arraystretch}{0.9}
    \begin{tabular}{l|r|rrr|rr}
    \hline
     & Full sample & \multicolumn{3}{c|}{Subsamples ($n=50{,}000$, $50$ draws)} & \multicolumn{2}{c}{Share of draws} \\
    Cohort & Estimate (S.E.) & Mean & S.D. & Mean S.E. & Same sign & $p<0.05$ \\
    \hline
    \multicolumn{7}{l}{\emph{Panel A: graduation cohort $\times$ LLM exposure}} \\
    2017 & -0.585 (0.048) & -0.624 & 0.361 & 0.355 & 94\% & 40\% \\
    2018 & -0.632 (0.048) & -0.741 & 0.418 & 0.355 & 96\% & 48\% \\
    2019 & -0.982 (0.048) & -1.022 & 0.301 & 0.358 & 100\% & 90\% \\
    2020 & -0.436 (0.049) & -0.435 & 0.290 & 0.364 & 92\% & 20\% \\
    2021 & -0.637 (0.051) & -0.642 & 0.322 & 0.376 & 98\% & 40\% \\
    2022 & -0.911 (0.057) & -0.898 & 0.357 & 0.420 & 100\% & 64\% \\
    2023 & 0.486 (0.074) & 0.401 & 0.380 & 0.549 & 90\% & 4\% \\
    2024 & 0.582 (0.146) & 0.541 & 0.367 & 1.076 & 92\% & 0\% \\
    \hline
    \multicolumn{7}{l}{\emph{Panel B: graduation cohort $\times$ log salary}} \\
    2017 & 1.112 (0.019) & 1.150 & 0.155 & 0.144 & 100\% & 100\% \\
    2018 & 1.361 (0.019) & 1.373 & 0.183 & 0.144 & 100\% & 100\% \\
    2019 & 1.095 (0.020) & 1.088 & 0.151 & 0.145 & 100\% & 100\% \\
    2020 & 0.568 (0.020) & 0.573 & 0.101 & 0.150 & 100\% & 100\% \\
    2021 & 0.409 (0.022) & 0.406 & 0.138 & 0.160 & 100\% & 68\% \\
    2022 & 0.332 (0.025) & 0.330 & 0.176 & 0.185 & 100\% & 38\% \\
    2023 & 0.898 (0.034) & 0.903 & 0.172 & 0.252 & 100\% & 100\% \\
    2024 & 0.872 (0.070) & 0.864 & 0.187 & 0.512 & 100\% & 26\% \\
    \hline
    \end{tabular}
\end{table}

\begin{table}[H]
    \centering
    \caption{\revise{\textbf{Distributions of LLM exposure and of occupational wages.} Panel A covers the $767$ detailed occupations that carry both an LLM exposure score~\cite{eloundou2024gpts} and a mean annual wage in the May 2021 BLS Occupational Employment and Wage Statistics, together employing $130.6$ million workers. Panel B restricts to the $321$ of those occupations entered by graduates in our sample, which account for $95.2\%$ of graduates, and adds the median Revelio first-job salary of the graduates entering each occupation. Wages and salaries are in real 2015 dollars. Each row describes the distribution across occupations, not across workers.}}
    \label{si:tbl:exposureWageDist}
    \footnotesize
    \setlength{\tabcolsep}{4pt}
    \renewcommand{\arraystretch}{0.9}
    \begin{tabular}{l|rrrrrrr}
    \hline
    & Mean & S.D. & Min & $p_{25}$ & Median & $p_{75}$ & Max \\
    \hline
    \multicolumn{8}{l}{\emph{Panel A: all $767$ occupations}} \\
    \quad LLM exposure ($llm$) & 0.326 & 0.220 & 0.000 & 0.127 & 0.333 & 0.500 & 1.000 \\
    \quad Mean annual wage (real 2015 \$) & 60,619 & 38,735 & 21,993 & 37,207 & 50,070 & 71,604 & 309,414 \\
    \hline
    \multicolumn{8}{l}{\emph{Panel B: the $321$ occupations entered by graduates}} \\
    \quad LLM exposure ($llm$) & 0.418 & 0.214 & 0.000 & 0.259 & 0.462 & 0.550 & 0.955 \\
    \quad Mean annual wage (real 2015 \$) & 62,350 & 32,525 & 22,727 & 38,383 & 54,764 & 83,444 & 258,925 \\
    \quad Median first-job salary (real 2015 \$) & 52,726 & 17,354 & 22,264 & 40,018 & 48,775 & 64,248 & 110,452 \\
    \hline
    \end{tabular}
\end{table}

\begin{table}[H]
    \centering
    \caption{\revise{\textbf{LLM exposure and occupational wages across the $767$ matched occupations.} Panel A groups occupations by their LLM exposure score~\cite{eloundou2024gpts} and Panel B by decile of their mean annual wage. Employment and wages are from the May 2021 BLS Occupational Employment and Wage Statistics; wages are converted to real 2015 dollars. Median wage is the median across the occupations in the group, so each occupation counts once regardless of size; the employment column reports how many workers those occupations employ. Median wage rises with exposure through the $0.4$--$0.5$ band and falls thereafter, and mean exposure rises across wage deciles through the ninth and falls in the tenth.}}
    \label{si:tbl:exposureWage}
    \footnotesize
    \setlength{\tabcolsep}{5pt}
    \renewcommand{\arraystretch}{0.9}
    \begin{tabular}{l|rrrr}
    \hline
    Group & Occupations & Employment (millions) & Median wage & Mean $llm$ \\
    \hline
    \multicolumn{5}{l}{\emph{Panel A: by LLM exposure band}} \\
    0.0--0.1 & 165 & 20.0 & 38,575 & 0.041 \\
    0.1--0.2 & 94 & 20.2 & 39,449 & 0.147 \\
    0.2--0.3 & 95 & 14.1 & 44,003 & 0.247 \\
    0.3--0.4 & 87 & 18.0 & 55,411 & 0.350 \\
    0.4--0.5 & 132 & 18.3 & 75,302 & 0.445 \\
    0.5--0.6 & 121 & 24.1 & 70,271 & 0.528 \\
    0.6--0.7 & 38 & 9.9 & 51,503 & 0.637 \\
    0.7--1.0 & 35 & 6.1 & 54,510 & 0.824 \\
    \hline
    \multicolumn{5}{l}{\emph{Panel B: by decile of the occupation's mean annual wage}} \\
    1 & 77 & 30.3 & 27,054 & 0.182 \\
    2 & 77 & 19.6 & 32,334 & 0.219 \\
    3 & 76 & 9.6 & 37,207 & 0.237 \\
    4 & 77 & 12.2 & 41,364 & 0.270 \\
    5 & 77 & 7.1 & 47,098 & 0.253 \\
    6 & 76 & 7.5 & 53,536 & 0.315 \\
    7 & 77 & 13.4 & 60,376 & 0.389 \\
    8 & 76 & 10.0 & 71,604 & 0.449 \\
    9 & 77 & 8.9 & 86,329 & 0.501 \\
    10 & 77 & 11.9 & 120,629 & 0.446 \\
    \hline
    \end{tabular}
\end{table}

\clearpage
\section{Tracking Significant Difference in Graduation Cohort Job Gaps}
\label{si:sec:jobGap}
To determine whether the relative change in job-start shares for LLM-exposed jobs differs significantly across graduation cohorts, we construct a month-by-month test based on a Delta-method approximation of the sampling variability of the statistic of interest. 
For a given graduation cohort $g$ in month $t$, let $p(g \mid t, llm)$ denote the share of LLM-exposed jobs started by graduates of class $g$, and let $p(g \mid t)$ denote the corresponding share across all jobs. The quantity of interest is the proportional difference
$$
    \text{jobGap}_{g,llm}(t) = \frac{p(g \mid t, llm) - p(g \mid t)}{p(g \mid t)},
$$
which measures how over- or under-represented graduates of class $g$ are in newly started LLM-exposed jobs relative to the overall job-start distribution. 
Because $p(g \mid t, \text{llm})$ and $p(g \mid t)$ are estimated from finite samples---specifically, binomial proportions based on the counts of LLM-exposed jobs and all jobs, respectively---we approximate their joint sampling distribution using a multivariate normal form and apply the Delta method to obtain an analytic expression for $\operatorname{Var}(jobGap_{g,llm}(t))$. 
Under an independence assumption between the two estimates, this variance reduces to a closed-form expression involving the binomial variances of $\hat p(g \mid t, \text{llm})$ and $\hat p(g \mid t)$.

Having obtained $\widehat{\operatorname{Var}}(jobGap_{g,llm}(t))$ for each graduation cohort and month, we test whether two cohorts $g_1$ and $g_2$ differ significantly by evaluating the null hypothesis $H_0: jobGap_{g_1,llm}(t) = jobGap_{g_2,llm}(t)$ for each month $t$. 
The test statistic is a standard $z$-score,
\[
    z_{g_1,g_2,llm}(t) = \frac{jobGap_{g_1,llm}(t) - jobGap_{g_2,llm}(t)}
    {\sqrt{\widehat{\operatorname{Var}}(jobGap_{g_1,llm}(t)) + \widehat{\operatorname{Var}}(jobGap_{g_2,llm}(t))}},
\]
which compares the estimated difference in representation against its sampling uncertainty. 
The resulting two-sided $p$-value identifies the months in which the two graduation cohorts differ significantly in how strongly they are represented among new LLM-exposed job starts. 
This approach provides a transparent, nonparametric month-by-month comparison that directly targets the statistic used in descriptive figures while maintaining principled statistical inference through variance estimation and standard error propagation.
\revise{We report the resulting $p$-values separately for each decile of first-job LLM exposure, from the least exposed decile to the most exposed (see Fig.~\ref{fig:gradJobGapPvalue_1} through Fig.~\ref{fig:gradJobGapPvalue_10}). Significant differences between the 2021 and 2022 graduation cohorts and earlier reference cohorts concentrate in the most exposed deciles.}

\foreach [count=\i] \n in {1,...,10} {%
    \begin{figure}[H]
        \centering
        \includegraphics[width=\textwidth]{figures/siFigures/gradJobGapPvalues/gradYearJobGap_pvalue_\n.pdf}    
        \caption{
            \textbf{Two-proportion z-tests per month (with multiple-testing correction) comparing graduate cohort's $\text{jobGap}$ by month after May of their graduation year for jobs with LLM exposure decile \n.}
            Lines are colored by reference graduation cohort which are selected to be ones that did not experience the launch of ChatGPT during their early years in the workforce post-graduation.
        }
        \label{fig:gradJobGapPvalue_\n}
    \end{figure}
}

\clearpage
\section{Graduation Cohort Job Gaps by Occupations' AI-Exposure Decile}
\label{si:sec:jobGapAi}
\begin{figure}[H]
    \centering
    \includegraphics[width=\textwidth]{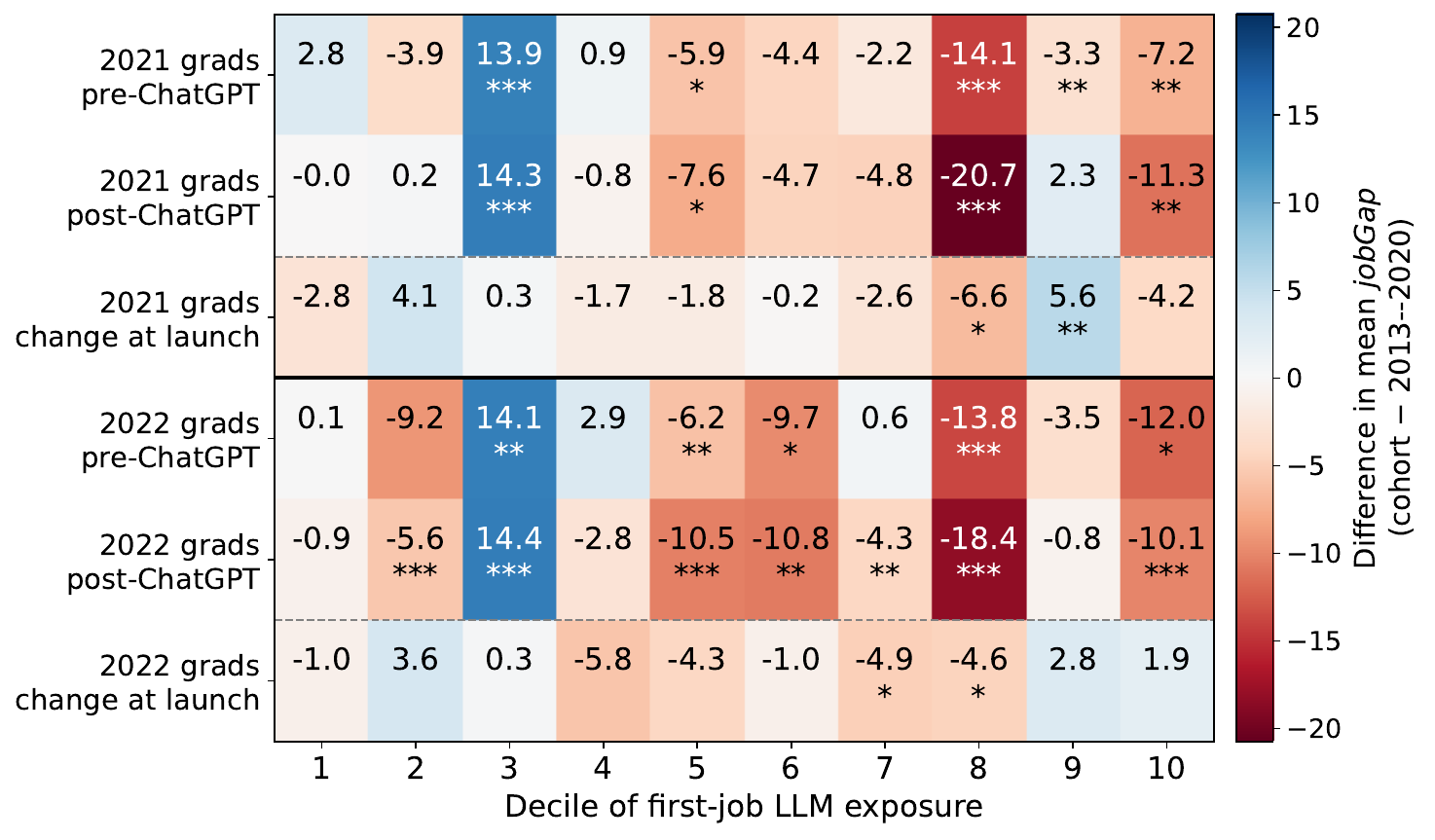}
    \caption{\revise{\textbf{Summary of the decile figures in this section.} Each cell reports the difference in mean $jobGap_{g,llm}(t)$ between a graduation cohort and the 2013--2020 cohorts averaged over the same months since their own graduation, in percentage points, for the decile named on the horizontal axis. The first two rows of each block are the two windows shown in panel B of the corresponding decile figure, and the third is their difference, that is, how much the cohort's gap moves at the launch of ChatGPT. Stars mark two-sample $t$-tests of the cell against zero at the $5\%$, $1\%$, and $0.1\%$ levels. Each column carries a fixed occupational composition, so a column's level reflects which occupations fall in that decile as much as its exposure; the comparison to read is within a column, before against after, and the change rows report exactly that.}}
    \label{si:fig:jobGapHeatmap_exposure}
\end{figure}

\revise{This section disaggregates the job gap statistic $jobGap_{g,llm}(t)$ defined in SI Section~\ref{si:sec:jobGap} by decile of first-job LLM exposure~\cite{eloundou2024gpts} rather than by quintile. For each decile, we plot the share of jobs started by each graduation cohort in each month since May of the graduation year against the overall share of jobs started in that month (see Fig.~\ref{fig:gradJobGap_exposure_1} through Fig.~\ref{fig:gradJobGap_exposure_10}). The under-representation of the 2021 and 2022 cohorts grows across deciles, and is largest in the most LLM-exposed occupations. Each figure pairs the month-by-month series in panel A with a summary in panel B that averages the same statistic over the months before and after the launch of ChatGPT, so that the timing of a change can be read directly rather than traced along the series. Figure~\ref{si:fig:jobGapHeatmap_exposure} collects those summaries for all ten deciles in one place.}

\foreach \n/\m in {1/0,2/1,3/2,4/3,5/4,6/5,7/6,8/7,9/8,10/9} {%
    \begin{figure}[H]
        \centering
        \begin{overpic}[width=\textwidth]{figures/siFigures/jobGaps/jobExposure/gradYearJobShare_exposure_\n.pdf}
            \put(7,24.3){\fbox{\small A}}
        \end{overpic}
        \begin{overpic}[width=\textwidth]{figures/siFigures/jobGaps/jobExposure/gradYearJobShare_exposure_bar_\m.pdf}
            \put(7,26.7){\fbox{\small B}}
        \end{overpic}
        \caption{
            {\bf (A)} For each graduation cohort by month from May of their graduation year, the share of jobs with AI-exposure decile \n~started compared to the overall share of jobs.
            Vertical lines indicate 95\% confidence intervals.
            Dashed lines indicate the launch of ChatGPT for graduation cohorts 2021 and 2022.
            \revise{{\bf (B)} The same statistic averaged over the months before and after the launch of ChatGPT, separately for the 2021 and 2022 graduation cohorts, each compared against the 2013--2020 cohorts averaged over the same months since their own graduation. Error bars are 95\% confidence intervals.}
        }
        \label{fig:gradJobGap_exposure_\n}
    \end{figure}
}

\clearpage
\section{Graduation Cohort Job Gaps by Occupations' Salary Decile}
\label{si:sec:jobGapSalary}
\begin{figure}[H]
    \centering
    \includegraphics[width=\textwidth]{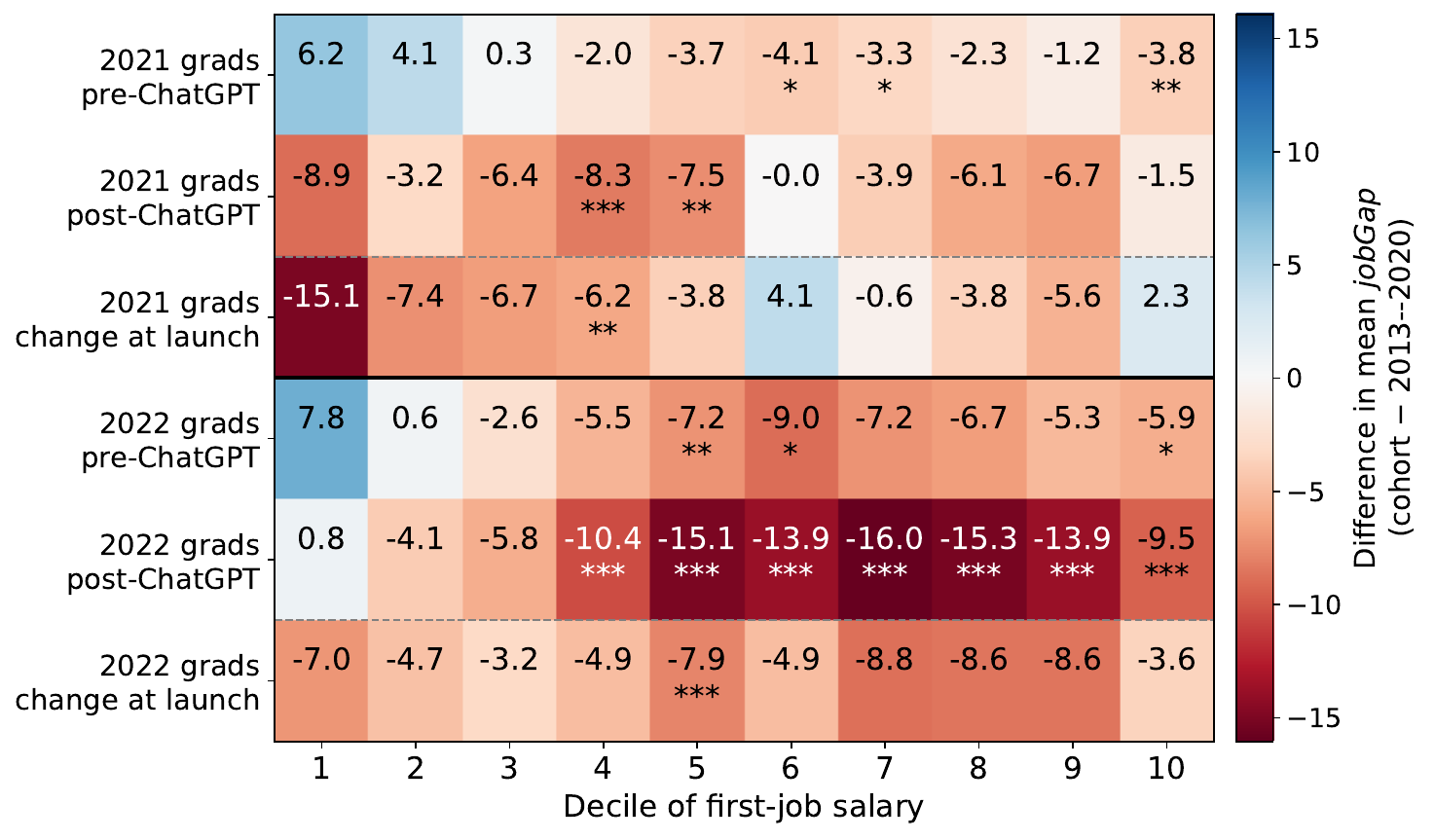}
    \caption{\revise{\textbf{Summary of the decile figures in this section.} As in Figure~\ref{si:fig:jobGapHeatmap_exposure}, but for deciles of first-job salary. Each cell reports the difference in mean $\text{jobGap}$ between a graduation cohort and the 2013--2020 cohorts over the same months since their own graduation, in percentage points; the third row of each block is the change in that difference at the launch of ChatGPT. Stars mark two-sample $t$-tests at the $5\%$, $1\%$, and $0.1\%$ levels.}}
    \label{si:fig:jobGapHeatmap_salary}
\end{figure}

\revise{As a placebo comparison for SI Section~\ref{si:sec:jobGapAi}, we repeat the same disaggregation using deciles of first-job salary (in real 2015 dollars) in place of deciles of LLM exposure (see Fig.~\ref{fig:gradJobGap_salary_1} through Fig.~\ref{fig:gradJobGap_salary_10}). If recent graduation cohorts were simply losing ground in better-paying jobs, the pattern would track the salary deciles as clearly as it tracks the exposure deciles. As in SI Section~\ref{si:sec:jobGapAi}, panel A of each figure gives the month-by-month series and panel B averages the same statistic over the months before and after the launch of ChatGPT, and Figure~\ref{si:fig:jobGapHeatmap_salary} collects those summaries for all ten deciles. It does track the salary deciles: in the months after the launch the 2022 cohort is significantly under-represented in salary deciles four through ten, by between $9.5$ and $16.0$ percentage points, a cleaner gradient than the exposure deciles produce. The cohort shortfall is therefore not specific to LLM-exposed work, a point we return to in the main text.}

\foreach \n/\m in {1/0,2/1,3/2,4/3,5/4,6/5,7/6,8/7,9/8,10/9} {%
    \begin{figure}[H]
        \centering
        \begin{overpic}[width=\textwidth]{figures/siFigures/jobGaps/salary/gradYearJobShare_salary_\n.pdf}
            \put(7,24.3){\fbox{\small A}}
        \end{overpic}
        \begin{overpic}[width=\textwidth]{figures/siFigures/jobGaps/salary/gradYearJobShare_salary_bar_\m.pdf}
            \put(7,26.7){\fbox{\small B}}
        \end{overpic}
        \caption{
            {\bf (A)} For each graduation cohort by month from May of their graduation year, the share of jobs with salary decile \n~started compared to the overall share of jobs.
            Vertical lines indicate 95\% confidence intervals.
            Dashed lines indicate the launch of ChatGPT for graduation cohorts 2021 and 2022.
            All salaries are converted to real 2015 dollars to account for inflation.
            \revise{{\bf (B)} The same statistic averaged over the months before and after the launch of ChatGPT, separately for the 2021 and 2022 graduation cohorts, each compared against the 2013--2020 cohorts averaged over the same months since their own graduation. Error bars are 95\% confidence intervals.}
        }
        \label{fig:gradJobGap_salary_\n}
    \end{figure}
}

\clearpage
\section{Monthly Graduation Cohort Job Shares}
\label{si:sec:gradJobShare}
The pointwise difference test is a simple and widely used method for determining whether two time series of probabilities differ at specific points in time. 
We use it here as a robustness check to compare the job market performance by graduation cohort based on jobs started each month since May of their graduation year (i.e., similar to Fig.~\ref{fig:jobDelay}).
For a given graduation cohort $g$, we create a time series $p(g|t,llm)$ representing the share of jobs started in a given month $t$ (i.e., number of months since May graduation year) with LLM exposure~\cite{eloundou2024gpts} quintile $llm$.
That is, $p(g|t,llm) = x_{g,t,llm} / n_{t_g,llm}$ where $x_{g,t,llm}$ is the number of jobs started in month $t$ within LLM exposure quintile $llm$ by members of graduation cohort $g$ and $n_{t_g,llm}$ is the total number of jobs within LLM exposure quintile $llm$ started in month $t_g$ by anybody.
Here, we index the month $t_g$ with graduation cohort $g$ because the month $t_g$ is relative to the graduation cohort $g$; that is, $t_{2016}=2$ is June 2016, two months from May of the graduation year 2016, while $t_{2022}=2$ is June 2022.
We construct a time series in this way for each graduation cohort from 2013 through 2023.

At each month from May graduation year, we compare the time series for each pre-ChatGPT graduation cohort (i.e., 2013 through 2020) to graduation cohorts that experienced ChatGPT's launch while they were new to the labor market (i.e., 2021, 2022, 2023). 
For each pair of time series $p(g_1|t,llm)$ and $p(g_2|t,llm)$, we calculate the pointwise difference statistic according to
\begin{equation}
    z_{g_1,g_2,t,llm} = \frac{p(g_1|t,llm) - p(g_2|t,llm)}{\sqrt{p_{t,llm}(1 - p_{t,llm})\left(\frac{1}{n_{t_{g1},llm}} + \frac{1}{n_{t_{g_2},llm}}\right)}}
    \label{eq:pointwise}
\end{equation}
where 
\begin{equation}
    p_{t,llm} = \frac{x_{g_1,t,llm}+x_{g_2,t,llm}}{n_{t_{g_1},llm}+n_{t_{g_2},llm}}
\end{equation}
is the pooled proportion.
$z_{g_1,g_2,t,llm}$ measures how many standard errors apart the two observed probabilities are. 
A value of $z_{g_1,g_2,t,llm}$ near zero indicates that the difference between the two estimates is small relative to sampling noise, while large positive or negative values indicate stronger evidence against the null.

Interpreting the statistic proceeds in the same way as any standard normal $z$-test: one computes a one-sided $p$-value to test if $p(g_2|t,llm)$ (here, $g_2\in\{2021,2022,2023\}$) is statistically significantly lower than $p(g_1|t,llm)$ according to $p_{value} = 1 - \Phi(|z_t|)$, where $\Phi(\cdot)$ is the standard normal cumulative distribution function. 
If this $p$-value falls below a predetermined significance level (e.g., $0.05$), then the difference at time $t$ is deemed statistically significant. Because the test is conducted independently at each time point, we plot the sequence of $p_{value}$ values to visualize when the series diverge (see Fig.~\ref{fig:gradSharePvalue_1} through Fig.~\ref{fig:gradSharePvalue_5}). 
In Figure~\ref{fig:gradSharePvalue_5}, we examine graduate cohort performance in the occupations with the greatest AI-exposure and find that graduation cohorts 2021 and 2022 are often significantly under-performing compared to earlier graduation cohorts in their first year post-graduation.
As a kind of placebo test, Figure~\ref{fig:gradSharePvalue_1} compares performance in the jobs with least AI-exposure.
This analysis serves as robustness check on our main analysis in Figure~\ref{fig:jobDelay}.
But this alternative approach does not control for the differences in labor market conditions in month $t_{g_1}$ and $t_{g_2}$ which are in fact different points in time. 

\foreach [count=\i] \n in {1,...,5} {%
    \begin{figure}[H]
        \centering
        \includegraphics[width=\textwidth]{figures/siFigures/gradSharePvalues/gradYearJobShare_pvalue_\n.pdf}    
        \caption{
            \textbf{One-sided pointwise difference p-values testing if graduates from recent years secured a significantly lower share of jobs with LLM exposure quintile \n.}
            Lines are colored by reference graduation cohort which are selected to be ones that did not experience the launch of ChatGPT during their early years in the workforce post-graduation.
        }
        \label{fig:gradSharePvalue_\n}
    \end{figure}
}

\clearpage
\section{Workers' Education Exposure to AI}
\label{si:sec:aiedu}

We augment worker profiles with a dataset~\cite{javadian2024course} of 3 million US higher education course syllabi from the Open Syllabus Project.
\revise{The syllabus data are the third and final observation period in this study: the most recent courses available are those offered in 2017, so education exposure is measured from syllabi offered up to and including 2017, as described among the limitations below.}
For each syllabus $s$, we compute a vector $\vec{t}_s\in\mathbb{R}^{|T|}$ of semantic similarities between the learning objectives of the course and each O*NET task $t\in T$.
Let $\vec{e}\in[0,1]^{|T|}$ be the task–level exposure vector~\cite{eloundou2024gpts} under GPT-4, Rubric 1 $\beta$, containing a 1 if a task is exposed to automation from LLMs.
We define the syllabus–level education exposure as the dot product
$
    edu_s \;=\;  \vec{t}_s\cdot\vec{e}.
$
Finally, given a single worker profile, we estimate the worker's education exposure as the average $edu_s$ for syllabi from the profile's education field of study, degree-granting institution, and the four years leading up to and including their graduation year.

\subsection{\revise{How to read the education exposure score}}
\label{si:sec:eduScale}

\revise{Because $edu_s$ is a dot product of similarities with exposure weights, it is not a count of tasks, and we describe it here in units a reader can interpret.
Each element of $\vec{t}_s$ is a similarity, not an indicator: across the $3{,}074{,}606$ syllabi in the corpus these similarities are non-negative and never exceed $0.66$, with a median across tasks below $0.21$ for every syllabus.
A task therefore contributes a fraction of its exposure weight rather than a whole task, and $edu_s$ should be read as a similarity-weighted measure of how much of the LLM-exposed task space a course's stated objectives cover.}

\revise{Two reference points make the scale concrete.
The exposure vector sums to $\sum_\tau e_\tau = 6{,}164.6$ over the $17{,}992$ O*NET tasks, of which $9{,}869$ ($54.9\%$) carry non-zero exposure and $2{,}463$ ($13.7\%$) are fully exposed; this sum is the value a syllabus would attain if it matched every exposed task perfectly, and it is the natural denominator for expressing $edu_s$ as a share.
Observed scores occupy the lower half of that range: across syllabi, $edu_s$ has mean $1{,}745.2$ and standard deviation $386.1$, a median of $1{,}801$ ($29.2\%$ of the maximum) and a maximum of $3{,}170$ ($51.4\%$).
Tables~\ref{tab:two_CIPCodefull_dot___gpt4_rubric1_beta} and~\ref{tab:university_exposurefull_dot___gpt4_rubric1_beta} report both normalizations alongside the raw score, and the regression tables report $edu$ as a $z$-score so that its coefficients are per standard deviation.}

\revise{It follows that ratios of raw scores are not ratios of task counts.
The most exposed institution in Table~\ref{tab:university_exposurefull_dot___gpt4_rubric1_beta} scores $1.93$ times the least exposed, but this does not mean its graduates encountered twice as many LLM-exposed tasks.
It means that its courses resemble the exposed task space about $1.6$ standard deviations more than the average course, against about $1.4$ standard deviations less for the least exposed, and that the two differ by roughly $18$ percentage points of the attainable maximum.
The ordering of Table~\ref{tab:university_exposurefull_dot___gpt4_rubric1_beta} is informative in this light: community and technical colleges score highest and research universities lowest, because the measure rewards coursework whose objectives are phrased as occupational tasks, which is what the O*NET task statements describe.}

\revise{Table~\ref{si:tbl:syllabusDescriptives} reports the corpus underlying these scores.
The Open Syllabus extract covers $2{,}599$ institutions, $38$ two-digit CIP fields and cohort years $2000$ through $2024$, giving $42{,}828$ distinct university-by-field programs.
Evidence per program is highly uneven: the median program is backed by $6$ syllabi, the first quartile by $2$ and the ninety-fifth percentile by $253$, and $8{,}906$ programs ($20.8\%$) rest on a single syllabus.
Large programs dominate the graduate sample, but scores for thinly covered programs should be read as noisy estimates of curricular content.
Coverage across cohort years is sparser still.
Of the $875{,}955$ university-by-field-by-year cells, $410{,}927$ ($46.9\%$) contain no syllabus within the four-year window ending in that cohort year; every one of these carries the score that the same program takes in a year whose window is non-empty, and a quarter of programs ($25.6\%$) take a single value across all $25$ cohort years.
Educational exposure in this study should therefore be read as a stable characteristic of a program rather than as a year-by-year measurement, which is consistent with our use of it as a baseline attribute measured before generative AI.}

\revise{The derived extract we analyze records each syllabus only through its similarity vector and the resulting scores, so we cannot report syllabus length or word counts.
What it does support is a comparison of the distribution of course-level exposure across institutions and fields, which is what Table~\ref{si:tbl:syllabusDescriptives} provides.}

This approach has several limitations.
First, the syllabus data is a subset of all US higher education.
Second, the most recent syllabi available are for courses offered in 2017.
For workers who graduated after 2017, we default to the relevant syllabi from years 2014-2017 and we consider their education exposure in the context of \emph{status quo} educational activity from before the COVID pandemic and before generative AI.
Despite these limitations, to the best of our knowledge, there is no better data for estimating the difference in teaching across fields of study, educational institutions, and graduation years (see Table~\ref{tab:two_CIPCodefull_dot___gpt4_rubric1_beta} for average $edu_s$ across fields of study and Table~\ref{tab:university_exposurefull_dot___gpt4_rubric1_beta} for the most and least exposed universities.)

\subsection{\revise{Reading the three-way interaction}}
\label{si:sec:eduInteraction}

\begin{figure}[H]
    \centering
    \includegraphics[width=\textwidth]{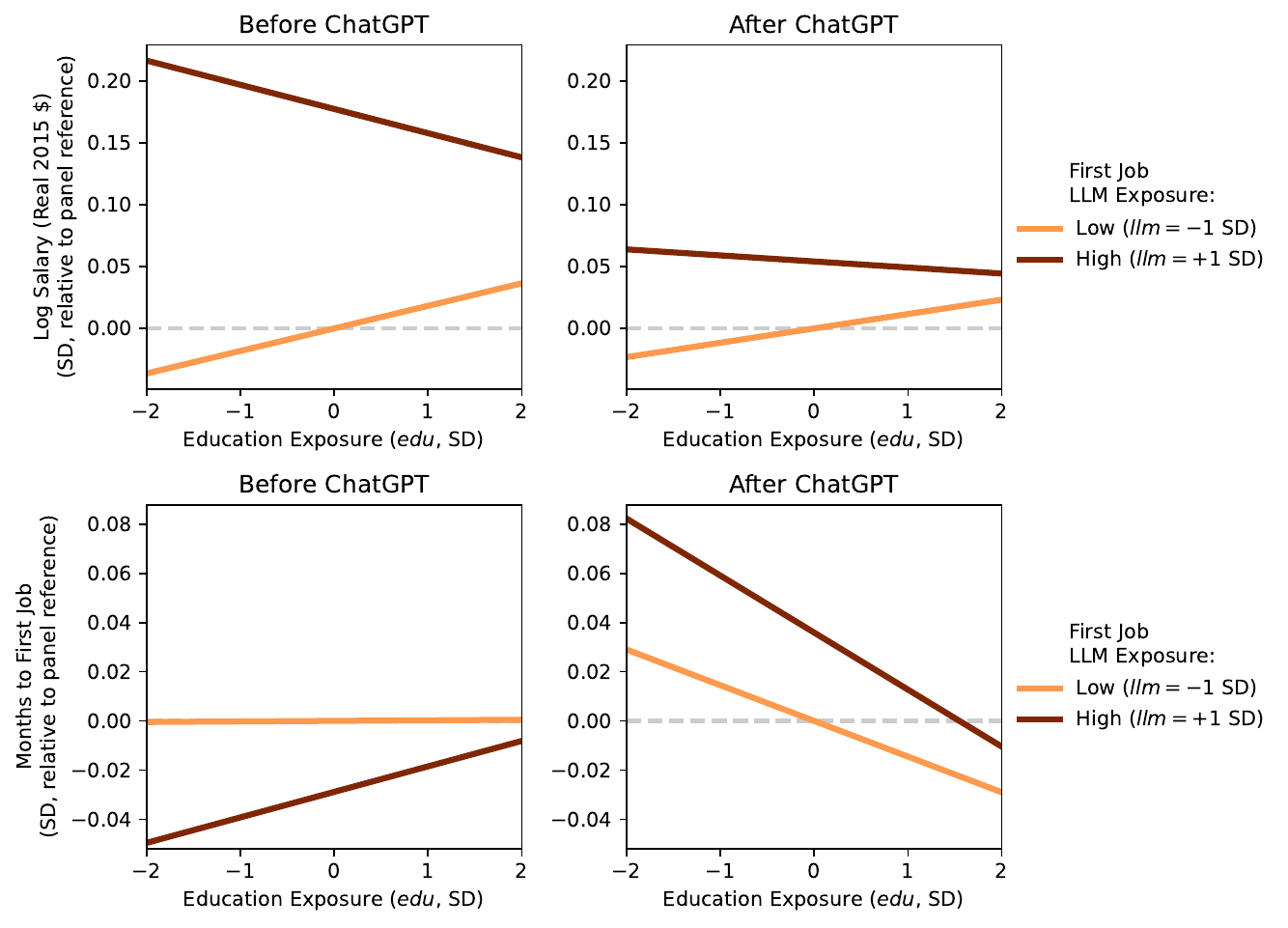}
    \caption{\revise{\textbf{What Models 2 and 4 of Table~\ref{tab:eduExposure} imply for a graduate's first job.} Rows are the two outcomes and columns the two periods, with $gpt=0$ for first jobs starting before December 2022 and $gpt=1$ for those starting on or after it. Within a panel the light line holds the first job's occupational LLM exposure at $llm=-1$ standard deviation and the dark line at $llm=+1$, while education exposure varies along the horizontal axis; all predictors and both dependent variables are $z$-scores, so a slope is the change in standard deviations of the outcome per standard deviation of $edu$. Each panel is drawn relative to its own reference graduate, who has average education exposure and enters a job at $llm=-1$ in the same period, which subtracts the level shift shared by every job starting after the launch and retains every term involving $edu$ or $llm$. The three-way interaction $edu\times llm\times gpt$ is the change between columns in the gap between the two slopes: taking that gap as the high-exposure slope minus the low-exposure slope, it narrows from $-0.038$ to $-0.016$ for salary and reverses from $+0.010$ to $-0.009$ for months to first job. See SI Section~\ref{si:sec:eduInteraction}.}}
    \label{si:fig:eduInteraction}
\end{figure}

\revise{The models in Table~\ref{tab:eduExposure} interact education exposure with both the LLM exposure of the first job and the launch of ChatGPT, and a three-way interaction is difficult to read off a coefficient table.
Figure~\ref{si:fig:eduInteraction} therefore plots what Models 2 and 4 imply for a graduate's first job.
Each panel fixes one outcome and one period, one line fixes the first job's LLM exposure at one standard deviation below the mean and the other one standard deviation above it, and education exposure varies along the horizontal axis.
Because every predictor and both dependent variables are $z$-scores, the slope of a line is the association between a one standard deviation increase in $edu$ and the outcome, measured in standard deviations of that outcome.}

\revise{Each panel is drawn relative to its own reference graduate, who has average education exposure and enters a job one standard deviation below the mean in LLM exposure in that same period.
This removes only the level shift common to every job that starts after the launch, which for months to first job is $2.338$ standard deviations and would otherwise compress everything else in the panel; the association of $llm$ with the outcome, its change at the launch, and every term involving $edu$ are all retained.
The three-way interaction is then the change, from the left panel to the right panel, in the gap between the two slopes.}

\revise{For salary, education exposure is associated with higher pay in jobs of low LLM exposure and lower pay in jobs of high LLM exposure, and the gap between those two slopes narrows from $0.038$ standard deviations before the launch ($+0.018$ against $-0.020$) to $0.016$ after it ($+0.012$ against $-0.005$).
For months to first job, the slopes are $+0.000$ and $+0.010$ before the launch and $-0.015$ and $-0.023$ after it, so search times fall with education exposure only in the later period and fall fastest in the most exposed jobs.
Every interaction term in Models 2 and 4 is significant at the $5\%$ level or better; the weakest is $edu\times gpt$ in Model 2, at $p=0.029$.
The figure also makes the size of these associations visible: across the four standard deviations of education exposure spanned by each panel, every line moves by less than a tenth of a standard deviation of its outcome, which is the sense in which we describe the three-way interactions as modest in the main text.}

\subsection{\revise{Sample size and effect size}}
\label{si:sec:eduEffectSize}

\revise{The models in Table~\ref{tab:eduExposure} are estimated on $1{,}258{,}068$ graduates.
That sample is the $1{,}732{,}499$ first jobs held by $2019$ through $2024$ graduates within three years of graduation, restricted to those whose program carries an education exposure score and whose first-job occupation carries a work-from-home share.
It is smaller than the $2{,}665{,}306$ graduates of SI Table~\ref{si:tbl:jobDelayRegression} because that analysis covers the $2017$ through $2024$ cohorts and requires neither a matched syllabus nor a work-from-home score.
At this size the standard error on a standardized coefficient is of order $0.001$, so statistical significance no longer discriminates between findings and the interpretive burden falls on magnitude.
This section therefore reports the estimates in the units of the dependent variables, the share of variance the education exposure terms add, and the same models re-estimated on small subsamples.}

\revise{The dependent variables have standard deviations of $0.488$ log points and $6.31$ months, so a coefficient of $b$ standard deviations corresponds to $100(e^{0.488b}-1)\%$ of salary or $189b$ days of job search.
Read this way, Model 2 says that a one standard deviation increase in the first job's LLM exposure is associated with $4.4\%$ higher salary before the launch of ChatGPT and $1.3\%$ after it, while the terms involving education exposure are $-0.03\%$ for $edu$, $+0.20\%$ for $edu\times gpt$, $-0.92\%$ for $edu\times llm$ and $+0.52\%$ for $edu\times llm\times gpt$.
Model 4 says that a one standard deviation increase in the first job's LLM exposure is associated with a search $2.7$ days shorter before the launch and $3.4$ days longer after it, with $+1.0$ day for $edu$, $-4.6$ days for $edu\times gpt$, $+1.0$ day for $edu\times llm$ and $-1.8$ days for $edu\times llm\times gpt$.
The largest association involving education exposure is therefore about half a percent of salary or a working week of search time, against $4.4\%$ of salary for the largest association in the table, the LLM exposure premium itself.}

\revise{The same point can be made as variance explained.
Dropping the four education exposure terms from Model 2 lowers $R^2$ from $0.4423$ to $0.4420$, an increment of $0.0003$ (Cohen's $f^2=0.0005$), while dropping the four LLM exposure terms instead lowers it to $0.4381$, an increment of $0.0042$ ($f^2=0.0075$).
In Model 4 the two increments are $0.0001$ and $0.0002$.
These are small by any convention: first-job salary and search time are overwhelmingly explained by the fixed effects for university, field of study, degree, graduation year, state and sector and by the occupation's wage level.
We report the education exposure terms as evidence about the sign and the timing of an association, not as an account of individual outcomes.}

\revise{Finally, we re-estimate both specifications on $50$ independent random subsamples of $50{,}000$ graduates each, drawn without replacement with seeded draws, using the identical specification including every university indicator present in the draw (Table~\ref{si:tab:eduSubsample}).
We match the design of SI Table~\ref{si:tbl:jobDelaySubsample} so that the two exercises can be compared.
The standard errors behave as sampling theory requires: a sample $25$ times smaller should inflate them by $\sqrt{1{,}258{,}068/50{,}000}=5.02$, and the mean within-draw standard error is $5.07$ times the full-sample standard error for both outcomes.
That is direct evidence that the small standard errors of Table~\ref{tab:eduExposure} reflect the size of the sample rather than an understatement of uncertainty.}

\revise{What survives resampling differs across the table, and we state it plainly rather than claim uniform robustness.
For months to first job every term except $edu$ reproduces its full-sample sign in at least $94\%$ of draws ($edu$ does so in $84\%$), and $edu\times gpt$, which carries the finding that education exposure is associated with shorter searches after the launch, does so in every draw and remains significant at the $5\%$ level in $76\%$ of them.
For log salary the two interactions with the job's LLM exposure are comparably stable, $edu\times llm$ reproducing its sign in every draw and $edu\times llm\times gpt$ in $96\%$.
The two weakest terms in the table do not survive: $edu$, which is indistinguishable from zero in the full sample, and $edu\times gpt$, the weakest interaction at $p=0.029$, reproduce their signs in $58\%$ and $60\%$ of draws and are significant in $6\%$ and $4\%$.
An estimate of $0.004$ standard deviations cannot be resolved against a standard error of $0.009$, so this is the expected consequence of the referee's test rather than a failure of it, and it is the reason we do not rest the argument on any single coefficient.}

\revise{Two caveats belong with the exercise.
Fitting roughly $1{,}230$ parameters to $50{,}000$ observations makes the design nearly singular in places, and $R^2$ rises from $0.442$ to a mean of $0.456$ for salary and from $0.375$ to $0.391$ for months, which is overfitting rather than improved explanation.
The draws are also not exactly centered on the full-sample estimates: the mean of $edu\times llm\times gpt$ in Model 4 is $-0.014$ against a full-sample $-0.009$, the largest such gap in either panel, which is smaller than the spread of that term across draws but larger than the standard error of their mean and so reflects the behavior of the term at $n=50{,}000$ rather than sampling noise alone.
And the draws are subsets of the same data, so they test whether the estimates are an artifact of precision, not whether they would replicate in a different sample.
Unlike the job seeking subsamples of SI Table~\ref{si:tbl:jobDelaySubsample}, all $800$ subsample estimates here had a computable standard error.}

\begin{table}[p]
\centering
\caption{\textbf{Average $edu_s$ by field of study (two-digit CIP code) rounded to the nearest integer}.
Fields are ordered from highest to lowest average $edu_s$.
\revise{$edu_s$ is not a count of tasks. It is the sum, over all $17{,}992$ O*NET tasks, of the semantic similarity between a course's learning objectives and the task, weighted by that task's LLM exposure $e_\tau$. The fourth column expresses it as a percentage of $\sum_\tau e_\tau = 6{,}164.6$, the value a course would attain if it matched every LLM-exposed task perfectly, which is the exposure-weighted average similarity between the field's courses and the exposed task space. The fifth column is a $z$-score against the distribution of $edu_s$ over all $3{,}074{,}606$ syllabi (mean $1{,}745.2$, SD $386.1$).}
}
\label{tab:two_CIPCodefull_dot___gpt4_rubric1_beta}
\resizebox{\textwidth}{!}{%
\begin{tabular}{clcccc}
\hline
\textbf{CIP Code} & \multicolumn{1}{c}{\textbf{CIP Title}} & \textbf{Avg. $edu_s$} & \textbf{Syllabus Count} & \textbf{\% of $\sum_\tau e_\tau$} & \textbf{$z$-score} \\ \hline
44 & PUBLIC ADMINISTRATION AND SOCIAL SERVICE PROFESSIONS. & 1996 & 22,171 & 32.4\% & +0.65 \\
52 & BUSINESS, MANAGEMENT, MARKETING, AND RELATED SUPPORT SERVICES. & 1970 & 229,512 & 32.0\% & +0.58 \\
41 & SCIENCE TECHNOLOGIES/TECHNICIANS. & 1899 & 5,970 & 30.8\% & +0.40 \\
51 & HEALTH PROFESSIONS AND RELATED PROGRAMS. & 1870 & 218,642 & 30.3\% & +0.32 \\
47 & MECHANIC AND REPAIR TECHNOLOGIES/TECHNICIANS. & 1862 & 25,364 & 30.2\% & +0.30 \\
19 & FAMILY AND CONSUMER SCIENCES/HUMAN SCIENCES. & 1849 & 54,320 & 30.0\% & +0.27 \\
46 & CONSTRUCTION TRADES. & 1843 & 26,303 & 29.9\% & +0.25 \\
04 & ARCHITECTURE AND RELATED SERVICES. & 1840 & 17,779 & 29.8\% & +0.25 \\
12 & CULINARY, ENTERTAINMENT, AND PERSONAL SERVICES. & 1838 & 16,256 & 29.8\% & +0.24 \\
29 & MILITARY TECHNOLOGIES AND APPLIED SCIENCES. & 1837 & 18,290 & 29.8\% & +0.24 \\
43 & HOMELAND SECURITY, LAW ENFORCEMENT, FIREFIGHTING AND RELATED PROTECTIVE SERVICES. & 1832 & 100,072 & 29.7\% & +0.22 \\
31 & PARKS, RECREATION, LEISURE, FITNESS, AND KINESIOLOGY. & 1832 & 79,198 & 29.7\% & +0.22 \\
09 & COMMUNICATION, JOURNALISM, AND RELATED PROGRAMS. & 1829 & 68,891 & 29.7\% & +0.22 \\
13 & EDUCATION. & 1813 & 391,918 & 29.4\% & +0.18 \\
15 & ENGINEERING/ENGINEERING-RELATED TECHNOLOGIES/TECHNICIANS. & 1812 & 58,488 & 29.4\% & +0.17 \\
25 & LIBRARY SCIENCE. & 1807 & 7,276 & 29.3\% & +0.16 \\
10 & COMMUNICATIONS TECHNOLOGIES/TECHNICIANS AND SUPPORT SERVICES. & 1795 & 34,791 & 29.1\% & +0.13 \\
42 & PSYCHOLOGY. & 1773 & 81,603 & 28.8\% & +0.07 \\
01 & AGRICULTURAL/ANIMAL/PLANT/VETERINARY SCIENCE AND RELATED FIELDS. & 1761 & 28,312 & 28.6\% & +0.04 \\
49 & TRANSPORTATION AND MATERIALS MOVING. & 1758 & 23,422 & 28.5\% & +0.03 \\
11 & COMPUTER AND INFORMATION SCIENCES AND SUPPORT SERVICES. & 1754 & 126,678 & 28.5\% & +0.02 \\
48 & PRECISION PRODUCTION. & 1741 & 4,244 & 28.2\% & -0.01 \\
03 & NATURAL RESOURCES AND CONSERVATION. & 1733 & 8,924 & 28.1\% & -0.03 \\
14 & ENGINEERING. & 1707 & 34,159 & 27.7\% & -0.10 \\
26 & BIOLOGICAL AND BIOMEDICAL SCIENCES. & 1693 & 142,118 & 27.5\% & -0.14 \\
50 & VISUAL AND PERFORMING ARTS. & 1687 & 185,898 & 27.4\% & -0.15 \\
24 & LIBERAL ARTS AND SCIENCES, GENERAL STUDIES AND HUMANITIES. & 1679 & 8,667 & 27.2\% & -0.17 \\
23 & ENGLISH LANGUAGE AND LITERATURE/LETTERS. & 1678 & 186,199 & 27.2\% & -0.17 \\
22 & LEGAL PROFESSIONS AND STUDIES. & 1660 & 44,008 & 26.9\% & -0.22 \\
39 & THEOLOGY AND RELIGIOUS VOCATIONS. & 1659 & 31,960 & 26.9\% & -0.22 \\
45 & SOCIAL SCIENCES. & 1642 & 133,580 & 26.6\% & -0.27 \\
30 & MULTI/INTERDISCIPLINARY STUDIES. & 1632 & 70,527 & 26.5\% & -0.29 \\
40 & PHYSICAL SCIENCES. & 1627 & 133,964 & 26.4\% & -0.31 \\
27 & MATHEMATICS AND STATISTICS. & 1620 & 204,848 & 26.3\% & -0.32 \\
38 & PHILOSOPHY AND RELIGIOUS STUDIES. & 1526 & 40,161 & 24.8\% & -0.57 \\
54 & HISTORY. & 1523 & 51,186 & 24.7\% & -0.58 \\
05 & AREA, ETHNIC, CULTURAL, GENDER, AND GROUP STUDIES. & 1510 & 72,981 & 24.5\% & -0.61 \\
16 & FOREIGN LANGUAGES, LITERATURES, AND LINGUISTICS. & 1497 & 85,926 & 24.3\% & -0.64 \\ \hline
\end{tabular}%
}
\end{table}

\begin{table}[p]
\centering
\caption{\textbf{The ten universities with highest and lowest average $edu_s$ rounded to the nearest integer}.
\revise{The last two columns are defined as in Table~\ref{tab:two_CIPCodefull_dot___gpt4_rubric1_beta}. Because $edu_s$ is a similarity-weighted sum rather than a tally of tasks, the ratio of two universities' scores is not a ratio of task counts: New Mexico Junior College scores $1.93$ times Princeton University, but the interpretation is that its courses sit about $1.6$ SD above the average course and Princeton's about $1.4$ SD below, a gap of roughly $3$ SD in the syllabus distribution. Community and technical colleges occupy the top of this table and research universities the bottom because the measure rewards coursework whose stated objectives resemble workplace tasks.}
}
\label{tab:university_exposurefull_dot___gpt4_rubric1_beta}
\resizebox{\textwidth}{!}{%
\begin{tabular}{lcccc}
\hline
\multicolumn{1}{c}{\textbf{University Name}} & \textbf{Avg. $edu_s$} & \textbf{Syllabus Count} & \textbf{\% of $\sum_\tau e_\tau$} & \textbf{$z$-score} \\ \hline
New Mexico Junior College & 2355 & 16,633 & 38.2\% & +1.58 \\
Modesto Junior College & 2159 & 10,779 & 35.0\% & +1.07 \\
Park University & 2032 & 27,096 & 33.0\% & +0.74 \\
Fresno City College & 2017 & 24,786 & 32.7\% & +0.70 \\
Kentucky Community and Technical College System & 2015 & 10,244 & 32.7\% & +0.70 \\
Rowan-Cabarrus Community College & 1997 & 33,403 & 32.4\% & +0.65 \\
Hartnell College & 1987 & 16,516 & 32.2\% & +0.63 \\
South Texas College & 1977 & 11,998 & 32.1\% & +0.60 \\
Clark State Community College & 1974 & 46,094 & 32.0\% & +0.59 \\
Santa Rosa Junior College & 1959 & 35,613 & 31.8\% & +0.55 \\
$\vdots$ & $\vdots$ & $\vdots$ & $\vdots$ & $\vdots$ \\
Loyola University New Orleans & 1636 & 15,398 & 26.5\% & -0.28 \\
Midwestern State University & 1622 & 25,202 & 26.3\% & -0.32 \\
Minnesota State Colleges and Universities System & 1592 & 22,869 & 25.8\% & -0.40 \\
Alamo Colleges & 1588 & 159,995 & 25.8\% & -0.41 \\
University of Alabama, Tuscaloosa & 1560 & 53,368 & 25.3\% & -0.48 \\
Galveston College & 1476 & 18,908 & 23.9\% & -0.70 \\
Wilkes University & 1338 & 10,008 & 21.7\% & -1.05 \\
University of Michigan–Ann Arbor & 1253 & 23,798 & 20.3\% & -1.27 \\
University of California, Irvine & 1252 & 11,834 & 20.3\% & -1.28 \\
Princeton University & 1223 & 11,667 & 19.8\% & -1.35 \\ \hline
\end{tabular}%
}
\end{table}

\begin{table}[p]
    \centering
    \caption{\revise{\textbf{The syllabus corpus behind the education exposure score.} Panel A reports coverage. Panel B reports how many distinct syllabi back a university-by-field program; because \texttt{num\_syllabi} counts syllabi within the four-year window ending in each cohort year, consecutive years overlap and the distinct count is the sum over years divided by four, which reproduces the syllabus counts in Table~\ref{tab:university_exposurefull_dot___gpt4_rubric1_beta} to within $0.4\%$. Panel C reports the distribution of $edu_s$ at each level of aggregation; the median corresponds to $29.2\%$ of $\sum_\tau e_\tau = 6{,}164.6$ per syllabus, $27.8\%$ per program and $28.9\%$ per graduate, and the per-graduate row is the estimation sample of Table~\ref{tab:eduExposure}. Of the $42{,}828$ programs, $8{,}906$ ($20.8\%$) rest on a single syllabus and $20{,}470$ ($47.8\%$) on five or fewer.}}
    \label{si:tbl:syllabusDescriptives}
    \renewcommand{\arraystretch}{0.9}
    \footnotesize
    \begin{tabular}{lrrrrrr}
        \hline
        \multicolumn{7}{l}{\textbf{Panel A. Coverage}} \\ \hline
        Syllabi & \multicolumn{6}{r}{3,074,606} \\
        Institutions & \multicolumn{6}{r}{2,599} \\
        Fields of study (2-digit CIP) & \multicolumn{6}{r}{38} \\
        University-by-field programs & \multicolumn{6}{r}{42,828} \\
        Cohort years & \multicolumn{6}{r}{2000--2024} \\
        Program-by-year cells & \multicolumn{6}{r}{875,955} \\
        \quad empty in their four-year window & \multicolumn{6}{r}{410,927 (46.9\%)} \\
        \quad programs with a single score across years & \multicolumn{6}{r}{10,946 (25.6\%)} \\
        \hline
        \multicolumn{7}{l}{\textbf{Panel B. Distinct syllabi per program}} \\ \hline
        & Mean & p25 & Median & p75 & p95 & Max \\
        Syllabi & 65.6 & 2 & 6 & 26 & 253 & 65,006 \\
        \hline
        \multicolumn{7}{l}{\textbf{Panel C. Distribution of $edu_s$}} \\ \hline
        & $N$ & Mean & SD & p5 & Median & p95 \\
        Per syllabus & 3,074,606 & 1,745.2 & 386.1 & 1,011.3 & 1,801.2 & 2,279.6 \\
        Per program & 42,828 & 1,682.1 & 316.0 & 1,123.9 & 1,710.9 & 2,134.6 \\
        Per graduate & 1,258,068 & 1,739.5 & 325.9 & 1,150.0 & 1,778.6 & 2,177.8 \\
        Per field of study & 38 & 1,747.4 & 121.9 & 1,523.8 & 1,759.7 & 1,899.8 \\
        \hline
    \end{tabular}
\end{table}

\begin{table}[p]
	\centering
	\caption{
        Education exposure to LLM exposed tasks is associated with shorter job searches after ChatGPT and partially offsets salary penalties in highly exposed occupations.$gpt=1$ if job starts on or after December 2022.
        \revise{$edu$ is education exposure averaged over matched syllabi, a similarity-weighted measure rather than a count of tasks (see SI Sections~\ref{si:sec:aiedu} and~\ref{si:sec:eduScale}).}
        $llm$ is first-job occupation six-digit SOC code LLM exposure~\cite{eloundou2024gpts}. 
        All models include fixed effects for the state and two-digit NAICS code of the job as well as for the terminal degree type (e.g., bachelors or masters), graduation year, college field of study (i.e., two-digit CIP Code), and degree-granting university.
    }
	\label{si:tab:eduExposure} 

    \renewcommand{\arraystretch}{.7}
	\begin{tabular}{l|ll|ll}
        \hline
        Dependent Variable: & \multicolumn{2}{c|}{Log Salary (Real 2015 \$)} & \multicolumn{2}{c}{Months to First Job} \\ \hline
        Variable & Model 1 & Model 2 & Model 3 & Model 4 \\
        \hline Job Opening Rate & 0.014$^{***}$ & 0.014$^{***}$ & 2.994$^{***}$ & 2.994$^{***}$ \\
        Major SOC Log Annual Wage (2020) & 0.330$^{***}$ & 0.312$^{***}$ & -0.224$^{***}$ & -0.202$^{***}$ \\
        Work-From-Home Score (\%) & 0.010$^{***}$ & 0.006$^{***}$ & -0.016$^{***}$ & -0.010$^{***}$ \\
        Post ChatGPT Indicator ($gpt$) & 0.017$^{***}$ & 0.015$^{***}$ & 14.743$^{***}$ & 14.764$^{***}$ \\
        \revise{Education Exposure} ($edu$) & -0.000 & -0.000 & 0.000$^{***}$ & 0.000$^{***}$ \\
        LLM Exposure ($llm$) &  & 0.214$^{***}$ &  & -0.450$^{***}$ \\
        $gpt\times llm$ &  & -0.030$^{***}$ &  & 0.205$^{***}$ \\
        $edu \times gpt$ & 0.003$^{***}$ & 0.002$^{*  }$ & -0.155$^{***}$ & -0.153$^{***}$ \\
        $edu \times llm$ &  & -0.009$^{***}$ &  & 0.032$^{***}$ \\
        $edu \times llm \times gpt$ &  & 0.005$^{***}$ &  & -0.060$^{***}$ \\
        \hline
        State FE & Yes & Yes & Yes & Yes \\
        Sector FE & Yes & Yes & Yes & Yes \\
        Terminal Degree Type FE & Yes & Yes & Yes & Yes \\
        Graduation Year FE & Yes & Yes & Yes & Yes \\
        College Major FE & Yes & Yes & Yes & Yes \\
        University FE & Yes & Yes & Yes & Yes \\
        \hline
        $R^2$ & 0.438 & 0.442 & 0.375 & 0.375 \\
        adj. $R^2$ & 0.437 & 0.442 & 0.374 & 0.374 \\
        \revise{Observations} & \revise{1,258,068} & \revise{1,258,068} & \revise{1,258,068} & \revise{1,258,068} \\
        \hline \multicolumn{5}{c}{$p_{val}<0.1^*$, $p_{val}<0.01^{**}$, $p_{val}<0.001^{***}$} \\ \hline
    \end{tabular}
\end{table}

\begin{table}[H]
    \centering
    \caption{\revise{\textbf{The education exposure models re-estimated on random subsamples of $50{,}000$ graduates.} Models 2 and 4 of Table~\ref{tab:eduExposure} are refit on $50$ independent subsamples of $50{,}000$ graduates drawn without replacement from the estimation sample, using the identical specification and seeded draws. For each term the table reports the full-sample estimate with its standard error, the mean and standard deviation of the estimates across draws, the mean of the standard errors estimated within draws, the share of draws whose estimate carries the same sign as the full-sample estimate, and the share of draws in which the estimate is significant at the $5\%$ level with that sign. All coefficients are in standard deviations of the outcome per standard deviation of the predictor. A sample $25$ times smaller implies standard errors $\sqrt{1{,}258{,}068/50{,}000}=5.02$ times larger; the realized ratio is $5.07$ in both panels. See SI Section~\ref{si:sec:eduEffectSize}.}}
    \label{si:tab:eduSubsample}
    \footnotesize
    \setlength{\tabcolsep}{4pt}
    \renewcommand{\arraystretch}{0.9}
    \begin{tabular}{l|r|rrr|rr}
    \hline
     & Full sample & \multicolumn{3}{c|}{Subsamples ($n=50{,}000$, $50$ draws)} & \multicolumn{2}{c}{Share of draws} \\
    Variable & Estimate (S.E.) & Mean & S.D. & Mean S.E. & Same sign & $p<0.05$ \\
    \hline
    \multicolumn{7}{l}{\emph{Panel A: log first-job salary (Model 2)}} \\
    Education Exposure ($edu$) & -0.001 (0.001) & -0.001 & 0.004 & 0.004 & 58\% & 6\% \\
    LLM Exposure ($llm$) & 0.089 (0.001) & 0.090 & 0.005 & 0.005 & 100\% & 100\% \\
    $edu \times gpt$ & 0.004 (0.002) & 0.004 & 0.009 & 0.009 & 60\% & 4\% \\
    $gpt\times llm$ & -0.062 (0.002) & -0.061 & 0.010 & 0.009 & 100\% & 100\% \\
    $edu \times llm$ & -0.019 (0.001) & -0.020 & 0.004 & 0.004 & 100\% & 100\% \\
    $edu \times llm \times gpt$ & 0.011 (0.002) & 0.012 & 0.007 & 0.009 & 96\% & 28\% \\
    \hline
    \multicolumn{7}{l}{\emph{Panel B: months to first job (Model 4)}} \\
    Education Exposure ($edu$) & 0.005 (0.001) & 0.005 & 0.004 & 0.005 & 84\% & 16\% \\
    LLM Exposure ($llm$) & -0.014 (0.001) & -0.016 & 0.005 & 0.005 & 100\% & 84\% \\
    $edu \times gpt$ & -0.024 (0.002) & -0.025 & 0.009 & 0.010 & 100\% & 76\% \\
    $gpt\times llm$ & 0.032 (0.002) & 0.030 & 0.010 & 0.010 & 100\% & 84\% \\
    $edu \times llm$ & 0.005 (0.001) & 0.005 & 0.004 & 0.004 & 94\% & 24\% \\
    $edu \times llm \times gpt$ & -0.009 (0.002) & -0.014 & 0.008 & 0.010 & 98\% & 30\% \\
    \hline
    \end{tabular}
\end{table}

\clearpage
\section{Weighting Revelio Graduates to Match NCSG Survey Graduates}
\label{si:sec:graduateReweight}
The National Survey of College Graduates (NSCG) is a biennial survey sponsored by the U.S. National Science Foundation's National Center for Science and Engineering Statistics (NCSES) and conducted by the U.S. Census Bureau.
The survey provides nationally oriented data on the characteristics of U.S. college graduates, with particular emphasis on the science and engineering workforce. In the 2023 cycle, the target population is non-institutionalized residents of the United States or Puerto Rico who have earned at least a bachelor's degree and are younger than 76 as of the survey reference date. 
The NSCG is drawn from the American Community Survey (ACS) sampling frame and uses a four-panel rotating panel design in which each new panel completes a baseline interview followed by up to three biennial follow-ups before rotating out. 
Data are collected via a tri-modal approach (web, mail, and CATI), and the public-use files include final survey weights designed to support unbiased population inference under the complex design; for the 2023 cycle, NCSES reports a weighted response rate of 61\% and notes that sampling errors are computed via successive difference replication.

In studies using LinkedIn worker profiles---which are typically a non-probability sample whose composition can differ systematically from the population of college graduates---the NSCG can serve as an external benchmark to re-weight analyses toward population-representative estimates for outcomes observable on LinkedIn. 
Concretely, we compare the distribution of graduates by college field of study in the Revelio/LinkedIn data to the distribution in the corresponding graduation year in NSCG (see Fig.~\ref{si:fig:sampleCompare_2013}-\ref{si:fig:sampleCompare_2023}).
Typically, compared to the NSCG, the Revelio worker profile data over samples Business and Communications majors and under samples Engineering and Education majors.

We can then weight our sample of LinkedIn profiles for the biennial graduation years in the NSCG so the distribution college fields of study are identical and repeat our analysis of earnings and job seeking time in graduates first job after degree completion (compare main text Table 1 to the weighted analysis on a subset of graduation cohorts in Table~\ref{si:tbl:eduExposureWeighted}).
In the weighted analysis, key results from the main regression remain unchanged or are perhaps stronger.
Education exposure ($edu$) is only weakly related to salaries and job seeking times before the launch of ChatGPT\revise{, at no more than $0.013$ SD in either outcome,} but becomes more predictive afterward.
In particular, the three-way interaction ($edu\times llm\times gpt$) continues to indicate the added value of educational exposure for graduates whose first job after degree completion is LLM-exposed after ChatGPT.

\foreach [count=\i] \n in {2013,2015,2017,2021,2023} {%
    \begin{figure}[p]
        \centering
        \includegraphics[width=.6\textwidth]{figures/siFigures/graduationShares/sampleCompare_\n.pdf}    
        \caption{
            \textbf{Comparing the Revelio distribuion of fields of study of bachelors recipients in graduation cohort \n to the distribuion according to the NSCG survey.}
        }
        \label{si:fig:sampleCompare_\n}
    \end{figure}
}

\begin{table}[p]
    \caption{
        \textbf{Similar to Table~\ref{tab:eduExposure} where we restrict the user sample to Bachelors degree recipients from graduation cohorts from 2015, 2019, 2021, and 2023 to match the NSCG survey and run OLS regression weighting individuals' by their major to match graduation distributions over field of study according to NSCG survey.}
        $gpt=1$ if job starts on or after December 2022.
        \revise{$edu$ is education exposure averaged over matched syllabi, a similarity-weighted measure rather than a count of tasks (see SI Sections~\ref{si:sec:aiedu} and~\ref{si:sec:eduScale}).}
        $llm$ is first-job occupation six-digit SOC code LLM exposure~\cite{eloundou2024gpts}. 
        Continuous predictors and dependent variables are centered and standardized (i.e., z-scores), \revise{as in Table~\ref{tab:eduExposure}}.
        All models include fixed effects for the state and two-digit NAICS code of the job as well as for graduation year \revise{and} college field of study (i.e., two-digit CIP Code).
        We do not include a terminal degree type fixed effect here because the sample is restricted to workers whose terminal degree is a bachelors degree.
        \revise{We also omit the university fixed effect used in Table~\ref{tab:eduExposure}, because the reweighting is defined over graduation year and field of study.}
    }
    \label{si:tbl:eduExposureWeighted}
    \centering
    \small
    \renewcommand{\arraystretch}{0.85}
    \begin{tabular}{l|ll|ll}
        \hline
        Dependent Variable: & \multicolumn{2}{c|}{Log Salary (Real 2015 \$)} & \multicolumn{2}{c}{Months to First Job} \\ \hline
        Variable & Model 1 & Model 2 & Model 3 & Model 4 \\
        \hline Job Opening Rate & \revise{0.041}$^{***}$ & \revise{0.047}$^{***}$ & \revise{0.513}$^{***}$ & \revise{0.512}$^{***}$ \\
        Major SOC Log Annual Wage (2020) & \revise{0.393}$^{***}$ & \revise{0.323}$^{***}$ & \revise{-0.038}$^{***}$ & \revise{-0.022}$^{***}$ \\
        Post ChatGPT Indicator ($gpt$) & -0.012 & 0.020$^{*  }$ & \revise{3.276}$^{***}$ & \revise{3.267}$^{***}$ \\
        Education Exposure ($edu$) & \revise{-0.013}$^{***}$ & \revise{-0.004}$^{***}$ & \revise{0.008}$^{***}$ & \revise{0.006}$^{***}$ \\
        $edu \times gpt$ & \revise{0.021}$^{***}$ & 0.006$^{*  }$ & \revise{-0.021}$^{***}$ & -0.017$^{***}$ \\
        LLM Exposure ($llm$) &  & \revise{0.175}$^{***}$ &  & \revise{-0.047}$^{***}$ \\
        $gpt\times llm$ &  & -0.078$^{***}$ &  & 0.052$^{***}$ \\
        $edu \times llm$ &  & -0.037$^{***}$ &  & 0.006$^{***}$ \\
        $edu \times llm \times gpt$ &  & \revise{0.027}$^{***}$ &  & \revise{-0.015}$^{***}$ \\
        \hline
        State FE & Yes & Yes & Yes & Yes \\
        Sector FE & Yes & Yes & Yes & Yes \\
        Graduation Year FE & Yes & Yes & Yes & Yes \\
        College Major FE & Yes & Yes & Yes & Yes \\
        \hline
        $R^2$ & 0.360 & 0.369 & 0.165 & 0.165 \\
        adj. $R^2$ & 0.360 & 0.369 & 0.165 & 0.165 \\
        \hline 
        \multicolumn{5}{l}{Number of Rows: \revise{1,005,481}}\\
        \multicolumn{5}{l}{$p_{val}<0.1^*$, $p_{val}<0.01^{**}$, $p_{val}<0.001^{***}$} \\ \hline
    \end{tabular}
\end{table}


\clearpage
\section{Graduation Cohort Job Gaps by Education Exposure Decile}
\label{si:sec:jobGapEdu}
\begin{figure}[H]
    \centering
    \includegraphics[width=\textwidth]{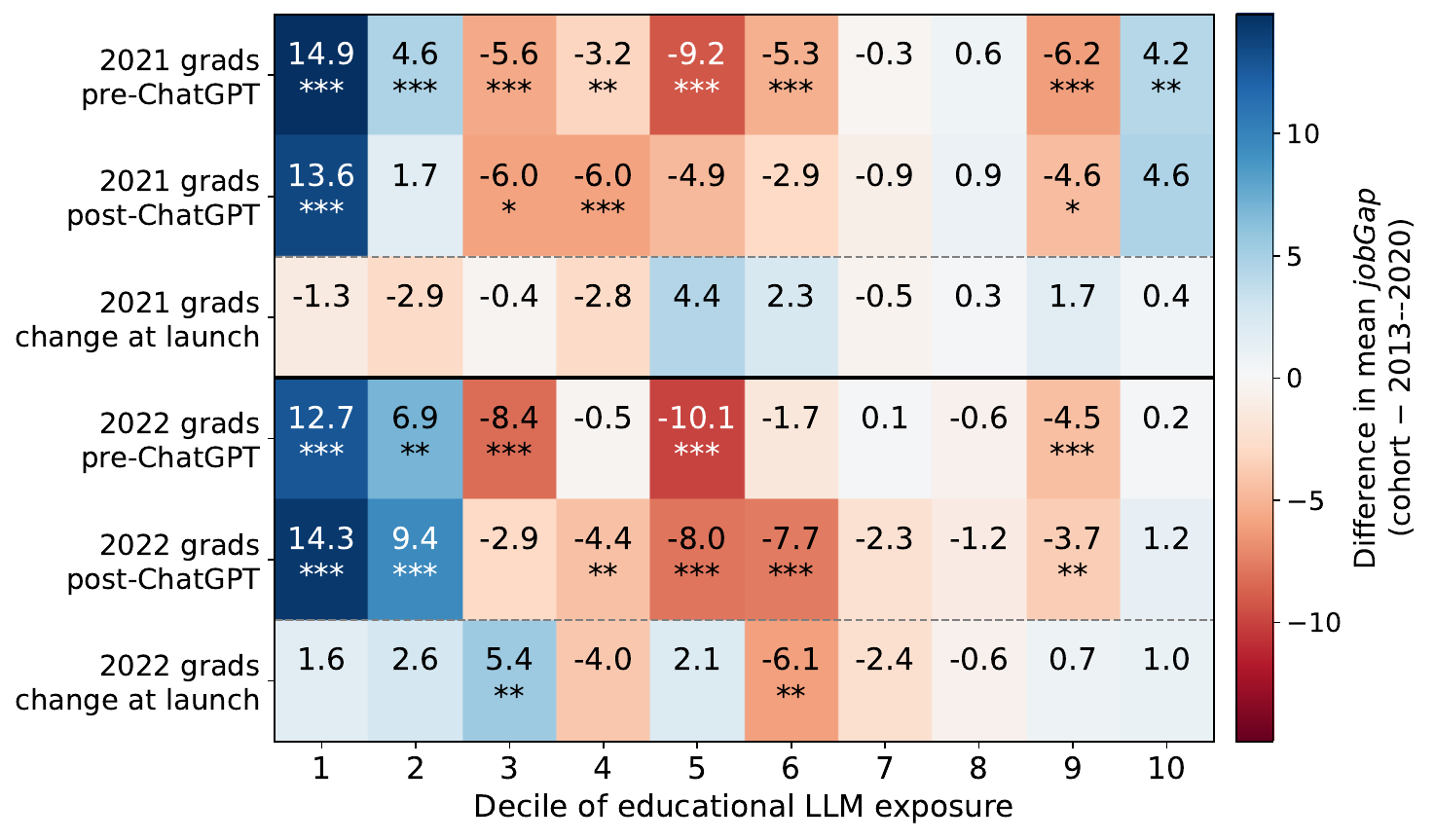}
    \caption{\revise{\textbf{Summary of the decile figures in this section.} As in Figure~\ref{si:fig:jobGapHeatmap_exposure}, but for deciles of educational LLM exposure. Each cell reports the difference in mean $\text{jobGap}$ between a graduation cohort and the 2013--2020 cohorts over the same months since their own graduation, in percentage points; the third row of each block is the change in that difference at the launch of ChatGPT. Stars mark two-sample $t$-tests at the $5\%$, $1\%$, and $0.1\%$ levels.}}
    \label{si:fig:jobGapHeatmap_edu}
\end{figure}

\revise{Finally, we repeat the disaggregation of SI Section~\ref{si:sec:jobGapAi} using deciles of the graduate's educational LLM exposure ($edu$, see SI Section~\ref{si:sec:aiedu}) in place of the LLM exposure of the first job (see Fig.~\ref{fig:gradJobGap_edu_1} through Fig.~\ref{fig:gradJobGap_edu_10}). This asks whether the cohort gap is organized by what graduates were trained on rather than by the jobs they entered. As in SI Section~\ref{si:sec:jobGapAi}, panel A of each figure gives the month-by-month series and panel B averages the same statistic over the months before and after the launch of ChatGPT, and Figure~\ref{si:fig:jobGapHeatmap_edu} collects those summaries for all ten deciles. The education deciles carry large and persistent differences, but almost none of them move at the launch, which is what we would expect if the cohort gap is organized by the jobs graduates enter rather than by what they studied.}

\foreach \n/\m in {1/0,2/1,3/2,4/3,5/4,6/5,7/6,8/7,9/8,10/9} {%
    \begin{figure}[H]
        \centering
        \begin{overpic}[width=\textwidth]{figures/siFigures/jobGaps/eduExposure/gradYearJobShare_edu_\n.pdf}
            \put(7,24.3){\fbox{\small A}}
        \end{overpic}
        \begin{overpic}[width=\textwidth]{figures/siFigures/jobGaps/eduExposure/gradYearJobShare_edu_bar_\m.pdf}
            \put(7,26.9){\fbox{\small B}}
        \end{overpic}
        \caption{
            {\bf (A)} For each graduation cohort by month from May of their graduation year, the share of jobs started by graduates with education AI-exposure decile \n~started compared to the overall share of jobs.
            Vertical lines indicate 95\% confidence intervals.
            Dashed lines indicate the launch of ChatGPT for graduation cohorts 2021 and 2022.
            \revise{{\bf (B)} The same statistic averaged over the months before and after the launch of ChatGPT, separately for the 2021 and 2022 graduation cohorts, each compared against the 2013--2020 cohorts averaged over the same months since their own graduation. Error bars are 95\% confidence intervals.}
        }
        \label{fig:gradJobGap_edu_\n}
    \end{figure}
}

\clearpage
\section{\revise{Time to First Job: Observation Windows and Treatment Timing}}
\label{si:sec:jobDelayRobust}
\revise{
This section examines two features of the time-to-first-job analyses that could shape their results: the observation window available to each graduation cohort, and the definition of the post-ChatGPT indicator in Table~\ref{tab:eduExposure}.
}

\subsection{\revise{A common observation window}}
\revise{
Job start dates in the Revelio extract end in December 2024, so the most recent cohorts cannot be followed for the full three years used in the main analysis: the 2023 cohort is observed for about 19 months after May of its graduation year and the 2024 cohort for about seven.
Truncation removes the slowest job searches, and it can remove more of them from one group of jobs than another.
Graduation-year fixed effects absorb differences across cohorts in the level of job delays, but not truncation that differs within a cohort between more and less exposed jobs.
We therefore re-estimate the model behind Fig.~\ref{fig:jobDelay}A (Eq.~\ref{si:eq:jobDelay}) on windows that every included cohort fully covers: first jobs starting within 12 months of May of the graduation year for the 2017--2023 cohorts ($2{,}161{,}375$ graduates), and within 6 months for the 2017--2024 cohorts ($1{,}707{,}969$ graduates).
Fixed effects and controls are unchanged.
}

\begin{figure}[H]
    \centering
    \includegraphics[width=.7\textwidth]{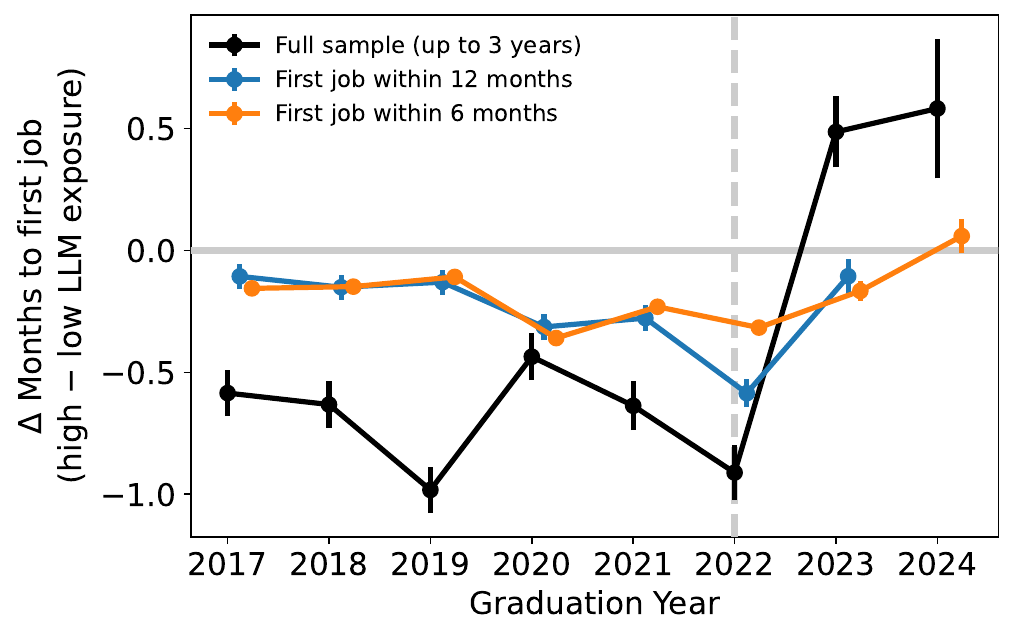}
    \caption{
        \revise{\textbf{On a common observation window, the job search advantage of LLM-exposed first jobs narrows after 2022 but does not reverse.}
        Cohort-specific coefficients $\beta_{g,llm}$ from Eq.~\ref{si:eq:jobDelay}: the difference in months to first job between graduates whose first job had high versus low LLM exposure.
        Black: the full sample of Fig.~\ref{fig:jobDelay}A (first jobs within three years of graduation). Blue: first jobs within 12 months of May of the graduation year, 2017--2023 cohorts. Orange: first jobs within 6 months, 2017--2024 cohorts.
        Bars are 95\% confidence intervals; the dashed line marks the launch of ChatGPT.}
    }
    \label{si:fig:jobDelayWindow}
\end{figure}

\revise{
Fig.~\ref{si:fig:jobDelayWindow} reports the cohort-specific LLM-exposure coefficients $\beta_{g,llm}$.
Coefficients are smaller in every cohort under the shorter windows because the range of possible delays is shorter, so they should be compared within a specification rather than across specifications.
Within the 12-month window, graduates whose first job was highly exposed found jobs faster than their peers in every cohort.
The advantage is largest for the 2020--2022 cohorts ($-0.314$, $-0.277$, and $-0.585$) and returns in 2023 ($-0.106$, SE $0.037$) to the range of the 2017--2019 cohorts ($-0.106$ to $-0.151$).
Within the 6-month window the 2023 coefficient is again negative ($-0.166$, SE $0.020$), and the 2024 coefficient is small and not statistically distinguishable from zero ($0.059$, SE $0.036$).
The positive coefficients for the 2023 and 2024 cohorts in the full sample ($0.486$ and $0.582$) therefore reflect the shorter observation window available to those cohorts rather than a reversal in the job search advantage of exposed jobs.
}

\subsection{\revise{A cohort-level post-ChatGPT indicator}}
\revise{
In Table~\ref{tab:eduExposure} the indicator $gpt$ equals one when the first job starts in or after December 2022.
For the time-to-first-job models (Models 3 and 4) this indicator is partly determined by the outcome: within a graduation cohort, a graduate whose first job starts after the launch has necessarily searched longer than a classmate whose first job starts before it.
Table~\ref{si:tbl:eduCohortGpt} instead assigns post-ChatGPT status by graduation cohort, setting $gpt_{cohort}=1$ for cohorts graduating in 2023 or later, whose job search began after the launch.
Because $gpt_{cohort}$ is collinear with the graduation-year fixed effects, only its interactions enter the model.
Models 1 and 2 use every cohort in the Table~\ref{tab:eduExposure} sample; Models 3 and 4 additionally restrict to first jobs within 12 months of May of the graduation year and to the 2019--2023 cohorts, a window that every included cohort fully covers.
All other terms and fixed effects are those of Table~\ref{tab:eduExposure}, and variables are standardized in the same way.
}

\revise{
The two associations the main text relies on hold under the cohort-level indicator.
Higher education exposure is associated with shorter job searches after the launch ($edu\times gpt_{cohort}$ between $-0.015$ and $-0.022$ across the four models, compared with $-0.024$ in Table~\ref{tab:eduExposure}), and more exposed first jobs are associated with longer searches after the launch ($gpt_{cohort}\times llm$ of $0.043$ in the full sample and $0.013$ on the common window, compared with $0.032$).
The three-way interaction is not robust: it is $-0.005$ ($p<0.1$) in the full sample and $0.001$ (not significant) on the common window, compared with $-0.009$ in Table~\ref{tab:eduExposure}.
We therefore do not rely on the three-way interaction for time to first job.
}

\begin{table}[H]
    \centering
    \caption{
        \revise{\textbf{Time to first job with post-ChatGPT status assigned by graduation cohort.}
        $gpt_{cohort}=1$ for graduation cohorts 2023 and later; it is absorbed by the graduation-year fixed effects, so only its interactions are reported.
        Models 1--2: all 2019--2024 cohorts in the Table~\ref{tab:eduExposure} sample. Models 3--4: first jobs within 12 months of May of the graduation year, 2019--2023 cohorts.
        Continuous predictors and the dependent variable are standardized within each sample, as in Table~\ref{tab:eduExposure}.}
    }
    \label{si:tbl:eduCohortGpt}
    \renewcommand{\arraystretch}{.7}
    \begin{tabular}{l|ll|ll}
        \hline
        Dependent Variable: & \multicolumn{4}{c}{Months to First Job} \\ \hline
        Sample & \multicolumn{2}{c|}{All cohorts} & \multicolumn{2}{c}{12-month window} \\
        Variable & Model 1 & Model 2 & Model 3 & Model 4 \\
        \hline Job Opening Rate & 0.485$^{***}$ & 0.485$^{***}$ & 0.185$^{***}$ & 0.185$^{***}$ \\
        Major SOC Log Annual Wage (2020) & -0.018$^{***}$ & -0.015$^{***}$ & -0.032$^{***}$ & -0.030$^{***}$ \\
        Work-From-Home Score & -0.013$^{***}$ & -0.008$^{***}$ & -0.004$^{***}$ & -0.001 \\
        Education Exposure ($edu$) & 0.003$^{***}$ & 0.003$^{***}$ & 0.005$^{***}$ & 0.005$^{***}$ \\
        LLM Exposure ($llm$) &  & -0.019$^{***}$ &  & -0.010$^{***}$ \\
        $gpt_{cohort}\times llm$ &  & 0.043$^{***}$ &  & 0.013$^{***}$ \\
        $edu \times gpt_{cohort}$ & -0.022$^{***}$ & -0.021$^{***}$ & -0.017$^{***}$ & -0.015$^{***}$ \\
        $edu \times llm$ &  & 0.003$^{***}$ &  & 0.006$^{***}$ \\
        $edu \times llm \times gpt_{cohort}$ &  & -0.005$^{*  }$ &  & 0.001 \\
        \hline
        State FE & Yes & Yes & Yes & Yes \\
        Sector FE & Yes & Yes & Yes & Yes \\
        Terminal Degree Type FE & Yes & Yes & Yes & Yes \\
        Graduation Year FE & Yes & Yes & Yes & Yes \\
        College Major FE & Yes & Yes & Yes & Yes \\
        University FE & Yes & Yes & Yes & Yes \\
        \hline
        $R^2$ & 0.195 & 0.196 & 0.067 & 0.067 \\
        adj. $R^2$ & 0.194 & 0.195 & 0.066 & 0.066 \\
        Observations & 1,258,068 & 1,258,068 & 1,036,282 & 1,036,282 \\
        \hline \multicolumn{5}{c}{$p_{val}<0.1^*$, $p_{val}<0.01^{**}$, $p_{val}<0.001^{***}$} \\ \hline
    \end{tabular}
\end{table}

\end{document}

%% file: beta_exposure_table.tex
{\footnotesize\linespread{1.0}\selectfont
\setlength{\LTcapwidth}{\textwidth}
\renewcommand{\arraystretch}{1.1}
\begin{longtable}{l p{0.45\textwidth} r r}
\caption{\revise{\textbf{Average LLM exposure by detailed occupation.}
    Average LLM exposure~\cite{eloundou2024gpts} of each detailed (six-digit) O*NET-SOC occupation, sorted from least to most exposed.
    The final column reports the number of O*NET workplace tasks contributing to each average.
    Table~\ref{si:tab:majorSocExposure} aggregates these scores to the major occupation group level.}}
\label{si:tab:detailedSocExposure} \\
\hline
Detailed SOC Code & Title & \shortstack[r]{Avg. LLM\\Exposure} & Num. Tasks \\[2pt]
\hline
\endfirsthead
\multicolumn{4}{l}{\itshape Table \thetable\ (continued from previous page)} \\
\hline
Detailed SOC Code & Title & \shortstack[r]{Avg. LLM\\Exposure} & Num. Tasks \\[2pt]
\hline
\endhead
\hline
\multicolumn{4}{r}{\itshape continued on next page} \\
\endfoot
\hline
\endlastfoot
47-2072.00 & Pile Driver Operators & 0.000 & 5 \\
51-3023.00 & Slaughterers and Meat Packers & 0.000 & 15 \\
47-2053.00 & Terrazzo Workers and Finishers & 0.000 & 26 \\
51-9123.00 & Painting, Coating, and Decorating Workers & 0.000 & 13 \\
35-9021.00 & Dishwashers & 0.000 & 13 \\
31-1132.00 & Orderlies & 0.000 & 24 \\
51-3022.00 & Meat, Poultry, and Fish Cutters and Trimmers & 0.000 & 13 \\
47-2071.00 & Paving, Surfacing, and Tamping Equipment Operators & 0.000 & 20 \\
\end{longtable}
}